%% file: main.tex
\documentclass[twocolumn]{bmcart}

\usepackage[utf8]{inputenc} %unicode support

\usepackage{cite}
\usepackage{amsmath,amssymb,amsfonts}
\usepackage{graphicx}
\usepackage{textcomp}
\usepackage[colorinlistoftodos]{todonotes}
\usepackage{subfigure}
\usepackage{comment}
\usepackage{algorithm}
\usepackage{algpseudocode}
\usepackage{multirow}
\usepackage{siunitx} % for scientific unit

\begin{document}

%%% Start of article front matter
\begin{frontmatter}

\begin{fmbox}

\dochead{Research}

%%%%%%%%%%%%%%%%%%%%%%%%%%%%%%%%%%%%%%%%%%%%%%
%%                                          %%
%% Enter the title of your article here     %%
%%                                          %%
%%%%%%%%%%%%%%%%%%%%%%%%%%%%%%%%%%%%%%%%%%%%%%

\title{Using Weaker Consistency Models with Monitoring
and Recovery for Improving Performance of
Key-Value Stores}

%%%%%%%%%%%%%%%%%%%%%%%%%%%%%%%%%%%%%%%%%%%%%%
%%                                          %%
%% Enter the authors here                   %%
%%                                          %%
%% Specify information, if available,       %%
%% in the form:                             %%
%%   <key>={<id1>,<id2>}                    %%
%%   <key>=                                 %%
%% Comment or delete the keys which are     %%
%% not used. Repeat \author command as much %%
%% as required.                             %%
%%                                          %%
%%%%%%%%%%%%%%%%%%%%%%%%%%%%%%%%%%%%%%%%%%%%%%

\author[
   addressref={aff1},                   % id's of addresses, e.g. {aff1,aff2}
   corref={aff1},                       % id of corresponding address, if any
   %noteref={n1},                        % id's of article notes, if any
   email={nguye476@cse.msu.edu}   % email address
]{\inits{DN}\fnm{Duong} \snm{Nguyen}}
\author[
   %addressref={aff1,aff2},
   addressref={aff2},
   %corref={aff2},                       % id of corresponding address, if any
   email={acharapk@buffalo.edu}
]{\inits{AC}\fnm{Aleksey} \snm{Charapko}}
\author[
   %addressref={aff1,aff2},
   addressref={aff1},
   %corref={aff1},                       % id of corresponding address, if any
   email={sandeep@cse.msu.edu}
]{\inits{SK}\fnm{Sandeep S} \snm{Kulkarni}}
\author[
   %addressref={aff1,aff2},
   addressref={aff2},
   %corref={aff2},                       % id of corresponding address, if any
   email={demirbas@buffalo.edu}
]{\inits{MD}\fnm{Murat} \snm{Demirbas}}

%%%%%%%%%%%%%%%%%%%%%%%%%%%%%%%%%%%%%%%%%%%%%%
%%                                          %%
%% Enter the authors' addresses here        %%
%%                                          %%
%% Repeat \address commands as much as      %%
%% required.                                %%
%%                                          %%
%%%%%%%%%%%%%%%%%%%%%%%%%%%%%%%%%%%%%%%%%%%%%%

\address[id=aff1]{%                           % unique id
  \orgname{Michigan State University}, % university, etc
  %\street{Waterloo Road},                     %
  \postcode{MI 48824}                                % post or zip code
  \city{East Lansing},                              % city
  \cny{USA}                                    % country
}
\address[id=aff2]{%
  \orgname{University at Buffalo, SUNY},
  \postcode{NY 14260}
  \city{Buffalo},
  \cny{USA}
}

%%%%%%%%%%%%%%%%%%%%%%%%%%%%%%%%%%%%%%%%%%%%%%
%%                                          %%
%% Enter short notes here                   %%
%%                                          %%
%% Short notes will be after addresses      %%
%% on first page.                           %%
%%                                          %%
%%%%%%%%%%%%%%%%%%%%%%%%%%%%%%%%%%%%%%%%%%%%%%

% \begin{artnotes}
% %\note{Sample of title note}     % note to the article
% \note[id=n1]{Equal contributor} % note, connected to author
% \end{artnotes}

%\end{fmbox}% comment this for two column layout

%%%%%%%%%%%%%%%%%%%%%%%%%%%%%%%%%%%%%%%%%%%%%%
%%                                          %%
%% The Abstract begins here                 %%
%%                                          %%
%% Please refer to the Instructions for     %%
%% authors on http://www.biomedcentral.com  %%
%% and include the section headings         %%
%% accordingly for your article type.       %%
%%                                          %%
%%%%%%%%%%%%%%%%%%%%%%%%%%%%%%%%%%%%%%%%%%%%%%

\begin{abstractbox}

\begin{abstract} % abstract
Consistency properties provided by \textit{most} key-value stores can be classified into sequential consistency and eventual consistency. 
The former is easier to program with but suffers from lower performance whereas the latter suffers from potential anomalies while providing higher performance.   
%Limitations of the CAP theorem imply that if availability is desired in the presence of network partitions, one must sacrifice sequential consistency, a consistency model that is more natural for system design. 
We focus on the problem of what a designer should do if he/she has an algorithm that works correctly with sequential consistency but is faced with an underlying key-value store that provides a weaker (e.g., eventual or causal) consistency. We propose a detect-rollback based approach: The designer identifies a correctness predicate, say $P$, and continues to run the protocol, as our system monitors $P$. If $P$ is violated (because the underlying key-value store provides a weaker consistency), the system rolls back and resumes the computation at a state where $P$ holds.

We evaluate this approach with graph-based applications running on the Voldemort key-value store. Our experiments with deployment on Amazon AWS EC2 instances shows that using eventual consistency with monitoring can provide a $50\%$ -- $80\%$ 
increase in throughput when compared with sequential consistency. We also observe that the overhead of the monitoring itself was low (typically less than $4\%$) and the latency of detecting violations was small. In particular, in a scenario designed to intentionally cause a large number of violations, more than $99.9\%$ of violations were detected in less than \SI{50}{milliseconds}
%$50$ milliseconds 
in regional networks (all clients and servers in the same Amazon AWS region), and in less than \SI{3}{seconds}
%$3$ seconds 
in global networks. 

We find that for some applications, frequent rollback can cause the program using eventual consistency to effectively \textit{stall}. We propose alternate mechanisms for dealing with re-occurring rollbacks. Overall, for applications considered in this paper, we find that even with rollback, eventual consistency provides better performance than using sequential consistency. 

%{\color{red} The final benefit that an application can achieve after accounting for the cost of monitoring and rollback is between $8\%-45\%$, depending on the application.}

\end{abstract}

%%%%%%%%%%%%%%%%%%%%%%%%%%%%%%%%%%%%%%%%%%%%%%
%%                                          %%
%% The keywords begin here                  %%
%%                                          %%
%% Put each keyword in separate \kwd{}.     %%
%%                                          %%
%%%%%%%%%%%%%%%%%%%%%%%%%%%%%%%%%%%%%%%%%%%%%%


% MSC classifications codes, if any
%\begin{keyword}[class=AMS]
%\kwd[Primary ]{}
%\kwd{}
%\kwd[; secondary ]{}
%\end{keyword}

\end{abstractbox}
\end{fmbox}% uncomment this for twcolumn layout

\end{frontmatter}

%%%%%%%%%%%%%%%%%%%%%%%%%%%%%%%%%%%%%%%%%%%%%%
%%                                          %%
%% The Main Body begins here                %%
%%                                          %%
%% Please refer to the instructions for     %%
%% authors on:                              %%
%% http://www.biomedcentral.com/info/authors%%
%% and include the section headings         %%
%% accordingly for your article type.       %%
%%                                          %%
%% See the Results and Discussion section   %%
%% for details on how to create sub-sections%%
%%                                          %%
%% use \cite{...} to cite references        %%
%%  \cite{koon} and                         %%
%%  \cite{oreg,khar,zvai,xjon,schn,pond}    %%
%%  \nocite{smith,marg,hunn,advi,koha,mouse}%%
%%                                          %%
%%%%%%%%%%%%%%%%%%%%%%%%%%%%%%%%%%%%%%%%%%%%%%

\section{Introduction}\label{sec:intro}
\input{Intro.tex}

\section{System Architecture}\label{sec:sys-arch}
\input{SystemArchitecture.tex}

\section{The Problem of Predicate Detection in Distributed Systems}
\label{sec:problem}
\input{pred}

\section{A Framework for Optimistic Execution}
\label{sec:framework}
\input{framework}

\section{Monitoring Module}\label{sec:pred-detect-module}
\input{PredicateDetectionModule.tex}

\section{Rollback from Violations}
\label{sec:rollback}
\input{rollback.tex}

\section{Evaluation Results and Discussion}\label{sec:evaluations}
\input{Experiment-setup.tex}

\input{Experiment-results.tex}

\section{Related Work}\label{sec:related-work}
\input{related.tex}

\section{Conclusion}\label{sec:conclusion}
\input{conclusion.tex}

%%%%%%%%%%%%%%%%%%%%%%%%%%%%%%%%%%%%%%%%%%%%%%
%%                                          %%
%% Backmatter begins here                   %%
%%                                          %%
%%%%%%%%%%%%%%%%%%%%%%%%%%%%%%%%%%%%%%%%%%%%%%

\begin{backmatter}

\section*{Abbreviation}
AWS: Amazon Web Service.  
CAP: CAP theorem, also known as Brewer's theorem; C=Consistency, A = Availability, P = Partition tolerance.
CPU: Central Processing Unit. 
EC2: Amazon Elastic Compute Cloud.  
GP2: Amazon General Purpose SSD storage volume. 
HDFS: Hadoop Distributed File System. 
HLC: Hybrid Logical Clock. 
HVC: Hybrid Vector Clock. 
I/O: Input/Output.
JBCS: Journal of the Brazilian Computer Society.
LADC: Latin-American Symposium on Dependable Computing. 
\si{\ms}: The time unit millisecond.
NP in NP-complete and NP-hard: Non-deterministic Polynomial time. 
\si{ops}: Operations per second.
$P$: we often denote a predicate as $P$ in this manuscript (this is not an abbreviation). 
PC: Personal Computer. 
PT: Physical Time. 
RAM: Random Access Memory. 
\si{\s}: The time unit second.
TAO: Facebook's The Associations and Objects distributed data store. 
VC: Vector Clock. 
VLS: Virtual Lightweight Snapshots. 
XML: Extensible Markup Language. 

\section*{Availability of data and material}
      The source code and experimental results supporting the conclusions of this article are at https://doi.org/10.5281/zenodo.3338381.
      
\section*{Competing interests}
   The authors declare that they have no competing interests.

\section*{Author's contributions}
    All authors have contributed to the methodological and research aspects of the research. The authors have also read and approved the final manuscript.

% \section*{Funding}
%   This work is supported in part by NSF CNS-1329807, NSF CNS-1318678, NSF XPS-1533870, and NSF XPS-1533802.

\section*{Acknowledgement}
    We thank the reviewers of the 8th Latin-American Symposium on Dependable Computing (LADC 2018) for their suggestions on our work.

\section*{Authors' information}
    Duong Nguyen, Michigan State University, nguye476@msu.edu; Aleksey Charapko, University at Buffalo, SUNY, acharapk@buffalo.edu; Sandeep S. Kulkarni, Michigan State University, sandeep@cse.msu.edu; Murat Demirbas, University at Buffalo, SUNY, demirbas@buffalo.edu.

%%%%%%%%%%%%%%%%%%%%%%%%%%%%%%%%%%%%%%%%%%%%%%%%%%%%%%%%%%%%%
%%                  The Bibliography                       %%
%%                                                         %%
%%  Bmc_mathpys.bst  will be used to                       %%
%%  create a .BBL file for submission.                     %%
%%  After submission of the .TEX file,                     %%
%%  you will be prompted to submit your .BBL file.         %%
%%                                                         %%
%%                                                         %%
%%  Note that the displayed Bibliography will not          %%
%%  necessarily be rendered by Latex exactly as specified  %%
%%  in the online Instructions for Authors.                %%
%%                                                         %%
%%%%%%%%%%%%%%%%%%%%%%%%%%%%%%%%%%%%%%%%%%%%%%%%%%%%%%%%%%%%%

% if your bibliography is in bibtex format, use those commands:
\bibliographystyle{bmc-mathphys} % Style BST file (bmc-mathphys, vancouver, spbasic).
\bibliography{duong,acharapk}      % Bibliography file (usually '*.bib' )
% for author-year bibliography (bmc-mathphys or spbasic)
% a) write to bib file (bmc-mathphys only)
% @settings{label, options="nameyear"}
% b) uncomment next line
%\nocite{label}

% or include bibliography directly:
% \begin{thebibliography}
% \bibitem{b1}
% \end{thebibliography}

%%%%%%%%%%%%%%%%%%%%%%%%%%%%%%%%%%%
%%                               %%
%% Figures                       %%
%%                               %%
%% NB: this is for captions and  %%
%% Titles. All graphics must be  %%
%% submitted separately and NOT  %%
%% included in the Tex document  %%
%%                               %%
%%%%%%%%%%%%%%%%%%%%%%%%%%%%%%%%%%%

%%
%% Do not use \listoffigures as most will included as separate files

\section*{Figures}
\setcounter{figure}{0}    

\begin{figure}[h!]
  \caption{\csentence{The detect-rollback approach.}
      When the predicate of interest is violated, system state is restored to the most recent consistent snapshot and the computation resumes from there.}
\end{figure}

\begin{figure}[h!]
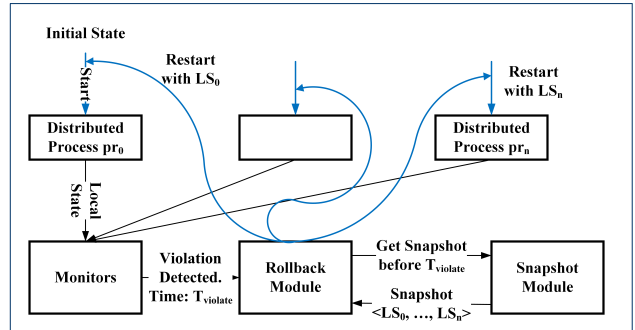

  \caption{\csentence{An overall framework for optimistic execution in key-value store.}
      }
\end{figure}

\begin{figure}[h!]
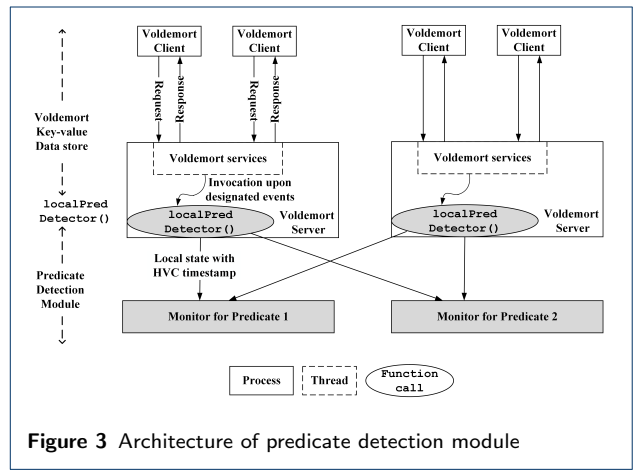

  \caption{\csentence{Architecture of predicate detection module.}
      }
\end{figure}

\begin{figure}[h!]
  \caption{\csentence{Example of predicate in XML format.}
      XML specification for $\neg P \equiv (x_1=1 \land y_1=1) \lor z_2=1$.}
\end{figure}

\begin{figure}[h!]
  \caption{\csentence{Illustration of candidates sent from a server to monitors.}
      Illustration of candidates sent from a server to monitors corresponding to three conjunctive predicates. If the predicate is semilinear, the candidate is always sent upon a PUT request of relevant variables..}
\end{figure}

\begin{figure}[h!]
  \caption{\csentence{Illustration of causality relation under HVC interval
perspective.}
      }
\end{figure}

\begin{figure}[h!]
  \caption{\csentence{Two phases of a task.}
      Two client tasks involved in a violation. Since detection latency is much smaller than the Read phase time, violation will be notified within Read phase of the current task of at least one client..}
\end{figure}

\begin{figure}[h!]
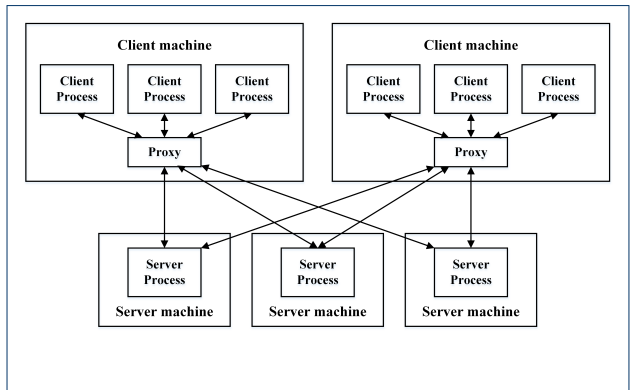

  \caption{\csentence{Simulating network delay using proxies.}
      }
\end{figure}

\begin{figure}[h!]
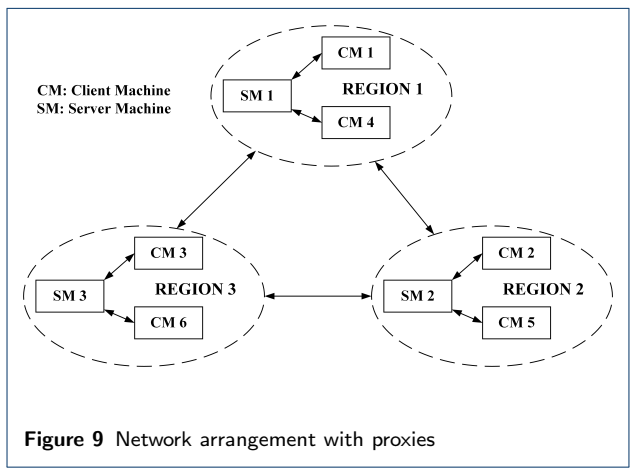

  \caption{\csentence{Network arrangement with proxies.}
      }
\end{figure}

\begin{figure}[h!]
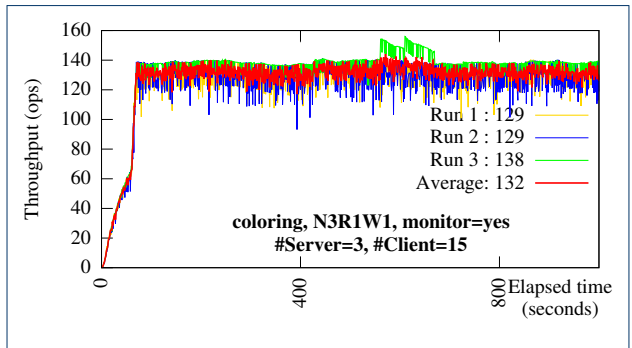

  \caption{\csentence{Illustration of result stabilization.}
      The Social Media Analysis application is run three times on Amazon AWS with monitoring enabled. Number of servers (N ) = 3. Number of clients per server (C/N ) = 5. Aggregated throughput measured by Social Media Analysis application in three different runs and their average is shown. This average is used to represent the stable value of the application throughput.}
\end{figure}

\begin{figure}[h!]
  \caption{\csentence{The benefit and overhead of monitoring in Social Media Analysis application.}
      (AWS) Social Media Analysis application, 3 servers, 15 clients. Throughput improvement compared to R1W3 and R2W2 is \SI{57}{\percent} and \SI{78}{\percent}, respectively, and the overhead of running monitors on each consistency setting (the overhead is less than \SI{2}{\percent}.}
\end{figure}

\begin{figure}[h!]
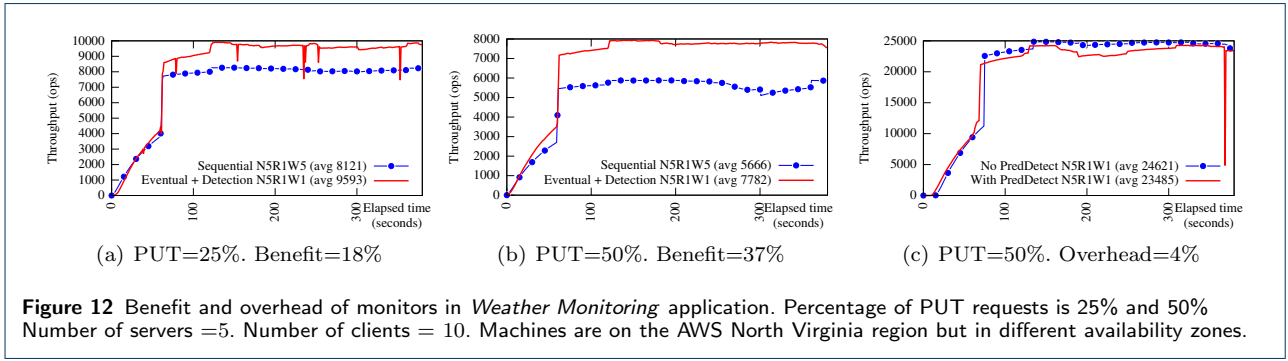

  \caption{\csentence{Benefit and overhead of monitors in Weather Monitoring application.}
      Percentage of PUT requests is \SI{25}{\percent} and \SI{50}{\percent}. Number of servers=5. Number of clients = 10. Machines are on the AWS North Virginia region but in different availability zones.}
\end{figure}

\begin{figure}[h!]
  \caption{\csentence{Effectiveness of livelock handling mechanisms.}
      Number of servers=3, number of clients=30. We observed that adaptive mechanism worked best for \textit{Social Media Analysis} (Figure \ref{fig:fig:livelock-social}), and backoff mechanism worked best for \textit{Weather Monitoring} (Figure \ref{fig:livelock-weather})}
\end{figure}

\begin{figure}[h!]
  \caption{\csentence{The benefit and overhead of Eventual consistency+Rollback vs. Sequential consistency in \textit{Weather Monitoring} application.}
      The inset figure within Figure \ref{fig:node-progress-weather-line-put0.50-delay10ms-lineSize500} is a close-up view showing the impact of rollback. The larger points near the end of each data sequence are where we choose the representative values for the data sequences.}
\end{figure}

\begin{figure}[h!]
  \caption{\csentence{Comparing the completion time of Sequential Consistency (R1W3) vs. Eventual Consistency with rollback and adaptive consistency (R1W1+adaptive) in \textit{Social Media Analysis} application.}
      On a power-law clustering graph, before 90\% of the nodes are processed, R1W1+adaptive progresses about 18\% faster than R1W3. Overall, R1W1+adaptive is $9.5\%$ faster than R1W3. On a regular random graph, the benefit before $90\%$ of the nodes are processed is $26\%$ and the overall benefit is $20.8\%$.}
\end{figure}

%%%%%%%%%%%%%%%%%%%%%%%%%%%%%%%%%%%
%%                               %%
%% Tables                        %%
%%                               %%
%%%%%%%%%%%%%%%%%%%%%%%%%%%%%%%%%%%

%% Use of \listoftables is discouraged.
%%

% \section*{Tables}
% \begin{table}[h!]
% \caption{Sample table title. This is where the description of the table should go.}
%       \begin{tabular}{cccc}
%         \hline
%           & B1  &B2   & B3\\ \hline
%         A1 & 0.1 & 0.2 & 0.3\\
%         A2 & ... & ..  & .\\
%         A3 & ..  & .   & .\\ \hline
%       \end{tabular}
% \end{table}

%%%%%%%%%%%%%%%%%%%%%%%%%%%%%%%%%%%
%%                               %%
%% Additional Files              %%
%%                               %%
%%%%%%%%%%%%%%%%%%%%%%%%%%%%%%%%%%%

% \section*{Additional Files}
%   \subsection*{Additional file 1: Summary of changes between this manuscript and its previous version (reference \cite{NCKD2018LADC})}
%     This file highlights the main difference and the extension between this manuscript and its previous version. This file is provided in pdf format.
    % Additional file descriptions text (including details of how to
    % view the file, if it is in a non-standard format or the file extension).  This might
    % refer to a multi-page table or a figure.

%   \subsection*{Additional file 2 --- Sample additional file title}
%     Additional file descriptions text.

\end{backmatter}

\end{document}

%% file: Intro.tex
Distributed key-value data stores have gained increasing popularity due to their simple data model and high performance \cite{DHJKLPSVV07SOSP}. 
A distributed key-value data store, according to CAP theorem \cite{Brewer00PODC, GL02SIGACT}, cannot simultaneously achieve sequential consistency and availability while tolerating network partitions.
Since fault-tolerance, especially the provision of an acceptable level of service in the presence of node or channel failures,
is a critical dependability requirement of any system, 
network partition tolerance is a necessity.
Hence, it is inevitable to make trade-offs between availability and consistency, resulting in a spectrum of weaker consistency models such as causal consistency and eventual consistency \cite{DIRZ14SOCC, LFKA11SOSP, RDK17SRDS, DHJKLPSVV07SOSP, LM10SIGOPS, Voldemort, SKGFSS12FAST}.

Weaker consistency models are attractive because they have the potential to provide higher throughput and higher customer satisfaction.
On the other hand, weaker consistency models suffer from data conflicts.
Although such data conflicts are infrequent \cite{DHJKLPSVV07SOSP}, such incidences will affect the correctness of the computation and invalidate subsequent results.

Furthermore, developing algorithms for the sequential consistency model is easier than developing those for weaker consistency models. Moreover, since the sequential consistency model is \textit{more natural}, the designer may already have access to an algorithm that is correct only under sequential consistency. Thus, in this case, the question for the designer is what to do \textit{ if the underlying system provides a weaker consistency} or \textit{if the underlying system provides better performance under weaker consistency models}?

As an illustration of such a scenario, consider a distributed computation that relies on a key-value store to arrange exclusive access to a critical resource for the clients. %, as shown in Figure \ref{fig:mutual-exclusion-violation}. 
If the key-value store employs sequential consistency and the clients use Peterson's algorithm, mutual exclusion is guaranteed \cite{BW02PPAM}, but the performance would be impeded due to the strict requirement of sequential consistency. If eventual consistency is adopted, then mutual exclusion is violated.
% when the request from a client to a server is delayed or lost, making the computation results hereinafter unreliable. 

In this case, the designer has two options: (1) Either develop a brand new algorithm that works under eventual consistency, or (2) Run the algorithm by pretending that the underlying system satisfies sequential consistency but monitor it to detect violations of the mutual exclusion requirement. 
In case of the first option, we potentially need to develop a new algorithm for every consistency model used in practice, whereas in case of the second option, the underlying consistency model is irrelevant although we may need to rollback the system to an earlier state if a violation is found. While the rollback in general distributed systems is a challenging task, existing approaches have provided rollback mechanisms for key-value stores with low overhead \cite{CADK17ICDCS}. Moreover, it is possible to develop efficient application-specific rollback algorithms by exploiting the properties of applications.

\begin{figure}[tbp]%[h]
% \vspace{-10pt}
\centering
\includegraphics[width=0.45\textwidth]{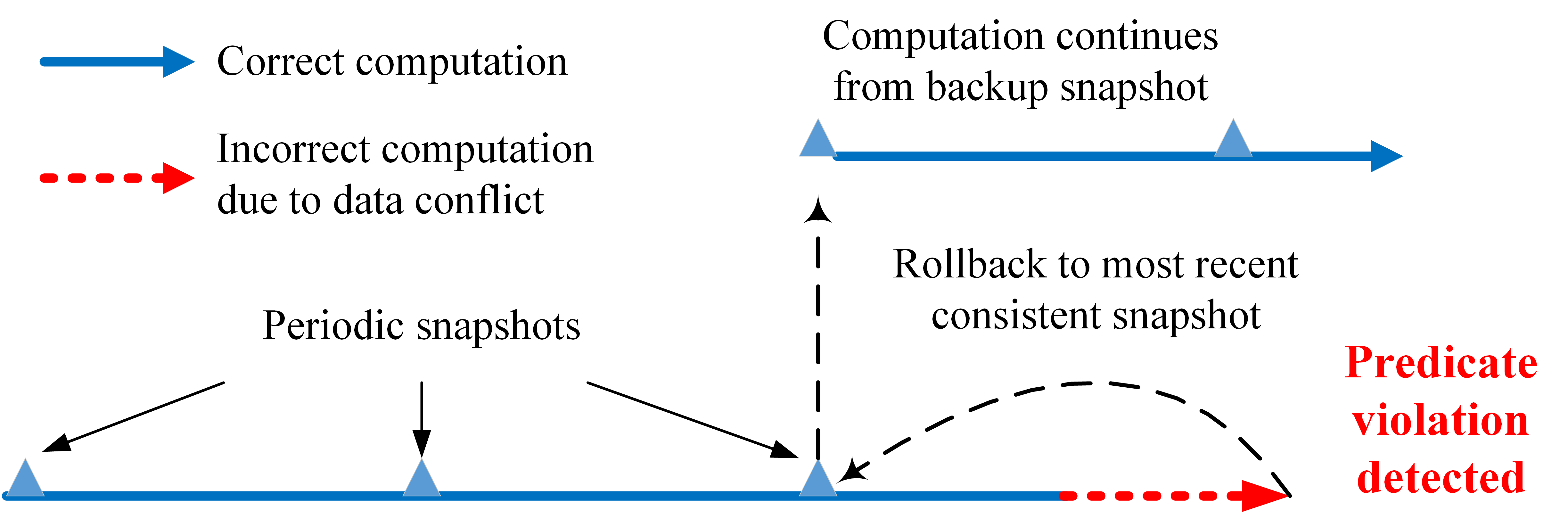}
\caption{The detect-rollback approach: when the predicate of interest is violated, system state is restored to the most recent consistent snapshot and the computation resumes from there.}
%\vspace{-10pt}
\label{fig:detect-recovery-continue}
\end{figure}

The predicate $P$ to monitor depends on the application. For the mutual exclusion application we alluded to above, $P$ might be exclusive access to the shared resource. 
As another example, consider the following. For many distributed graph processing applications, the clients process a given set of graph nodes. Since the state of a node depends on its neighbors, the clients need to coordinate to avoid updating two neighboring nodes simultaneously, otherwise they may read inconsistent information. In this case, predicate $P$ is the conjunction of  smaller predicates and each smaller predicate proscribes the concurrent access to one pair of neighboring graph nodes (Note that pairs of neighboring nodes belonging to the same client do not need monitoring). We note that in a general problem, a smaller predicate may involve any number of processes.
The application will continue executing as long as predicate $P$ is true. If $P$ is violated, the system will be rolled back to an earlier correct state from where subsequent execution will resume (cf. Figure \ref{fig:detect-recovery-continue}).

We require that the monitoring module is non-intrusive, i.e., it allows the underlying system to execute unimpeded. To evaluate the effectiveness of the monitors, we need to identify three parameters: 
(1) The benefit of using the monitors instead of relying on sequential consistency,
(2) The overhead of the monitors, i.e., how the performance is affected when we introduce the monitoring module, and (3) Detection latency of the monitors, i.e., how long the monitors take to detect violation of $P$. (Note that since the monitoring module is non-intrusive, it cannot prevent violation of $P$.)
%
%To evaluate the effectiveness of monitoring, we need to identify two parameters: (1) overhead of the monitor, i.e., how is the performance affected when we introduce the monitor, and (2) latency of the monitor, i.e., how long does the monitor take to detect violation of $P$. (Note that since the monitor is non-intrusive, it cannot prevent violation of $P$.)

\if false
\begin{figure}[tbhp]%[h]
 \vspace{-10pt}
\centering
\includegraphics[width=0.8\textwidth]{mutualexclusion.jpg}
\caption{An example of mutual exclusion violation in eventual consistency model. The request \emph{PUT x2=1} from client 2 to server 1 is also delayed, but not shown in the figure due to limited space.}
\vspace{-10pt}
\label{fig:mutual-exclusion-violation}
\end{figure}

\fi

\textbf{Contributions of the paper. }
We implement the monitors for linear and semilinear predicates based on the algorithms in \cite{Garg96, GC95ICDCS, CG98DC} and develop a rollback algorithm for some graph-based applications. We integrate our prototype into LinkedIn's Voldemort key-value store and run experiments on Amazon AWS network. Besides Amazon AWS network, we also run experiments on our local lab network where we can control network condition such as network latency.
%We implement a monitoring module prototype for the Voldemort key-value data store and run experiments on the Amazon AWS EC2 instances. For the module, we develop monitoring algorithms for linear and semilinear predicates based on the algorithms in \cite{Garg96, GC95ICDCS, CG98DC}. 
%
%Our algorithms use Hybrid Vector Clock \cite{DK13LADIS} to help save resources from examining false positive cases thanks to its loose synchronization with physical clock \cite{YNVSD16RV}. 
%
We evaluate our approach by running graph-based applications motivated by the task of \textit{Social Media Analysis} on social graphs and \textit{Weather Monitoring} on planar graphs. 
%
%Besides these experiments, we design some synthetic test cases and set up a local lab network where we can control network condition in order to evaluate the monitoring module in some other aspects such as the impact of workload characteristics and network latency.
%
The source code and experiment results are available at \cite{Nguyen2019jbcs-code}.
The observations from the experiments are as follows:

\begin{itemize}
\item On Amazon AWS network, we run both sequential consistency without the monitors and eventual consistency with the monitors. We observe that --even with the overhead of the monitors-- eventual consistency achieves a higher throughput than sequential consistency does. Specifically, the aggregate client throughput was improved by $50\%$ -- $80\%$ when running \textit{Social Media Analysis} motivated applications, and by \SI{37}{\percent} on \textit{Weather Monitoring} motivated applications.
%
%On Amazon AWS network, we run the social network analysis application both on sequential consistency without the monitoring module and on eventual consistency with the monitoring module. We observe that --even with the overhead of the monitors-- eventual consistency achieves a throughput $50\%$ to $80\%$ higher than that of sequential consistency. 
%
Furthermore, in those experiments, we find that violation of mutual exclusion is not frequent. For example, on \textit{Social Media Analysis}, a violation occurred every \SI{4500}{\s} 
%$4,500$ seconds 
on average, and was detected within
\SI{3}{\s}.
%Hence, the cost of predicate detection and state rollback is outweighed by the benefit of a boosted throughput while the reliability of the computation is still preserved.

\item We also evaluate the overhead of the monitoring module if it is intended solely for debugging or runtime monitoring. We find that when the monitors were used with sequential consistency, the overhead was at most \SI{8}{\percent}. And, for eventual consistency, the overhead was less than \SI{4}{\percent}. 

\item We design test cases with a large number of violations to stress the monitors.
In those test cases, more than \SI{99.9}{\percent} of violations were detected within \SI{50}{\ms} for Amazon AWS regional network (all machines in the same region), and within \SI{3}{\s} for the global network (machines in multiple regions).
%\footnote{Remove:In all cases, the detection latency is within $17$ seconds.} % Yes.

\item 
%We develop an efficient rollback algorithm for distributed graph-based application with the assumption that violations are detected quickly enough. 
To evaluate the final benefit the applications can achieve after accounting for the cost of the monitors and rollback, we run graph-based applications with our rollback algorithm on the local lab network.
We observe the final benefit varies depending on the properties of applications. Specifically, on non-terminating applications such as \textit{Weather Monitoring}, the progress of the application running on eventual consistency with monitors and rollback was $45\%$ -- $47\%$ faster than running on sequential consistency. On the other hand, on terminating applications such as \textit{Social Media Analysis}, the final application progress benefit was $10\%$ -- $20\%$.
One of the reasons for the reduced benefit in terminating applications is that at the end of terminating execution, there are a few nodes to be processed, thus the chance of conflicts and recurring violations is increased during this time.
In fact, if the application keeps using eventual consistency, the computation may \textit{stall} due to repeated rollbacks (livelocks). We use some strategies such as backoff and adaptive consistency to handle the livelock issue.
%We use random backoff and adaptive consistency to address this issue.
%To deal with livelocks, we use adaptive consistency, i.e. clients switch to sequential consistency when the computation approaches about 90\% completion of the work (the clients use feedback about violations from the monitors as the signal for switching to sequential consistency). 
We also observe that terminating applications using our approach progressed $16\%$ -- $28\%$ faster than using sequential consistency during the first \SI{90}{\percent} of the work, and $10\%$ -- $20\%$ faster overall (because it has to switch from eventual consistency to sequential consistency during the end of the execution).

\end{itemize}

To the best of our knowledge, our work is the first to experimentally quantify and analyze the benefits of eventual consistency with monitoring and rollback (compared to sequential consistency) on key-value stores. We also propose an efficient rollback algorithm for graph-based applications. Our results suggest that several correctness-sensitive applications are able to take advantage of weaker consistency models from the underlying data store to improve their performance while still preserving the correctness/safety properties. This opens an alternate design option and gives more flexibility to the application designer.

\textbf{Organization of the paper: } Section \ref{sec:sys-arch} describes the architecture of the key-value store used in this paper. 
In section \ref{sec:problem}, we define the notion of causality and identify how the uncertainty of event ordering in distributed systems affects the problem of predicate detection.
Section \ref{sec:framework} describes the overall architecture of the system using monitors. 
Section \ref{sec:pred-detect-module} explains the design of the predicate detection module used in this paper. 
In section \ref{sec:rollback}, we discuss rollback approaches when a violation is detected, and develop a rollback algorithm for some graph-based applications.
Section \ref{sec:evaluations} presents experimental results and discussion.
Section \ref{sec:related-work} compares our paper with related work and we conclude the paper in Section \ref{sec:conclusion}.

%% file: SystemArchitecture.tex
\subsection{Distributed Key-Value Store}

We utilize the standard architecture for key-value stores. Specifically, the data consists of (one or more) tables with two fields, a unique key and the corresponding value. 
The field value consists of a list of $<version, value>$ pairs. A version is a vector clock that describes the origin of the associated value. It is possible that a key has multiple versions when different clients issue PUT (write) requests for that key independently. % concurrently
When a client issues a GET (read) request for a key, all existing versions of that key will be returned. The client could resolve multiple versions for the same key on its own or use the resolver function provided from the library.
To provide efficient access to this table, it is divided into multiple partitions. Furthermore, to provide redundancy and ease of access, the table is replicated across multiple replicas.

To access the entries in this table, the client utilizes two operations, GET and PUT. The operation GET($x$) provides the client with the value (or values if multiple versions exist) associated with key $x$.  The operation PUT($x, val$) changes the value associated with key $x$ to $val$. The state of the servers can be changed only by PUT requests from clients.

\subsection{Voldemort Key Store}

Voldemort is LinkedIn's open source equivalence of Amazon's Dynamo key-value store. In Voldemort, clients are responsible for handling replication. When connecting to a server for the first time, a client receives meta-data from the server. The meta-data contains the list of servers and their addresses, the replication factor ($N$), required reads ($R$), required writes ($W$), and other configuration information.

When a client wants to perform a PUT (or GET) operation, it sends PUT (GET) requests to $N$ servers and waits for the responses for a predefined amount of time (timeout). If at least $W$ ($R$) acknowledgments (responses) are received before the timeout, the PUT (GET) operation is considered successful. If not, the client performs one more round of requests to other servers to get the necessary number of acknowledgments (responses). After the second round, if still less than $W$ ($R$) replies are received, the PUT (GET) operation is considered unsuccessful.

Since the clients do the task of replication, the values $N$, $R$, $W$ specified in the meta-data is only a suggestion. The clients can change those values for their needs.
By adjusting the value of $W$, $R$, and $N$, the client can tune the consistency model. For example, if $W + R > N$ and $W > \frac{N}{2}$ for every client, then they run on sequential consistency. On the other hand, if $W + R \le N$ then they have eventual consistency.

%% file: pred.tex
%In distributed systems, events form a partial order, denoted as the happened-before relation. 

Each process execution in a distributed system results in changing its local state, sending messages to other processes or receiving messages from other processes. In turn, this creates a partial order among local states of the processes in distributed systems. This partial order, the happened-before relation \cite{Lamport78CACM}, is defined as follows:

Given two local states $a$ and $b$, we say that $a$ happened before $b$ (denoted as $a \rightarrow b$) iff
\begin{itemize}
\item $a$ and $b$ are local states of the same process and $a$ occurred before $b$,
\item There exists a message $m$ such that $a$ occurred before sending message $m$ and $b$ occurred after receiving message $m$, or
\item There exists a state $c$ such that $a \rightarrow c$ and $c \rightarrow b$. 
\end{itemize}

We say that states $a$ and $b$ are concurrent (denoted as $a \| b$) iff
$\neg (a \rightarrow b) \ \  \wedge \ \ \neg (b \rightarrow a)$

The goal of a predicate detection algorithm is to ensure that the predicate of interest $P$ is always satisfied during the execution of the distributed system. In other words, we want monitors to notify us of cases where predicate $P$ is violated. 

To detect whether the given predicate $P$ is violated, we utilize the notion of \textit{possibility} modality \cite{Stoller00DC, MN91WDAG}. In particular, the goal is to find a set of local states $e_1, e_2, .. e_n$ such that
\begin{itemize}
\item One local state is chosen from every process,
\item All chosen states are pairwise concurrent. 
\item The predicate $\neg P$ is true in the global state $\langle e_1, e_2, \cdots, e_n \rangle$
\end{itemize} 

\subsection{Vector Clocks and Hybrid Vector Clocks}

To determine whether state $a$ happened before state $b$, we can utilize vector clocks or hybrid vector clocks. Vector clocks, defined by Fidge and Mattern \cite{Fidge87, Mattern89PDA}, are designed for asynchronous distributed systems that make no assumption about underlying speed of processes or about message delivery. Hybrid vector clocks \cite{DK13LADIS} are designed for systems where clocks of processes are synchronized within a given synchronization error (denoted as parameter $\epsilon$ in this paper). 
While the size of vector clocks is always $n$, the number of processes in the system, hybrid vector clocks have the potential to reduce the size to less than $n$.
%Hybrid vector clocks have the potential to reduce the size of the vector clock whereas the size of vector clocks is always $n$, the number of processes in the system. 

Our predicate detection module can work with either of these clocks. For simplicity, we recall hybrid vector clocks (HVC) below. 

Every process maintains its own HVC. HVC at process $i$, denoted as $HVC_i$, is a vector with $n$ elements such that $HVC_i[j]$ is the most recent information process $i$ knows about the physical clock of process $j$. $HVC_i[i] = PT_i$, the physical time at process $i$. Other elements $HVC_i[j], j \neq i$ is learned through the communication between processes. When process $i$ sends a message, it updates its HVC as follows: $HVC_i[i] = PT_i$, $HVC_i[j] = max(HVC_i[j], PT_i - \epsilon)$ for $j \neq i$. Then $HVC_i$ is piggy-backed with the outgoing message. Upon reception of a message $msg$, process $i$ will use the piggy-backed hybrid vector clock $HVC_{msg}$ to update its HVC: $HVC_i[i] = PT_i$, $HVC_i[j] = max(HVC_{msg}[j], PT_i - \epsilon)$ for $j \neq i$.

Hybrid vector clocks are vectors and can be compared as usual. Given two hybrid vector clock $HVC_i$ and $HVC_j$, we say $HVC_i$ is smaller than $HVC_j$, denoted as $HVC_i < HVC_j$, iff $HVC_i[k] \leq HVC_j[k] \forall k$ and $\exists l: HVC_i[l] < HVC_j[l]$. If $\neg (HVC_i < HVC_j) \wedge \neg(HVC_j < HVC_i)$, then the two hybrid vector clocks are concurrent, denoted as $HVC_i || HVC_j$. %\todo{I am not sure whether the notation $\simeq$ is correct or not}

If we set $\epsilon = \infty$, then hybrid vector clocks have the same properties as vector clocks. If $\epsilon$ is finite, certain entries in $HVC_i$ can have the default value $PT_i - \epsilon$ and their representation can be compressed. For example, if $n = 10, \epsilon = 20$, a hybrid vector clock $HVC_0 = [100, 80, 80, 95, 80, 80, 100, 80, 80, 80]$ could be represented by $n$($10$) bits $10010010001$ and a list of three integers \numlist{100;95;100}, instead of a list of ten integers.

We use HVC in our implementation to facilitate its use when the number of processes is very large. However, in the experimental results, we ignore this optimization and treat as if $\epsilon$ is $\infty$.

\subsection{Different Types of Predicate Involved in Predicate Detection}

In the most general form, predicate $P$ is an arbitrary boolean function on the global state and the problem of detecting $\neg P$ is NP-complete \cite{CG98DC}. However, for some classes of predicates such as linear predicates, semilinear predicates, and bounded sum predicates, there exist efficient detection algorithms \cite{GC95ICDCS, Garg96, CG98DC}. In this paper, we adapt these algorithms for monitoring applications running on key-value stores. Since the correctness of our algorithms follows from the existing algorithms, we omit the detailed discussion of the algorithms and focus on their effectiveness in key-value stores.
%develop our predicate detection algorithms based on those works.

%% file: framework.tex
\begin{figure}[t]
\includegraphics[width=0.45\textwidth]{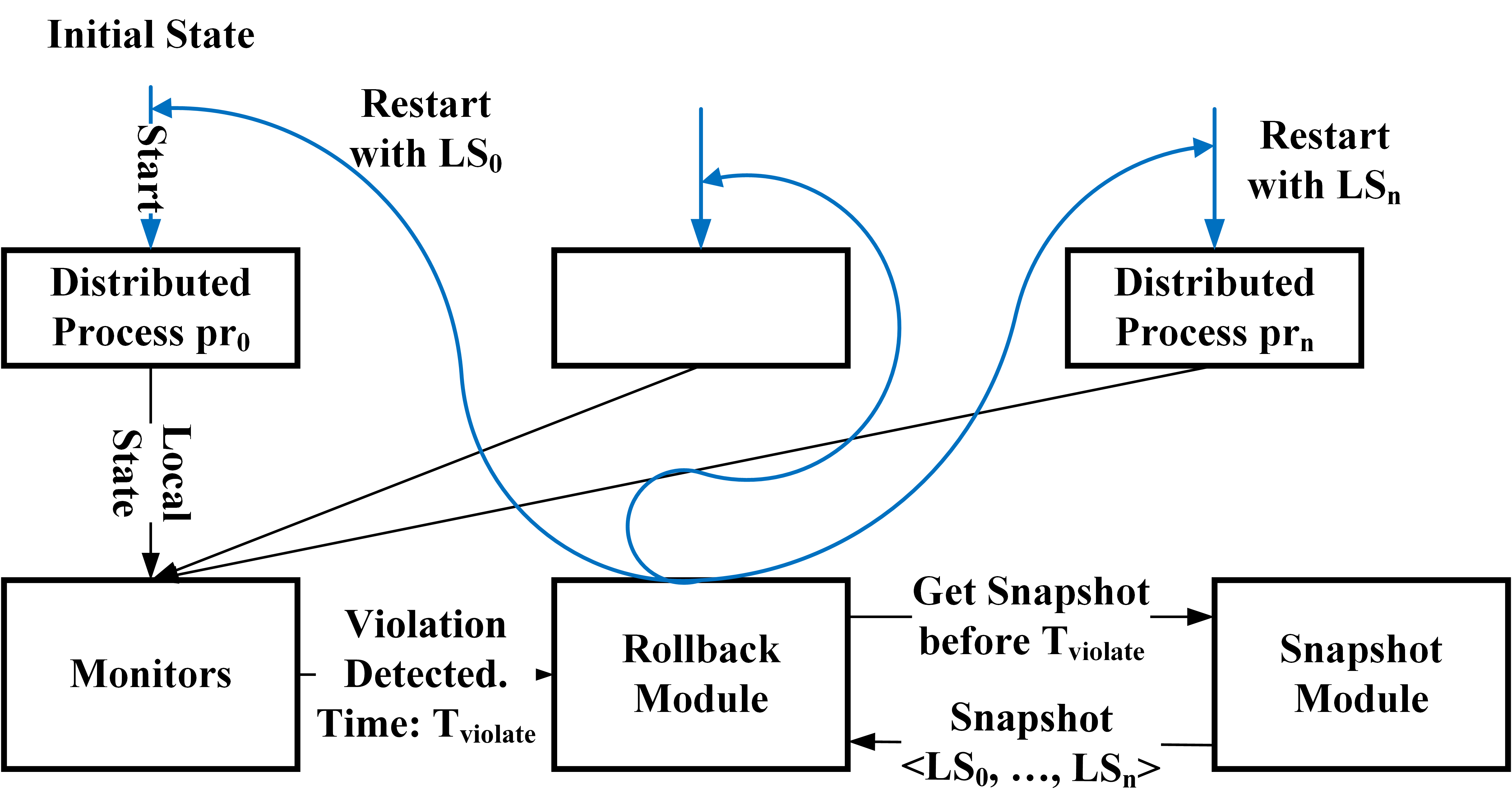}
\caption{An overall framework for optimistic execution in key-value store}
\label{fig:overal-framework}
\end{figure}

The overall framework for optimistic execution in key-value store (i.e. running eventual consistency with monitors and rollback) is as shown in Figure \ref{fig:overal-framework}.
In addition to the actual system execution in the key-value store, we include local detectors for every server (cf. Figure \ref{fig:architecture}). These local detectors provide information to the monitors. 
Note that the desired predicate $P$ can be a conjunction of several smaller predicates and
the monitors are designed to ensure that each smaller predicate, says $P_i$ (which involves one or more processes), continues to be true during the execution. In other words, the monitors are checking if a consistent snapshot where $\neg P_i$ is true (thus $\neg P$ is true) exists.

When the monitors detect violation of the desired property $P$, they notify the rollback module. 
The monitors also identify a safe estimate of the start time $T_{violate}$ at which the violation occurred, based on the timestamps of local states they received.

If the violation of predicate $P$ is rare and the overall system execution is short, we could simply restart the computation from the beginning. 

If the system computation is long, we can take periodic snapshots. Hence, when a violation is found, 
the rollback module notifies all clients and servers to stop the subsequent computation until the restoration to a checkpoint before $T_{violate}$ is complete. 
The exact length of intervals between the periodic snapshots would depend upon the cost of taking the snapshot and the probability of violating predicate $P$ in the intervals between snapshots. 

In case the violations are frequent, feedbacks from the monitor can help the clients to adjust accordingly. For example, if Voldemort clients are running in eventual consistency and find that their computations are restored too frequently, they can switch to sequential consistency by tuning the value of $R$ and $W$ without the involvement of the servers (Recall that in the Voldemort key-value store, the clients are responsible for replication).

Alternatively, we can utilize approach such as Retroscope \cite{CADK17ICDCS}.
Retroscope allows us to dynamically create a consistent snapshot that was valid just before $T_{violate}$ if $T_{violate}$ is within its window-log. This is possible if the predicate detection module is effective enough to detect the violation promptly. In \cite{CADK17ICDCS}, it authors have shown that it is possible to enable rollback for up to $10$ minutes while keeping the size of logs manageable.

The approach in Retroscope can be further optimized by identifying the cause of the rollback. For instance, recall the example from the Introduction that considers a graph application and requires that two clients do not operate on neighboring nodes simultaneously. Suppose a violation is detected due to clients $C_1$ and $C_2$ operating on neighboring nodes $V_1$ and $V_2$. In this case, we need to rollback $C_1$ and $C_2$ to states before they operated on $V_1$ and $V_2$. However, clients that do not depend upon the inconsistent values of nodes $V_1$ and $V_2$ need not be rolled back. Hence, unnecessary rollback can be avoided.

% \todo{Remove this paragraph since we have rollback?} The goal of this paper is to evaluate the effectiveness of the monitor. In particular, our goal is to determine the overhead of such a monitor and the benefit one could get by running the algorithm with a weaker consistency model. Since this benefit is independent of the strategy used for rollback, we only focus on the effectiveness and overhead of the monitor.\todo{S: I agree}
% %
% With this motivation, the properties of interest in this paper are 

% \begin{itemize}
% \item How much overhead occurs when monitors are introduced?  This will help us analyze the overhead when monitors are intended for debugging. 
% \item How does the performance of the system compare under the sequential consistency model (where $P$ was guaranteed to be true as the algorithm is correct) with the performance under a weaker consistency model with the monitors?
% \item How frequent are violations of $P$? This would identify the strategy that is suitable for rollback.
% \item How long does it take to detect violation of $P$? This would help determine whether logs would be sufficient to provide rollback using approaches such as those in \cite{CADK17ICDCS}.
% \end{itemize} 

%% file: PredicateDetectionModule.tex
\begin{figure}[t]%[tbhp]%[h]
% \vspace{-10pt}
\centering
\includegraphics[width=0.45\textwidth]{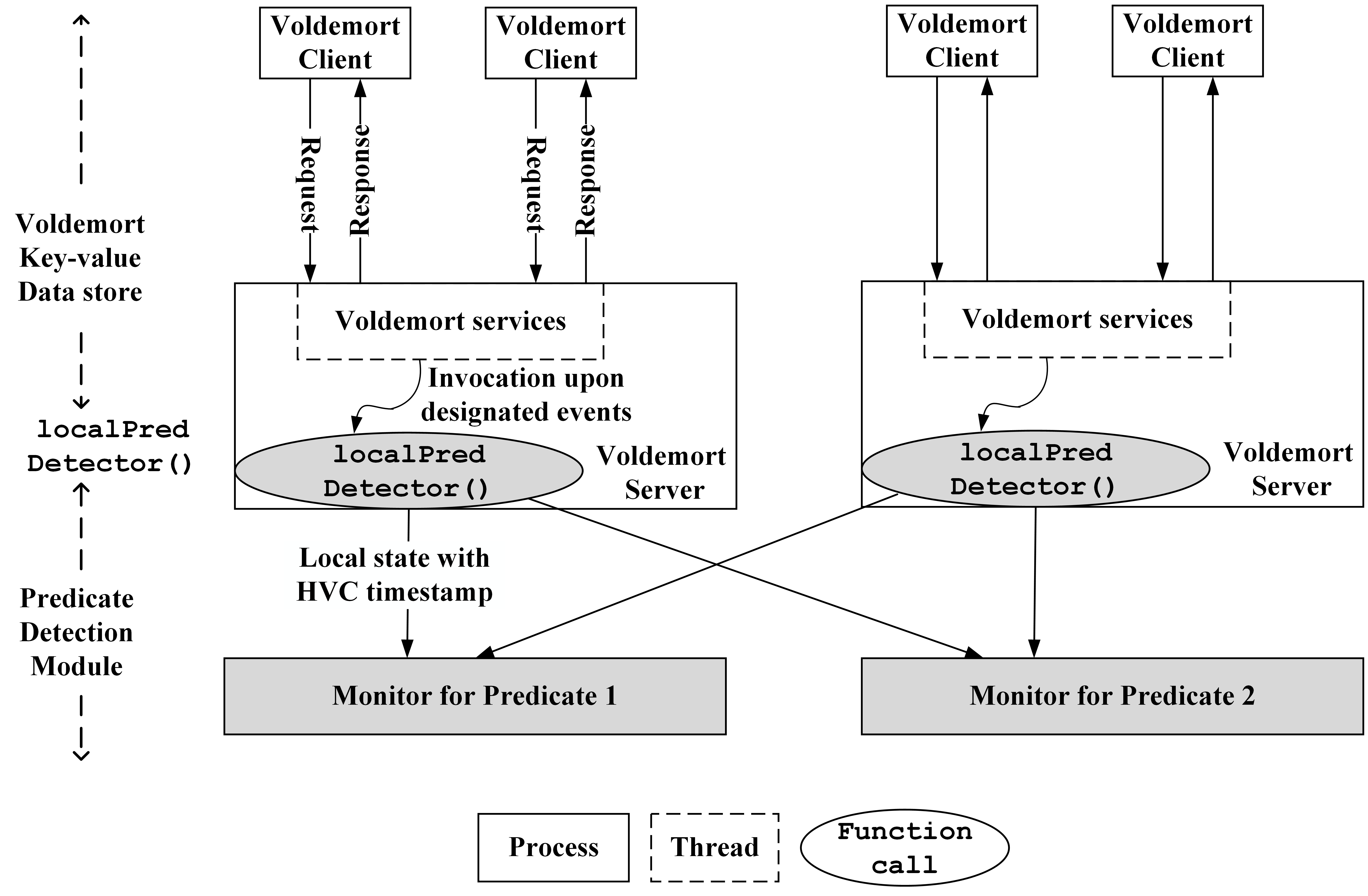}
\caption{Architecture of predicate detection module}
%\vspace{-10pt}
\label{fig:architecture}
\end{figure}

The monitoring module is responsible for monitoring and detecting violation of the global predicate of interest in a distributed system.
The structure of the module is as shown in Figure \ref{fig:architecture}. It consists of local predicate detectors attached to each server and the monitors independent of the servers.
The local predicate detector caches the state of its host server and sends information to the monitors. This is achieved by intercepting the PUT requests for variables that may affect the predicate being monitored. 
The monitors run predicate detection algorithm based on the information received to determine if the global predicate of interest $P$ has been violated.

We anticipate that the predicate of interest $P$ is a conjunction of all constraints that should be satisfied during the execution.
In other words, $P$ is of the form $P_1 \wedge P_2 \wedge \cdots P_l$ where each $P_i$ is a constraint (involving one or more processes) that the program is expected to satisfy. 
Each $P_i$ can be of different types (such as linear or semilinear).
The job of the monitoring module is to identify an instance where $P$ is violated, i.e., to determine if there is a consistent cut where $\neg P_1 \vee \neg P_2 \vee \cdots \neg P_l$ is true.
In order to monitor multiple predicates, the designer can have multiple monitors with one monitor for each predicate $P_i$ or one monitor for all predicates $P_i$'s.
In the former case, the detection latency is small but the overheads can be unaffordable when the number of predicates is large since we need many monitor processes.
In the latter case, the overhead is small but the detection latency is long.
We adopt a compromise: our monitoring module consists of multiple monitors and each monitor is responsible for multiple predicates.
The predicates are assigned to the monitors based on the hash of the predicate names in order to balance the monitors' workload.

The number of monitors equals the number of servers and the monitors are distributed among the machines running the servers.
We have done so to ensure that the cost of the monitors is accounted for in experimental results while avoiding overloading a single machine. 
An alternative approach is to have monitors on a different machine. In this case, the trade-off is between CPU cycles used by the monitors (when monitors are co-located with servers) and communication cost (when monitors are on a different machine). 
Our experiments suggest that in the latter approach (monitors on a different machine) monitoring is more efficient. 
However, since there is no effective way to compute the increased cost (of machines in terms of money), we report results where monitors are on the same machines as the servers.

%
%Note that in Figure \ref{fig:architecture}, each monitor is depicted as one process. In an implementation, a monitor may consist of multiple distributed processes collaborating to monitor a single (smaller) predicate. For simplicity, each monitor is one process in this paper discussion.
%Each predicate has its own monitor (the number of local predicate detectors is equal to the number of servers, the number of monitors is equal to the number of predicates).
%

Each (smaller) predicate $P_i$ is a boolean formula on the states of some variables. Since any boolean formula can be converted to a disjunctive normal form, users can provide the predicates being detected ($\neg P_i$'s) in disjunctive normal form. We use the XML format to represent the predicate. For example, the semilinear predicate, says $\neg P_1 \equiv (x_1=1 \land y_1=1) \lor z_2=1$, in XML format is shown in Figure \ref{fig:x1y1z1}. 
Observe that this XML format also identifies the type of the predicate (linear, semi-linear, etc.) so that the monitor can decide the algorithm to be used for detection.
%
%In this paper, we implement the monitors based on the predicate detection algorithms in \cite{GC95ICDCS, CG98DC}.

%Users provide the predicate being detected in \textit{xml} format representing a disjunctive normal form. For example, the predicate $(x_1=1 \land y_1=1) \lor z_2=1$ in \textit{xml} format is as shown in Figure \ref{fig:x1y1z1}. 

\begin{figure}
\begin{verbatim}
<predicate>
  <type>semilinear</type>
  <conjClause>
    <id>0</id>
    <var> 
      <name>x2</name> <value>1</value> 
    </var>
    <var> 
      <name>y2</name> <value>1</value> 
    </var>
  </conjClause>
  <conjClause>
    <id>1</id>
    <var> 
      <name>z2</name> <value>1</value> 
    </var>
  </conjClause>
</predicate>
\end{verbatim}

\caption{XML specification for $\neg P \equiv (x_1=1 \land y_1=1) \lor z_2=1$  }
\vspace{10pt}
\label{fig:x1y1z1}
\end{figure}

%\footnote{ that the predicate detection module considers $P$ is violated when it becomes true. Thus, in the above example, if $z_2=1$ then $P$ is violated. (wait, does that mean $P$ is violated, or $P$ is satisfied??? In the predicate detection module implementation, the conception is flipped to our usual conception as mentioned in the Introduction. We want $P$ to be false. If $P$ is true, that is something undesirable has happened. The reason I choose that approach is I think $P$ represents the bug. If $P$ is true, then the bug has happened. If the bug happens, I think that $P$ is violated. It is also because of that in conjunctive predicate, a global conjunctive predicate is true when all local predicate is true. In that case, the bug is detected. But if it is confusing, I will change back to the normal conception).}

\textbf{Implementation of local predicate detectors. }
Upon the execution of a PUT request, the server calls the interface function \texttt{localPredicateDetector} which examines the state change and sends a message (also known as a candidate) to one or more monitors if appropriate.
Note that not all state changes cause the \texttt{localPredicateDetector} to send candidates to the monitors.
The most common example of this is when the changed variable is not relevant to the predicates being detected.
Other examples depend upon the type of predicate being detected. 
As an illustration, if predicate $\neg P$ is of the form $x_1 \wedge x_2$ then we only need to worry about the case where $x_i$ changes from $false$ to $true$. 
%
%changes of the variables not relevant to the predicate, or, in linear predicates, changes that do not make local predicate become true, will not trigger a transmission). \todo{show overhead vs put \%}
%

%The local predicate detector maintains a cache of variables related to the predicates of interest to efficiently monitor the server state.  
%
A candidate sent to the monitor of predicate $P_i$ consists of an HVC interval and a partial copy of the server local state containing variables relevant to $P_i$. The HVC interval is the time interval on the server when $P_i$ is violated, and the local state has the values of variables which make $\neg P_i$ true. 
%
%\footnote{This is confusing. P is a global predicate. what is local monitor checking? The local monitor is useful when $\neg P$ is global conjunctive predicate, that is $\neg P$ is conjunctive of \textit{local} predicates. $\neg P$ is true when all local predicates are true. A local predicate is a predicate defined over local state only. Since the local monitor knows the local state, it also knows whether the local predicate is true or not. If the local predicate is true, then a candidate needs to be sent. If false, no candidate is sent, because a false local predicate would make $\neg P$ false. However, if $\neg P$ is a semilinear predicate such as mutual exclusion, then the local monitor would not do any checking, it has to send the candidate upon any PUT request.}

\begin{figure}[t]%[h]
% \vspace{-10pt}
\centering
\includegraphics[width=0.46\textwidth]{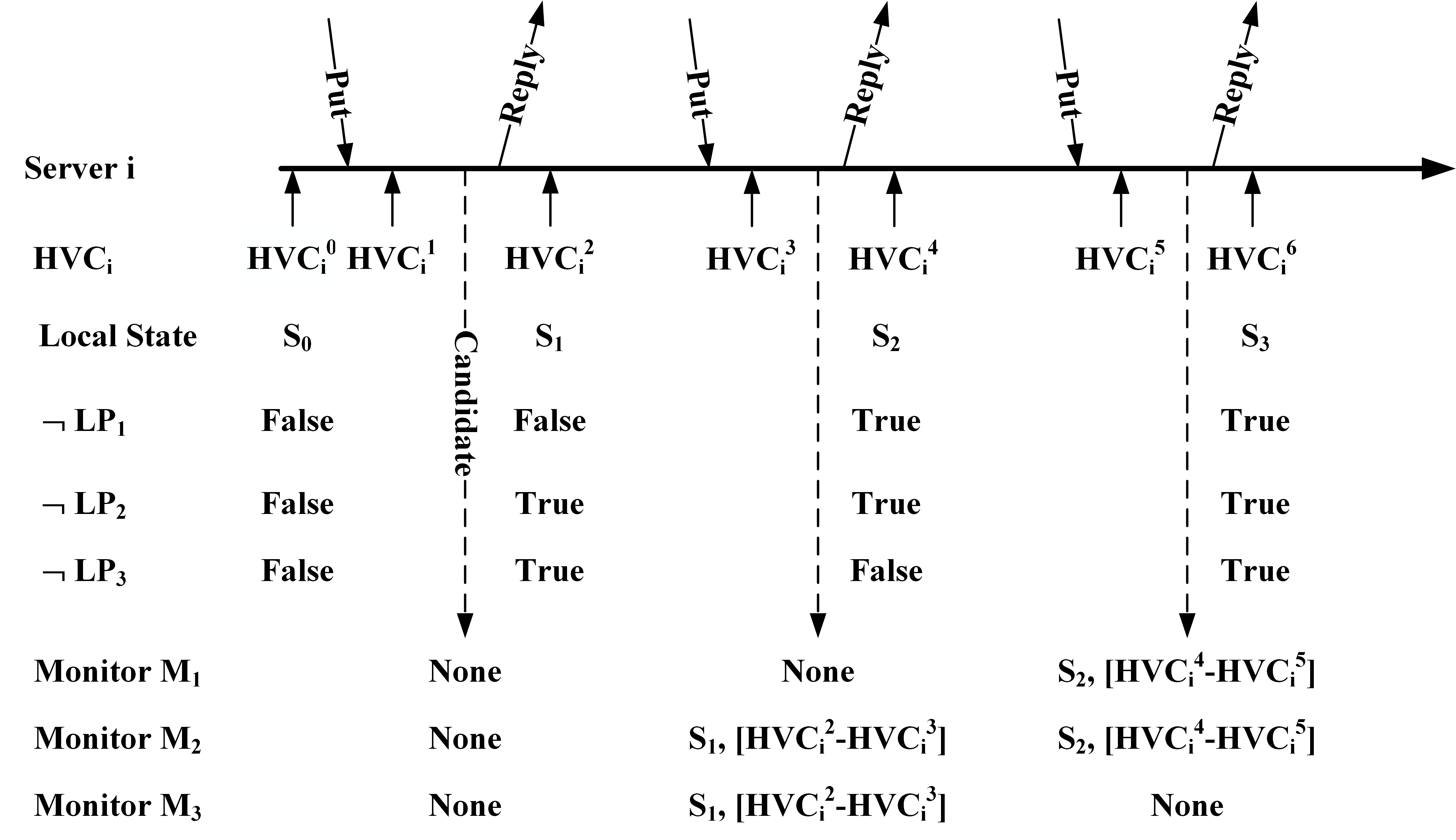}
\caption{Illustration of candidates sent from a server to monitors corresponding to three conjunctive predicates. If the predicate is semilinear, the candidate is always sent upon a PUT request of relevant variables.}
%\vspace{-5pt}
\label{fig:candidates}
\end{figure}

For example, assume the global predicate of interest to be detected is $\neg P \equiv \neg P_1 \vee \neg P_2 \cdots \vee \neg P_m$ where each $\neg P_j$ is a smaller global predicate. Assume that monitor $M_j$ is responsible for detection of predicate $\neg P_j$. Consider a smaller predicate, says $\neg P_2$, and for the sake of the example, assume that it is a conjunctive predicate, i.e. $\neg P_2 \equiv (\neg LP_2^1) \wedge (\neg LP_2^2) \wedge ... (\neg LP_2^n)$ where $n$ is the number of servers. We want to detect when $\neg P_2$ becomes true. On a server, say server $i$, the local predicate detector will monitor the corresponding local predicate $\neg LP_2^i$ (or $\neg LP_2$ for short, in the context of server $i$ as shown in Figure \ref{fig:candidates}). Since $\neg P_2$ is true only when all constituent local predicates are true, server $i$ only has to send candidates for the time interval when $\neg LP_2$ is true. In Figure \ref{fig:candidates}, upon the first PUT request, no candidate is sent to monitor $M_2$ because $\neg LP_2$ is false during interval $[HVC_i^0, HVC_i^1]$. After serving the first PUT request, the new local state makes $\neg LP_2$ true, starting from the time $HVC_i^2$. Therefore upon the second PUT request, a candidate is sent to monitor $M_2$ because $\neg LP_2$ is true during the interval $[HVC_i^2, HVC_i^3]$. This candidate transmission is independent of whether $\neg LP_2$ is true or not after the second PUT request is served. It depends on whether $\neg LP_2$ is true after execution of the previous PUT request. That is why, upon the second PUT request, a candidate is also sent to monitor $M_3$ but none is sent to $M_1$.
However, if the predicate is not a linear predicate, then upon a PUT request for a relevant variable, the local predicate detector has to send a candidate to the associated monitor anyway.

\color{black}

\textbf{Implementation of the monitors. }
The task of a monitor is to determine if some smaller predicate $P_i$ under its responsibility is violated, i.e., to detect if a consistent state on which $\neg P_i$ is true exists in the system execution. The monitor constructs a global view of the variables relevant to $P_i$ from the candidates it receives. The global view is valid if all candidates in the global view are pairwise concurrent. 
%Furthermore, we use HVC interval to eliminate false positive cases where candidates are concurrent from the perspective of causality but physically too far apart so that they are practically not concurrent. 

\begin{figure}[t]%[h]
% \vspace{-10pt}
\centering
\includegraphics[width=0.46\textwidth]
{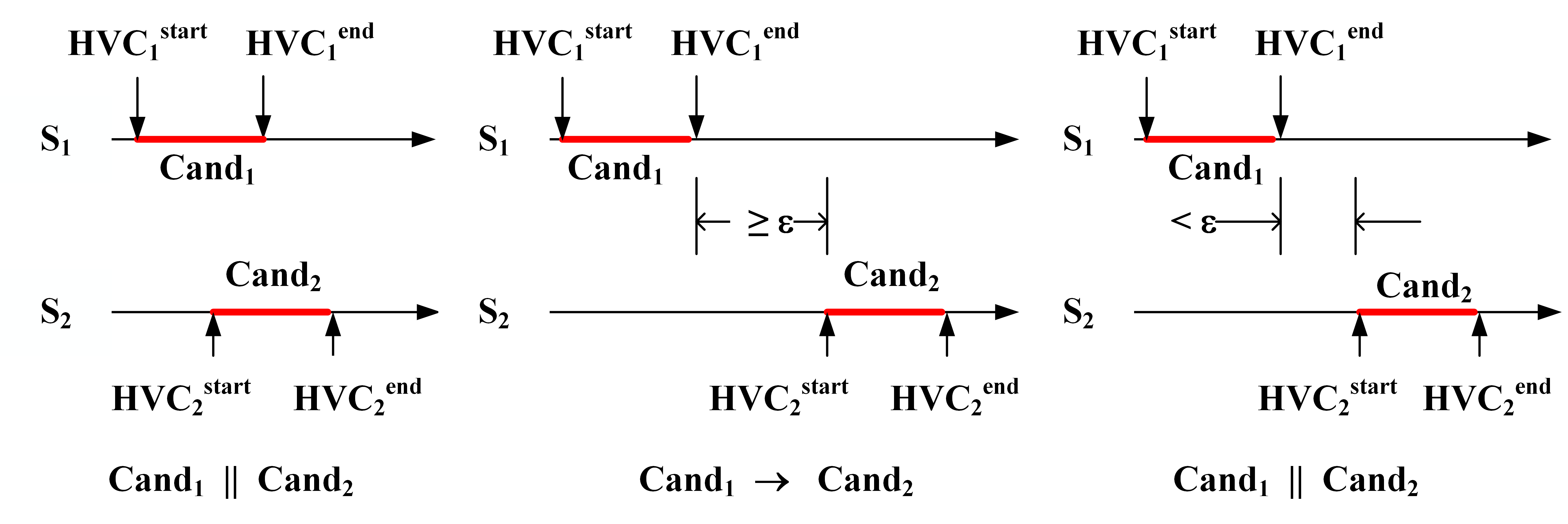}
\caption{Illustration of causality relation under HVC interval perspective}
%\vspace{5pt}
\label{fig:hvc-interval}
\end{figure}

The concurrence/causality relationship between a pair of candidates is determined as follows: 
suppose we have two candidates $Cand_1, Cand_2$ from two servers $S_1, S_2$ and their corresponding HVC intervals $[HVC_{1}^{start}, HVC_{1}^{end}], [HVC_{2}^{start}, HVC_{2}^{end}]$. 
Without loss of generality, assume that $\neg (HVC_{1}^{start} > HVC_{2}^{start})$ (cf. Figure \ref{fig:hvc-interval}).
\begin{itemize}
\item If $HVC_{2}^{start} < HVC_{1}^{end}$ then the two intervals have common time segment and $Cand_1 \| Cand_2$.
\item If $HVC_{1}^{end} < HVC_{2}^{start}$, and $HVC_{1}^{end}[S_1] \leq HVC_{2}^{start}[S_2] - \epsilon$ then interval one is considered happens before interval two. Note that $HVC[i]$ is the element corresponding to process $i$ in HVC. In this case $Cand_1 \rightarrow Cand_2$
\item If $HVC_{1}^{end} < HVC_{2}^{start}$, and $HVC_{1}^{end}[S_1] > HVC_{2}^{start}[S_2] - \epsilon$, this is the uncertain case where the intervals may or may not have common segment. In order to avoid missing possible violations, the candidates are considered concurrent.
\end{itemize}

%Recall that for the global view to be valid, all HVC intervals must be pairwise concurrent.

When a global predicate is detected, the monitor informs the administrator or triggers a designated process of recovery. We develop detection algorithms for the monitors of linear predicates and semilinear predicates based on \cite{GC95ICDCS, CG98DC} as shown in Algorithm \ref{algo:linear} and Algorithm \ref{algo:semilinear}. Basically, the algorithms have to identify the correct candidates to update the global state ($GS$) so that we would not have to consider all possible combinations of $GS$ as well as not miss the possible violations. In linear (or semilinear) predicates, these candidates are forbidden (or semi-forbidden) states. Forbidden states are states such that if we do not replace them, we would not be able to find the violation. Therefore, we must advance the global state along forbidden states. Semi-forbidden states are states such that if we advance the global state along them, we would find a violation if there exists any.
The procedure of advancing the global snapshot $GS$ along a local state $s$ ($s$ belongs to $GS$) means the successor of $s$ is added to $GS$. The successor of a local state $s$ is the next local state after $s$ on the same process.
As $s$ is replaced by its successor, the global snapshot $GS$ ``advances'' forward.
When advancing global state along a candidate (which contains a local state), that candidate may not be concurrent with other candidates existing in the global state. In that case, we have to advance the candidates to make them consistent. This is done by \texttt{consistent(GS)} in the algorithm. If we can advance global state along a candidate without calling \texttt{consistent(GS)}, that candidate is called an eligible state. The set of all eligible states in the global state is denoted as \texttt{eligible(GS)} in the algorithms.
For a more detailed discussion of linear and semi-linear predicates, we refer to \cite{CG98DC}.

\algdef{SE}[SUBALG]{Indent}{EndIndent}{}{\algorithmicend\ }%
\algtext*{Indent}
\algtext*{EndIndent}

\begin{algorithm}[ht]
\caption{Linear predicate monitor algorithm \cite{GC95ICDCS}}\label{algo:linear}
\begin{algorithmic}[1]
\\{\bf{Input:}}
\Indent
    \State $P$
    \Comment{global linear predicate to monitor}
\EndIndent
\\{\bf{Variable:}}
\Indent
    \State $GS$
    \Comment{global state}
\EndIndent
\\{\bf{Initialization:}}
\Indent
    \State $GS \leftarrow $ set of initial local states
\EndIndent

\While{P(GS)==true}
    \State Find forbidden local state $s \in GS$
    \State $GS \leftarrow GS \cup {succ(s)}$
    \Comment{advance $GS$ along $s$}
    \State consistent($GS$)
    \Comment{make $GS$ consistent}
\EndWhile{}
\State return $GS$
\end{algorithmic}
\end{algorithm}

\begin{algorithm}[h]
\caption{Semilinear predicate monitor algorithm \cite{CG98DC}}\label{algo:semilinear}
\begin{algorithmic}[1]
\\{\bf{Input:}}
\Indent
    \State $P$
    \Comment{global semilinear predicate to monitor}
\EndIndent
\\{\bf{Variable:}}
\Indent
    \State $GS$
    \Comment{global state}
\EndIndent
\\{\bf{Initialization:}}
\Indent
    \State $GS \leftarrow $ set of initial local states
\EndIndent

\While{P(GS)==true}
    \State Find a local state $s \in GS$ such that $s \in eligible(GS)$ and $s$ a semi-forbidden state of $P$ in $GS$.
    \State $GS \leftarrow GS \cup {succ(s)}$
    \Comment{advance $GS$ along $s$}
\EndWhile{}
\State return $GS$
\end{algorithmic}
\end{algorithm}

After a consistent global state $GS$ is obtained, we evaluate whether predicate $P$ is violated at this global state ($P(GS) = true$ means $P$ is satisfied, $P(GS) = false$ means $P$ is violated). If $P$ is violated, the algorithms return the global snapshot $GS$ as the evidence of the violation. Note that the monitors will keep running even after a violation is reported so that possible violations in the future will not be missed. This is the case when the applications, after being informed about the violation and rolling back to a consistent checkpoint before the moment when the violation occurred, continue their execution and violations occur again. Hence the monitors have to keep running in order to detect any violations of $P$.

The way we evaluate $P$ on global state $GS$ is slightly different from the algorithms in \cite{CG98DC,GC95ICDCS,Garg96, GW94TPDS}. In those algorithms, the candidates are sent directly from the clients containing the states of the clients. In our algorithms, the candidates are sent from the servers containing the information the servers know about the states of the clients that have been committed to the store by the clients. Note that, in a key-value store, the clients use the server store for sharing variables and committing updates. Therefore, the states of clients will eventually be reflected at the server store. Since the predicate $P$ is defined over the states of the clients, in order to detect violations of $P$ from the states stored at the server, we have to adapt the algorithms in \cite{CG98DC, GC95ICDCS, Garg96, GW94TPDS} to consider that difference.
Furthermore, the state of a client can be stored slightly differently at different servers. For example, a PUT request may be successful at the regional server but not successful at remote servers. In that case, assuming we are using eventual consistency, the regional server store will have the update while remote stores do not have the update. Our algorithms also consider this factor when evaluating $P$. For example, suppose variable $x$ has version $v_1$ at a server and version $v_2$ at another server. Suppose that if $x=v_1$ then $P$ is violated, and if $x=v_2$ then $P$ is satisfied. To avoid missing possible violations, our algorithms check all available versions of $x$ when evaluating $P$.

Since our algorithms are adapted from \cite{CG98DC, GC95ICDCS, Garg96, GW94TPDS}, the correctness of our algorithms follow from those existing algorithms. We refer to \cite{CG98DC, GC95ICDCS, Garg96, GW94TPDS} for more detailed discussion and proof of correctness of the algorithms.

\textbf{Handling a large number of predicates.} When the number of predicates to be monitored is large (e.g. hundreds of thousands, as in \textit{Social Media Analysis} application in next section or in graph-based applications discussed in the Introduction), it is costly to maintain monitoring resources (memory, CPU cycles) for all of them simultaneously. That not only slows down the detection latency but also consumes all the resources on the machines hosting the monitors (for example, we received \texttt{OutOfMemoryError} error when monitoring tens of thousands of predicates simultaneously).
However, we observe that not all predicates are active at the same time. Only predicates relevant to the nodes that the clients are currently working on are active.
A predicate is considered inactive when there is no activity related to that predicate for a predetermined period of time, and therefore the evaluation of that predicate is unchanged. Consequently, the monitors can clean up resources allocated for that predicate to save memory and processing time.

\textbf{Automatic inference of predicate from variable names. } This feature is also motivated by applications where the number of predicates to be monitored is large such as the graph-based applications. In this case, it is impossible for the users to manually specify all the predicates. However, if the variables relevant to the predicates follow some naming convention, our monitoring module can automatically generate predicates on-demand. 
For example, in graph-based applications, the predicates are the mutual exclusions on any edge whose endpoints are assigned to two different clients. Let $A$ and $B$ are two such nodes and $A\_B$ is the edge between them. Assume $A < B$. If the clients are using Peterson's mutual exclusion, the predicate for edge $A\_B$ will be 
\begin{align*}
\neg P_{A\_B} \equiv (flagA\_B\_A = true \wedge turnA\_B = "A") \\ \wedge (flagA\_B\_B = true \wedge turnA\_B = "B")
\end{align*}
When a server receives a PUT request from some client for a variable whose name is either \texttt{flagA\_B\_A}, or \texttt{flagA\_B\_B}, or \texttt{turnA\_B}, it knows that the client is interested in the lock for edge $A\_B$ and the local predicate detector will generate the predicate for edge $A\_B$ so that the monitors can detect if the mutual exclusion access on edge $A\_B$ is violated. On the other hand, if the servers never see requests for variables \texttt{flagA\_B\_A}, \texttt{flagA\_B\_B}, and \texttt{turnA\_B}, then both nodes $A$ and $B$ are assigned to the same client and we do not need the mutual exclusion predicate for edge $A\_B$.

%% file: rollback.tex
\subsection{Rollback Mechanism}
\label{subsec:rollback-mechanism}

While inconsistency is possible with eventual consistency, it is rare \cite{DHJKLPSVV07SOSP} given that networks are reliable and data conflicts are infrequent. However, such inconsistencies and data conflicts can arise and, hence, one needs to deal with these conflicts if we are using an application that relies on sequential consistency. We discuss the rollback approaches for such scenarios.

%Since the network is generally good \cite{BK2014Queue}, eventual consistency is reliable while providing significantly higher performance than sequential consistency. Data conflicts could happen but rarely, as shown in our experiments and in \cite{DHJKLPSVV07SOSP}. Therefore, the cost of recovery from inconsistent states is small compared to the benefit of eventual consistency.

\begin{figure}[tbp]%[h]
% \vspace{-10pt}
\centering
\includegraphics[width=0.45\textwidth]{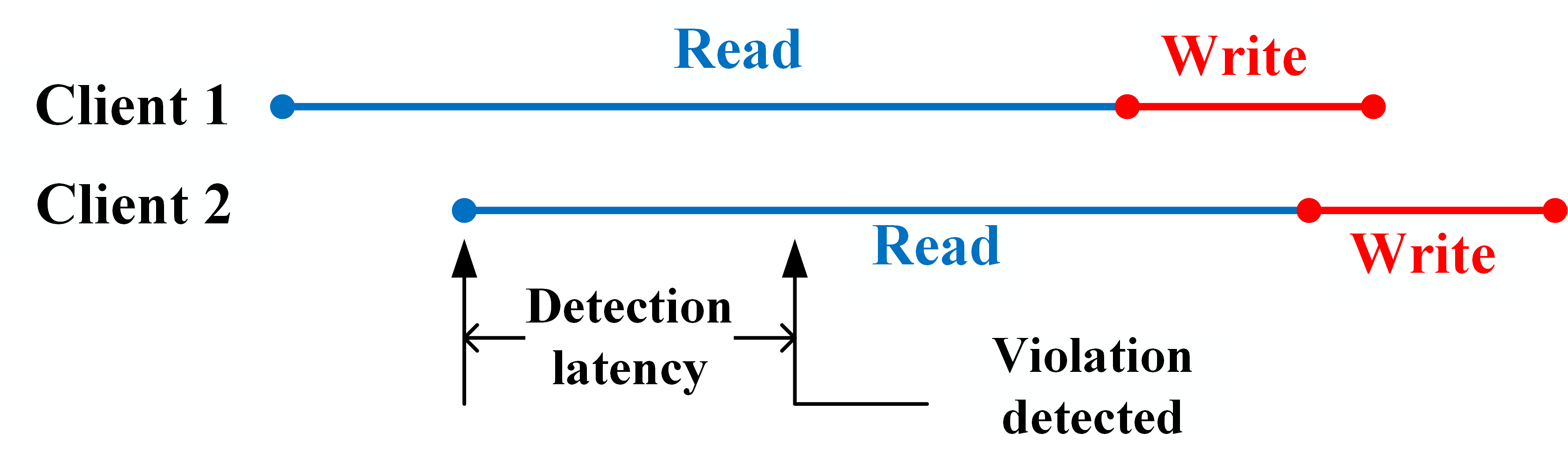}
\caption{Two client tasks involved in a violation. Since detection latency is much smaller than the \textit{Read} phase time, violation will be notified within \textit{Read} phase of the current task of at least one client.}
%\vspace{-10pt}
\label{fig:detection-latency}
\end{figure}

One possible approach for rollback, especially if violations can be detected quickly is as follows: 
We partition the work assigned to each client in terms of several tasks. Each task consists of two phases (cf. Figure \ref{fig:detection-latency}): (1) \textit{Read} phase: the client obtains all necessary locks for all nodes in the task, reading the necessary data, and identify the values that need to be changed. However, all updates in this phase are done in local memory. (2) \textit{Write} phase: the client writes the data that they are expected to change and reflect it in the data store. 

In such a system, a violation could occur if clients C1 and C2 are accessing the same data simultaneously. For sake of discussion, suppose that client C1 started accessing the data before C2. Now, if the detection of violation is quick then detection would occur before client C2 enters the write phase. In this case, client C2 has not performed any changes to the key-value store. In other words, client C2 can re-start its task (that involves reading the data from the key-value store) to recover from the violation. 

With this intuition, we can provide recovery as follows: When a violation is detected, if the client causing the violation is in the read phase, it aborts that task and starts that task again. On the other hand, if a client is in write phase (and this can happen to at most one task if detection is quick enough) then it continues its task normally. Note that with this approach, it is possible that two clients that result in a violation are both in the read phase. While one of the clients could be allowed to continue normally, this requires clients to know the status of other clients. We do not consider this option as it is expected that in most applications clients do not communicate directly. Rather, they communicate only via the key-value store. We utilize this approach in our rollback mechanism. In particular, when detection is quick, we use the Algorithm \ref{algo:rollback} for rollback (cf. Figure \ref{fig:detection-latency} and Algorithm \ref{algo:rollback}).

\iffalse
\begin{itemize}
\item If the client has not issued PUT requests in the current task, it will ignore the values read from the key-value store and restart the task.

%to update colors of nodes in the task yet, it will skip those updates and releases all the locks it is currently holding. Thus, the current task is aborted without the need for state rollback since no PUT request has been committed. 
%At least one client (among the two involved in the violation) is in this case.

\item If the client has issued some PUT requests, it will continue to finish all PUT requests and release the locks it holds. Thus, the current task is completed as normal. We note that if the violation is detected quickly then at most one client will continue along this path, thereby satisfying mutual exclusion requirement.
%at least one client will abort the updates, if the other client commits the updates, that does not cause coloring conflicts.
\end{itemize}
\fi 

Other approaches for rollback are as follows:

\begin{itemize}

    \item \textbf{Rollback via Retroscope\cite{CADK17ICDCS}.  } The most general approach is to utilize an algorithm such as RetroScope \cite{CADK17ICDCS}. Specifically, it allows one to rollback the state of the key-value store to an earlier state. The time, $t$, of rollback is chosen in such a way that there are no violations before time $t$. Upon such a rollback, we can determine the phases the clients are in at time $t$. If a client is in the read phase at time $t$, it will abort its current task and begin it again. And, if the client is in a write phase, it will finish that phase. Note that since there are no conflicts until time $t$, such write phases will not result in conflicts. 
    
    While this approach is most general, it is also potentially expensive. Hence, some alternate approaches are as follows:
    
    \item %stabilization
    \textbf{Use of Self-Stabilizing Algorithms. }
    One possibility is if we are using a self-stabilizing algorithm. An algorithm is self-stabilizing if it is guaranteed to recover to a legitimate state even in the presence of arbitrary state perturbation. In \cite{NKD2019ICDCN}, it is shown that if the underlying algorithm is self-stabilizing then we can simply ignore the violations as we can treat it as a state perturbation and the algorithm is already designed to handle it. In this case, there is neither a need for monitoring or rollback.  
    
    \item %application specific rollback (coloring 
    \textbf{Use of Application-Specific Rollbacks. }
    Another possibility is application specific rollback. To illustrate this, consider an example of graph coloring. For sake of illustration, consider that we have three nodes A, B, C, arranged in a line with node B in the middle. Each node may have additional neighbors as well.  Node A chooses its color based on the colors of its neighbors. Subsequently, node B chooses its color based on node A (and other neighbors of B). Afterward, C chooses its color based on B (and other neighbors of C). At this point, node B is required to rollback, it can still choose its color based on the new color of node C while still satisfying the constraints of graph coloring. In other words, in this application, we do not need to worry about cascading rollback. 
    
\end{itemize} 

\begin{algorithm}[t]%[ht]
\caption{Rollback algorithm at a client}\label{algo:rollback}
\begin{algorithmic}[1]
\For{taskId = clientFirstTask to clientLastTask}
    \While{(performTask(taskId) == False)}
    \EndWhile
\EndFor
\\
\Function{performTask}{$taskId$}
    \State Obtain relevant locks
    \State Read information from data-store
    \State Compute new values
    \If{Violation is received}
        \State Release locks
        \State \Return False
        \Comment{abort}
    \EndIf
    \State Write new values to data-store
    \State \Return True
    \Comment{success}
\EndFunction
\end{algorithmic}
\end{algorithm}

\subsection{Dealing with Potential of Livelocks}

One potential issue with rollback is a possibility of livelocks. Specifically, if two clients C1 and C2 rollback and continue their execution then the same violation is likely to happen again. We consider the following choices for dealing with such livelocks. 

\begin{itemize}
    \item \textbf{Random Backoff. } Upon rollback, clients perform a random backoff. 
    %Thus, in the subsequent execution, the possibility of the same violation is reduced. We note that the goal of backoff is simply to ensure that two clients cannot obtain the same lock simultaneously due to inconsistencies caused by eventual consistency. 
    With backoff, the requests for locks from clients arrive at different times in the key-value store. Hence, the second client is likely to observe locks obtained by the first client in a consistent manner. In turn, this will reduce the possibility of the same violation to recur. 
    \item \textbf{Reordering of Tasks. } If the work assigned to clients consists of several independent tasks, then clients can reorder the tasks upon detecting a violation. In this case, the clients involved in the rollback are likely to access different data and, hence, the possibility of another violation is reduced. 
    \item \textbf{Moving to Sequential Consistency. }
    If the number of violations is beyond a certain threshold, clients may conclude that the cost of rollback is too high and, hence, they can move to sequential consistency. While this causes one to lose the benefits of an eventual consistent key-value store, there would be no need for rollback or monitoring. 
\end{itemize}

%% file: Experiment-setup.tex
\subsection{Experimental Setup}
\label{subsec:experiment-setup}

\textbf{System configurations.} We ran experiments on Amazon AWS EC2 instances. The servers ran on M5.xlarge instances with 4 vCPUs, 16 GB RAM, and a GP2 general-purpose solid-state drive storage volume. The clients ran on M5.large instances with 2 vCPUs and 8 GB RAM. The EC2 instances were located in three AWS regions: Ohio, U.S; Oregon, U.S; Frankfurt, Germany.

We also ran experiments on our local lab network which is set up so that we can control network latency. We used 9 commodity PCs, 3 for servers, 6 for clients, with configurations as in Table \ref{table:local-lab-hardware}. Each client machine hosted multiple client processes, while each server machine hosted one Voldemort server process.

\begin{table}[ht]
\caption{Machine configuration in local lab experiments}
\begin{center}
\begin{tabular}{|l|l|l|}
    \hline
    Machine & CPU & RAM \\
    \hline
    Server machine 1, 2 & 4 Intel Core i5 3.33 GHz  & 4 GB \\
    \hline
    Server machine 3 & 4 Intel Core i3 3.70 GHz & 8 GB \\
    \hline
    Client machine 1, 2  & 4 Intel Core i5 3.33 GHz& 4 GB \\
    \hline
    Client machine 3, 4& Intel Core Duo 3.00 GHz & 4 GB \\
    \hline
    Client machine 5& 4 AMD Athlon II 2.8 GHz & 6 GB \\
    \hline
    Client machine 6 & 4 Intel Core i5 2.30 GHz & 4 GB \\
    \hline
\end{tabular}
\label{table:local-lab-hardware}
\end{center}
\end{table}

\begin{figure}[!t]%[h]
% \vspace{-10pt}
\centering
\includegraphics[width=0.46\textwidth]{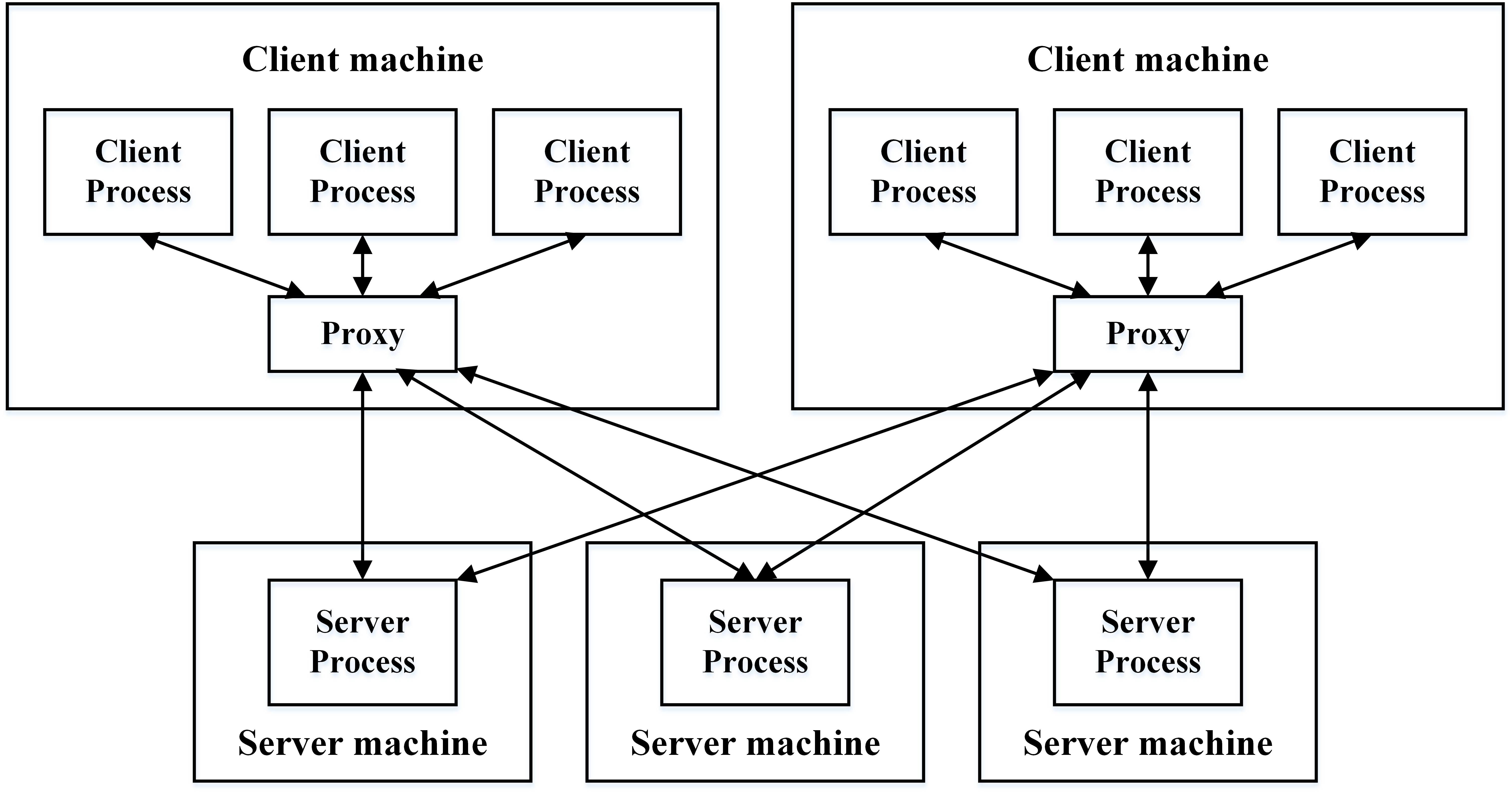}
\caption{Simulating network delay using proxies}
%\vspace{-10pt}
\label{fig:proxy}
\end{figure}

\begin{figure}[!t]%[h]
% \vspace{-10pt}
\centering
\includegraphics[width=0.46\textwidth]{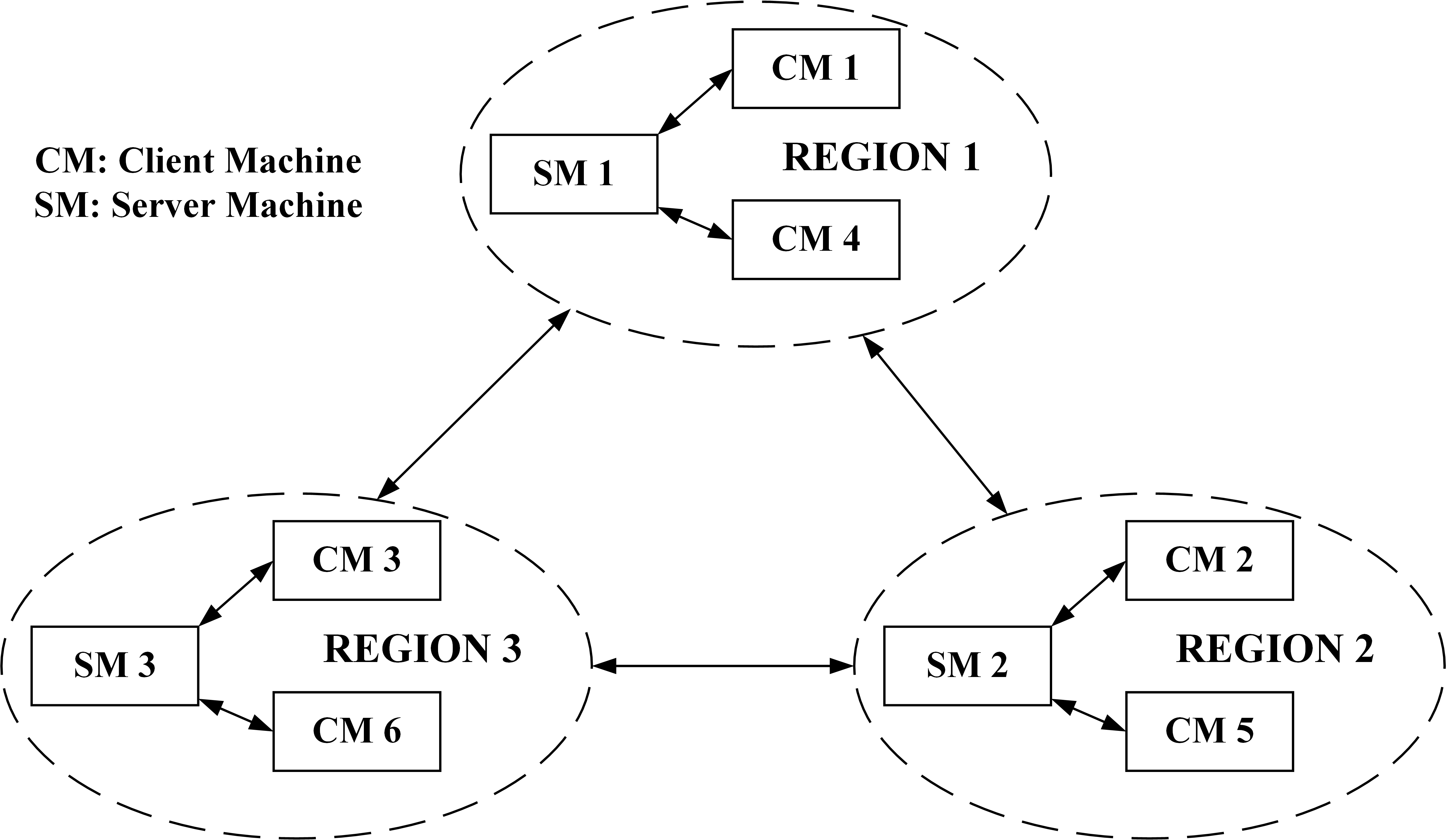}
\caption{Network arrangement with proxies}
%\vspace{-10pt}
\label{fig:local-lab-network}
\end{figure}

On the local network, we control the delay by placing proxies between the clients and the servers. For all clients on the same physical machine, there is one proxy process for those clients. All communication between those clients and any server is relayed through that proxy (cf. Figure \ref{fig:proxy}). Due to the proxy delays, machines are virtually arranged into three regions as in Figure \ref{fig:local-lab-network}. Latency within a region is small (\SI{2}{\ms}) while those across regions are high and tunable (e.g. \SIrange{50}{100}{\ms}). Since Voldemort uses active replication, we do not place proxies between servers. The latency in the proxies is simulated to follow the Gamma distribution \cite{BMHUV2002PAM,gammaDist2013}.

% \textbf{Latency Distribution.}
% The latency between clients and servers is controlled by proxies in local lab settings. The latency is simulated to follow Gamma distribution \cite{BMHUV2002PAM,gammaDist2013}.
% %
% With Gamma distribution, the latency between two nodes $A$ and $B$ is $D_{A,B} = D_{A,B}^d + D_{A,B}^s$. $D_{A,B}^d$ is the deterministic delay between $A$ and $B$ reflecting the topological distance between the two nodes. If the path between two nodes is fixed then $D_{A,B}^d$ is assumed to be a constant. $D_{A,B}^d$ is in the range of $[50,100]$ \footnote{According to https://www.cloudping.co/, the round trip latency between AWS regions could be between 40 ms to 300 ms}. $D_{A,B}^s$ is the stochastic delay reflecting the network condition between $A$ and $B$ at the time of transmission. $D_{A,B}^s$ is assumed to observe a Gamma distribution with shape factor between $0.6$ and $1.0$ \cite{BMHUV2002PAM}. In our experiments, we calculate $D_{A,B}^s = sample \times multiplier$ where $sample$ is a sample from Gamma distribution with the shape factor of $0.8$, and $multiplier$ is the multiplier to convert a scalar value to latency unit (i.e. milliseconds). We choose $multiplier = D_{A,B}^d \times 0.2$ by experiments.
% %We recall that in local lab experiments, we do not intend to precisely simulate delay in the real-world network. We would like to control the delay to test our approach. 
% So $$D_{A,B} = D_{A,B}^d \times (1 + sample \times 0.2)$$

\color{black}

We considered replication factors ($N$) of 3 and 5. The parameters $R$ (required reads) and $W$ (required writes) are chosen to achieve different consistency models as shown in Table \ref{table:consistency-models}.
The number of servers is equal to the replication factor $N$. The number of clients is varied between 15 and 90.

\begin{table}[!h]% [tbhp]
\begin{center}
\caption{Setup of consistency models with $N$ (replication factor), $R$ (required reads), and $W$ (required writes)}
\begin{tabular}{|l|l|l|l|l|}
    \hline
    N & R & W  & Abbreviation & Consistency model \\
    \hline
%    2 & 1 & 2 & N2R1W2 & Sequential \\
%     & 1 & 1 & N2R1W1 & Eventual \\
%    \hline
    3 & 1 & 3 & N3R1W3 & Sequential \\
      & 2 & 2 & N2R2W2 & Sequential \\
     & 1 & 1 & N3R1W1 & Eventual \\
    \hline
    5 & 1 & 5 & N5R1W5 & Sequential \\
%     & 2 & 4 & N5R2W4 & Sequential \\
     & 3 & 3 & N5R3W3 & Sequential \\
     & 1 & 1 & N5R1W1 & Eventual \\
    \hline
\end{tabular}
\label{table:consistency-models}
\end{center}
\end{table}

%
%Monitors are distributed among the machines running the servers. We have done so to ensure that the cost of the monitors is accounted for in experimental results while avoiding overloading a single machine. An alternative approach is to have monitors on a different server. In this case, the trade-off is between CPU cycles used by the monitors (when monitors are co-located with servers) and communication cost (when monitors are on a different machine). Our experiments suggest that the latter (monitors on a different machine) is more efficient. However, since there is no effective way to compute the increased cost (of machines in terms of money), we report results where monitors are on the same machines as the servers.

\textbf{Test cases.} In our experiments, we used 3 case studies: \textit{Social Media Analysis}, \textit{Weather Monitoring}, and \textit{Conjunctive}.

The application motivated by \textit{Social Media Analysis} considers a large graph representing users and their connections. The goal of clients is to update the state of each user (node) based on its connections. For the sake of illustration in our analysis, the attribute associated with each user is a color and the task is to assign each node a color that is different from its neighbors.
We use the tool \textit{networkx} \cite{networkx} to generate input graphs. There are two types of graph: (1) Power-law clustering graph that simulates the power-law degree and clustering characteristics of social networks, and (2) Random 6-regular graph in which each node has 6 adjacent edges and the edges are selected randomly. The reason we use random regular graphs is that they are the test cases where the workload is distributed evenly between clients and throughout the execution. The graphs have \num{50 000} nodes with about \num{150 000} edges.
Each client is assigned a set of nodes to be colored and run a distributed coloring algorithm \cite{Raynal2013distributed}. 

Since the color of a node is chosen based on its neighbors' colors, while a client $C_1$ is coloring node $v_1$, no other client is updating the colors of $v_1$'s neighbors. 
The goal of the monitors is to detect violation of this requirement. 
This requirement can be viewed as a mutual exclusion (semi-linear) predicate where a client going to update the color of $v_1$ has to obtain all the exclusive locks associated with the edges incident to $v_1$.
%a client $C_1$ has to obtain all the locks associated with edges incident to $v_1$ before updating the color for $v_1$.
%
Mutual exclusion is guaranteed if clients use Peterson's algorithm and the system provides sequential consistency \cite{BW02PPAM}. However, it may be violated in the eventual consistency model.
To avoid deadlock, clients obtain locks in a consistent order.
For example, let $A\_B$ and $C\_D$ are the locks associated with the edges between nodes $A$ and $B$, and $C$ and $D$ respectively. Assume $A < B$ and $C < D$. Then lock $A\_B$ is obtained before $C\_D$ when $A < C$ or when $A = C$ and $B < D$.
%

% Since the graph contains some nodes of very high degree, it is far sub-optimal if we use $maxDeg$ (the maximum degree in the graph) colors. Preprocessing high degree nodes reduces the number of colors used by the algorithm as well as the number of locks sought by the clients. 
% %
% A node is considered a high-degree node if its degree is greater than the threshold $q$. The value of $q$ is chosen such that the number of nodes with degree greater than $q$ is smaller than $q$. So high degree nodes could be colored by at most $q$ colors and the whole graph would use no more than $2q$ colors. In our experiment with various size graphs similar to our input graph, the number of nodes with degree $deg$ roughly follows the distribution
% $$count(deg) \approx 6.5 \times |V| \times deg ^ {-2.5}$$
% where $|V|$ is the number of nodes. Therefore the number of nodes with degree greater than $q$ is
% $$gCount(q) \approx \int_{q+1}^{|V|} count(deg) d{deg}$$
% By solving $q >= gCount(q)$ we roughly have:
% $$q \gtrapprox ({\frac{11 \times |V|}{3}})^{\frac{1}{2.5}}$$
% For example, $|V|=50000$, without (and with) preprocessing high degree nodes, the algorithm uses 1650 (and 255) colors.
%

The number of predicates being monitored in this test case is proportional to the number of edges.
%, even after preprocessing high degree nodes which account for less than $10\%$ of edges.

We note that the task performed by each client (i.e., choosing the color of a node) is just used as an example. It is easily generalized for other analysis of Social Media Graph (e.g., finding clusters, collaborative learning, etc.) 

The application motivated by \textit{Weather Monitoring} task considers a planar graph (e.g. a line or a grid) where the state of each node is affected by the state of its neighbors. 
In a line-based graph, all the nodes of the graph are arranged on a line and each client is assigned a segment of the line.
In a grid-based graph, the graph nodes are arranged on a grid. The clients are also organized as a grid and each client is responsible for a section of the grid of nodes.
In this application, we model a client that updates the state of each node by reading the state of its neighbors and updating its own state. This application can be tailored to vary the ratio of GET/PUT request.
This application is relevant to several practical planar graph problem such as weather forecasting \cite{FBMA2013PCS}, radio-coloring in wireless and sensor network \cite{PSNS1995PODC}, computing Voronoi diagram \cite{NXIA2008CCCG}.

Finally, the \textit{Conjunctive} application is an instance of distributed debugging where the predicate being detected (i.e., $\neg P$) is of the form $P_1 \wedge P_2 \wedge \cdots \wedge P_l$. Each local predicate $P_i$ becomes true with a probability $\beta$ and the goal of the monitors is to determine if the global conjunctive predicate $\neg P$ becomes true.
In this application, we monitor multiple conjunctive predicates simultaneously.
Since we can control how frequently these predicates become true by varying $\beta$, we can use it mainly to assess monitoring latency and stress the monitors. Conjunctive predicates are also useful in distributed testing such as to specify breakpoints.

%\textit{COLORING} is the main test case motivated by real world applications while \textit{GRAPH} and \textit{CONJUNCTIVE} are synthetic test cases designed to test our algorithm in aspects not covered by \textit{COLORING} as described in Sections \ref{subsec:result-aws-additional} and \ref{subsec:result-local}.

\textbf{Performance metrics and measurement.} We use throughput as the performance metrics in our experiments. Throughput can be measured at two perspectives: application, and Voldemort server. 
The two perspectives are not the same but related. One application request triggers multiple requests at Voldemort client. For example, one application PUT request is translated into one GET\_VERSION request (to obtain the last version of the key) and one PUT request (with a new incremented version) at the Voldemort client library. Then each Voldemort client request causes multiple requests at servers due to replication. Failures and timeout also make the counts at the applications and the servers differ. For example, an application request is served and counted at a server but if the server response is lost or arrives after the timeout, the request is considered unsuccessful and thus not counted at the application. Generally, servers' counts are greater than applications' counts.
In our experiments, we use the aggregated measurement at servers to assess the overhead of our approach since the monitors directly interfere with the operation of the server, and use aggregated measurement at applications to assess the benefit of our approach because that measurement is close to users' perspective. Hence, in the following sections, for the same experiment, we note that the measurements used for overhead and benefit evaluation are different.

\begin{figure}[t] %[tbhp]%[h]
% \vspace{-10pt}
\centering
\includegraphics[width=0.46\textwidth]{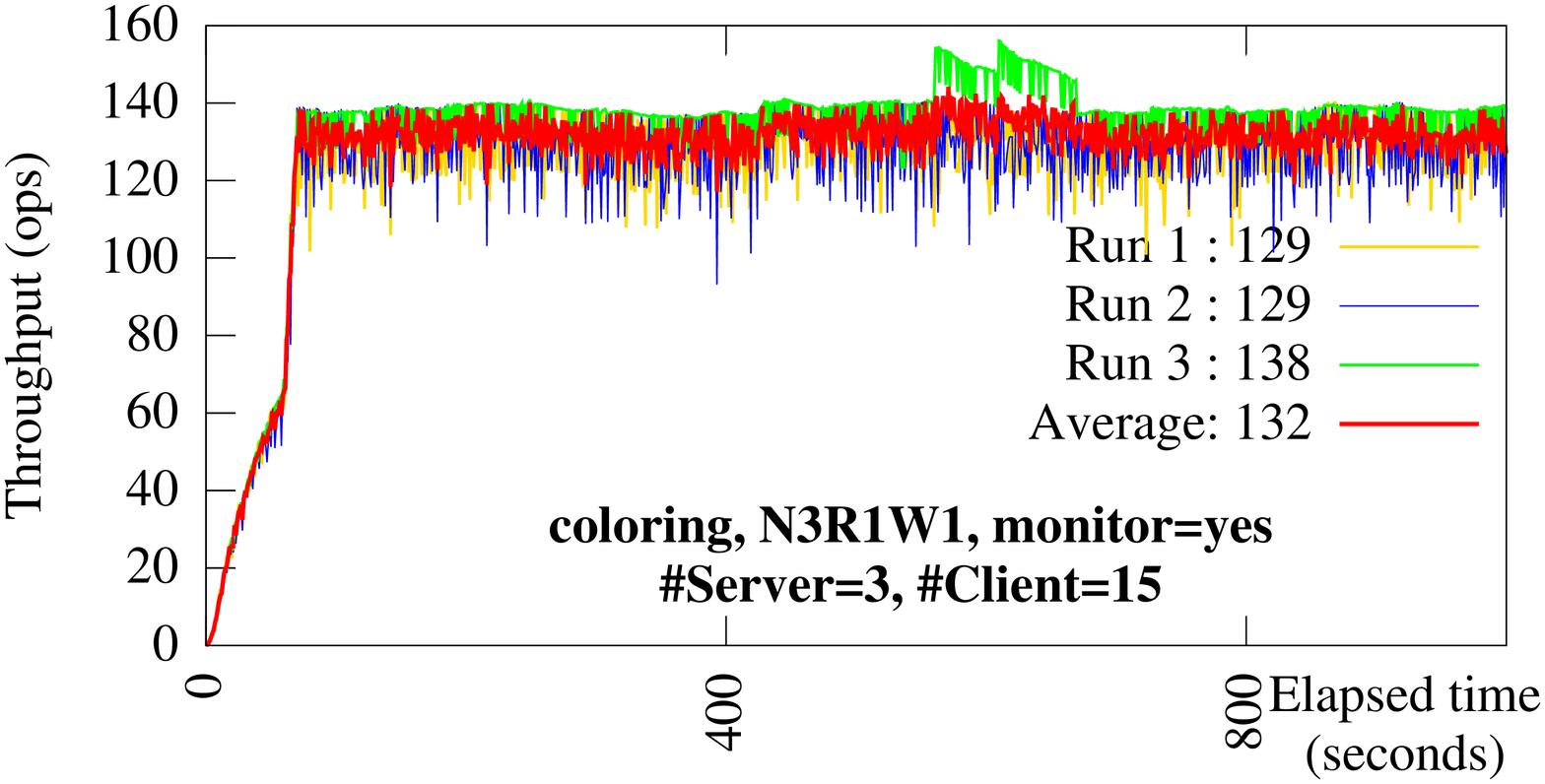}
    \caption{Illustration of result stabilization. The \textit{Social Media Analysis} application is run three times on Amazon AWS with monitoring enabled. Number of servers ($N$) = $3$. Number of clients per server ($C/N$) = $5$. Aggregated throughput measured by \textit{Social Media Analysis} application in three different runs and their average is shown. This average is used to represent the stable value of the application throughput.}
    \label{fig:rundiff}
\end{figure}

\textbf{Results stabilization.} We ran each experiment three times and used the average as the representative results for that experiment. Figure \ref{fig:rundiff} shows the stabilization of different runs of an experiment. Note that the values are aggregated from all applications. We observe that in every run, after a short period of initialization, the measurements converge on a stable value. When evaluating our approach, we use the values measured at the stable phase.
We also note that the aggregated throughput in Figure \ref{fig:rundiff} is not very high but expected. The pairwise round-trip latency between three AWS regions (Ohio, Oregon, Frankfurt) were \SIlist{76;103;163}{\ms}. The average round-trip latency was \SI{114}{\ms}. On M5.xlarge EC2 instances with a GP2 storage volume, the average I/O latency for a read and a write operation was roughly \SI{0.3}{\ms} and \SI{0.5}{\ms}, respectively. We will roughly estimate the cost of a GET request since in \textit{Social Media Analysis}, most operations are GET requests to read lock availability and colors of neighbors. Assume eventual consistency R1W1 is used, a GET request is executed by Voldemort client in two steps:

\begin{itemize}
\item[1.] Perform parallel request: client simultaneously sends GET requests to all servers (N = 3) and wait for responses with a timeout of \SI{500}{\ms}. The wait is over when either client gets responses from all servers or the timeout expires. In this case, the client will get all responses in about \SI{114.3}{\ms} (\SI{114}{\ms} for communication delay, and \SI{0.3}{\ms} for the read operation processing time at the server). 
\item[2.] Perform serial request: client checks if it has received enough required responses. If not, it has to send addition GET requests to servers to get enough number of responses. If after the additional requests, the required number of responses is not met, the GET request is considered unsuccessful. Otherwise, the result is returned. In the current case, the number of responses received (3) is greater than the required (R = 1). Thus this step is skipped.
\end{itemize}

From this discussion, a GET request takes roughly \SI{115}{\ms} to complete, on average. Since GET is the dominating operation in the \textit{Social Media Analysis} application, with 15 clients, the expected aggregated throughput is $\frac{15}{0.1143} \approx$ \SI{131}{ops}. The average throughput measured in experiments was \SI{132}{ops} (cf. Figure \ref{fig:rundiff}). 

If we run experiments where all machines are in the same region but in different availability zones, the aggregated throughput will be higher (cf. Figure \ref{fig:graph-aws}). For example, in the AWS North Virginia region, the average round-trip latency within an availability zone was about \SI{0.5}{\ms}, and between different availability zones was about \SI{1.4}{\ms}. Based on the discussion about GET request above, a GET request takes roughly \SI{0.8}{\ms} (\SI{0.5}{\ms} for network latency within an availability zone plus \SI{0.3}{\ms} for processing read request at the server). Similarly, a GET\_VERSION request takes \SI{0.8}{\ms}. Since we are using R1W1 configuration, an actual PUT request can be satisfied by the server within the same availability zone. Thus, an actual PUT request takes roughly \SI{1}{\ms} (\SI{0.5}{\ms} for network latency within an availability zone plus \SI{0.5}{\ms} for write operation processing time at the server).
A PUT request (consisting of a GET\_VERSION request and an actual PUT request) takes roughly \SI{1.8}{\ms}.
Assume the workload consists of \SI{50}{\percent} GETs and \SI{50}{\percent} PUTs, then on average, a request takes $0.5 \times 0.8 + 0.5 \times 1.8 =$ \SI{1.3}{\ms} = \SI[group-separator={}]{0.0013}{\s}. With 10 clients, the expected aggregate throughput is $\frac{10}{0.0013} =$ \SI{7692}{ops}. If the workload consists of \SI{75}{\percent} GETs and \SI{25}{\percent} PUTs, a request takes $0.75 \times 0.8 + 0.25 \times 1.8 =$ \SI{1.05}{\ms}  =  \SI[group-separator={}]{0.00105}{\s}, and the expected aggregate throughput is $\frac{10}{0.00105} =$ \SI{9524}{ops}. In our experiments, the aggregate throughput measured for \SI{50}{\percent} PUT and \SI{25}{\percent} PUT was \SI{7782}{ops} and \SI{9593}{ops}, respectively (cf. Figure \ref{fig:graph-aws-1-benefit} and \ref{fig:graph-aws-2-benefit}).
%\todo{D: I have added the expected throughput and actual throughput}

%% file: Experiment-results.tex
\subsection{Analysis of Throughput}\label{subsec:result-analysis-throughput}
%\subsection{Experimental Results on Amazon AWS}\label{subsec:result-aws-main}

\textbf{Comparison of Eventual Consistency with Monitors vs. Sequential Consistency}.
As discussed in the introduction, one of the problems faced by the designers is that they have access to an algorithm that is correct under sequential consistency but the underlying key-value store provides a weaker consistency. In this case, one of the choices is to pretend as if sequential consistency is available but monitor the critical predicate $P$. If this predicate is violated, we need to rollback to an earlier state and resume the computation from there. Clearly, this approach would be feasible if the monitored computation with eventual consistency provides sufficient benefit compared with sequential consistency. In this section, we evaluate this benefit. 

\begin{figure*}[t]
    %\vspace{-10pt}
    \begin{center}
        \subfigure[Benefit=57\% and 78\%]{%
            \label{fig:coloring-benefit}
            \includegraphics[width=0.46\textwidth]{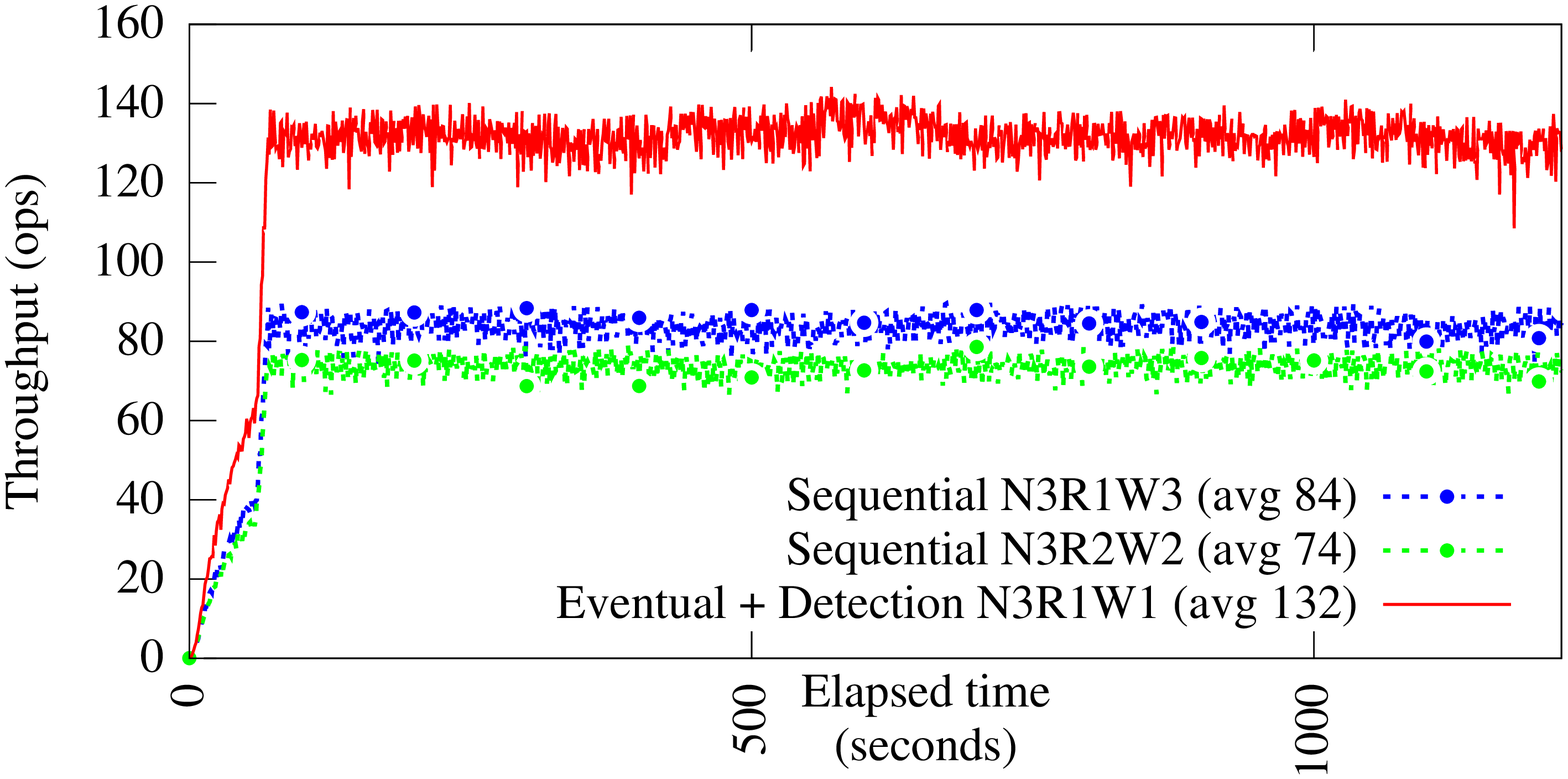}
        }
        \subfigure[Overhead on R1W1=1.4\%]{%
            \label{fig:coloring-overhead-1}
            \includegraphics[width=0.46\textwidth]{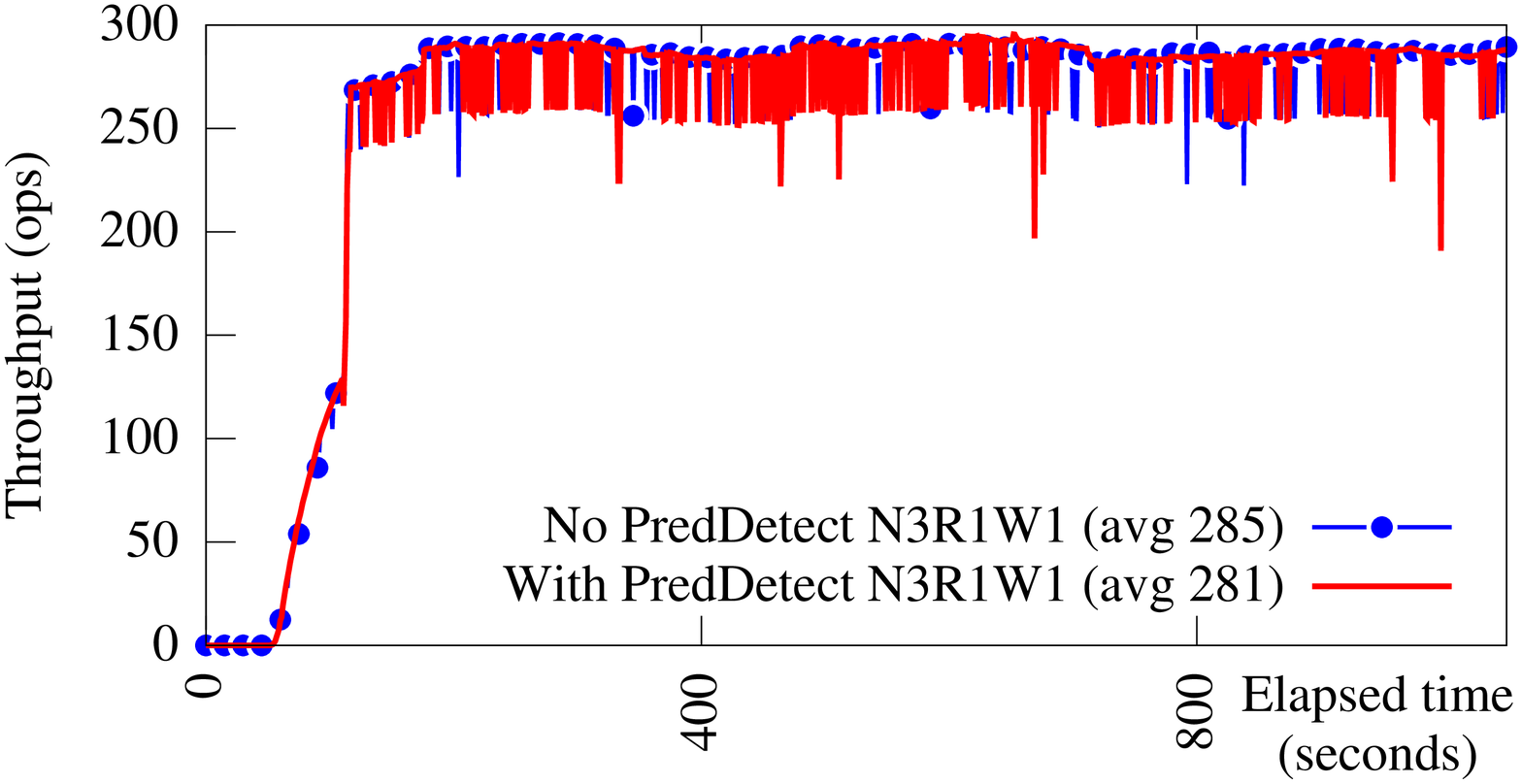}
        } \\%
        \subfigure[Overhead on R1W3=1.7\%]{%
           \label{fig:coloring-overhead-2}
            \includegraphics[width=0.46\textwidth]{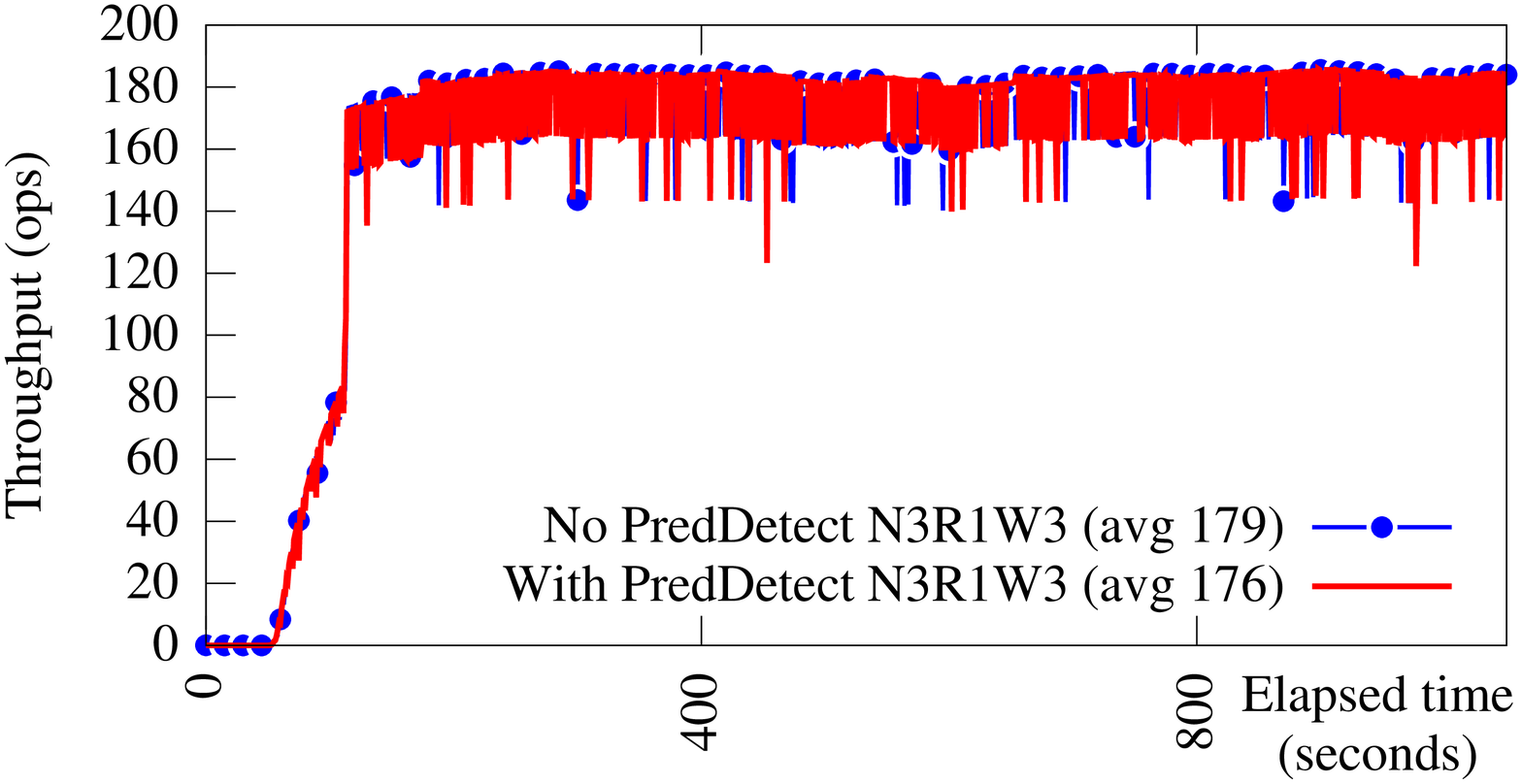}
        }
        \subfigure[Overhead on R2W2=1.3\%]{%
            \label{fig:coloring-overhead-3}
            \includegraphics[width=0.46\textwidth]{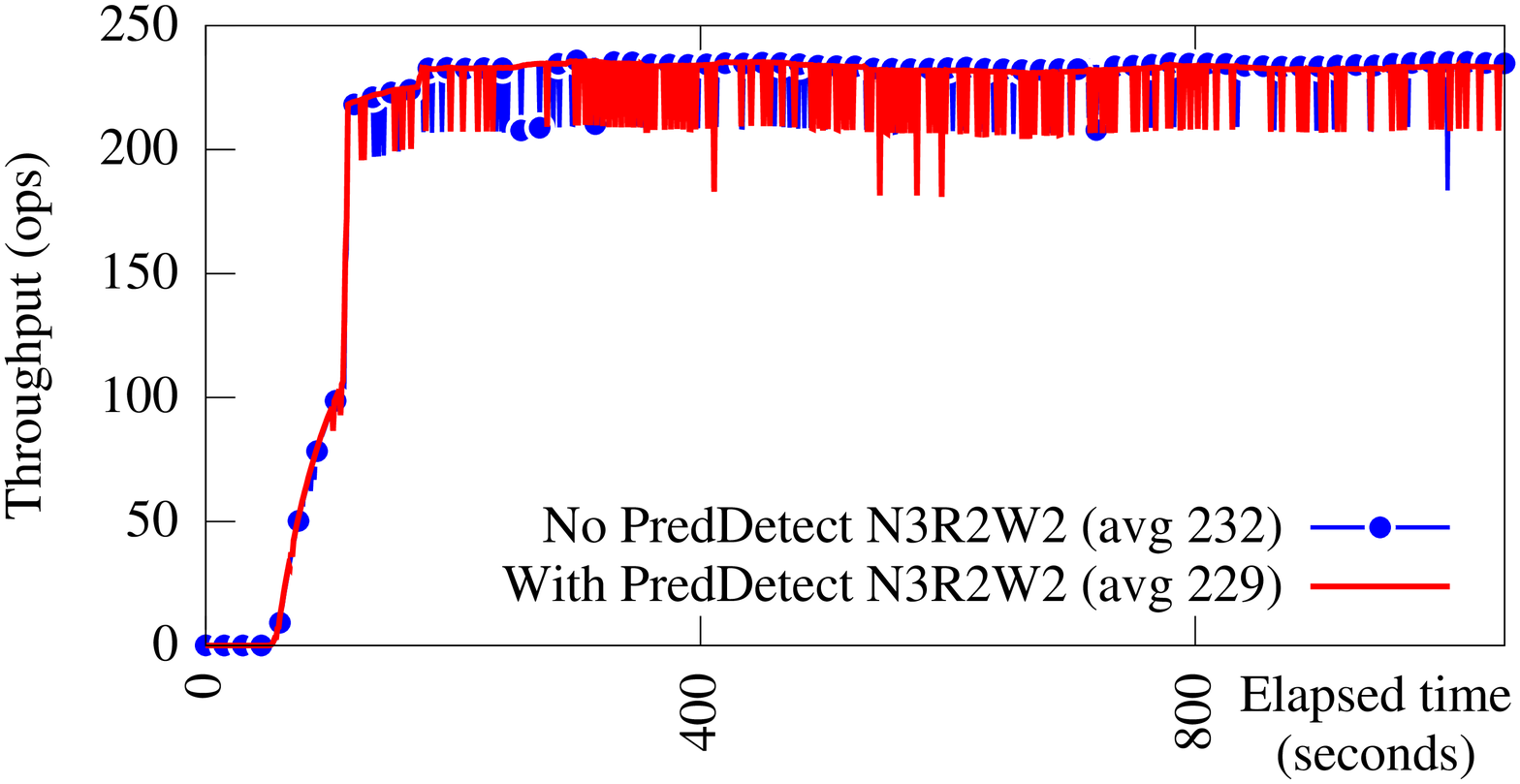}
        } %\\ %  ------- End of the first row ----------------------%
    \end{center}
    \vspace{-10pt}
    \caption{(AWS) \textit{Social Media Analysis} application, 3 servers, 15 clients. The benefit of eventual consistency with monitors vs. sequential consistency without monitors (throughput improvement compared to R1W3 and R2W2 is 57\% and 78\%, respectively), and the overhead of running monitors on each consistency setting (the overhead is less than $2\%$).}
    \label{fig:coloring-benefit-overhead}
\end{figure*}

Figure \ref{fig:coloring-benefit} compares the performance of our algorithms for eventual consistency with monitors and sequential consistency without monitors in the \textit{Social Media Analysis} application on the AWS environment. Using our approach, the client throughput was increased by \SI{57}{\percent} (for N3R1W3) and \SI{78}{\percent} (for N3R2W2). Note that the cost of a GET request is more expensive in N3R2W2 (the required number of positive acknowledgment is 2) than in N3R1W3 (the required acknowledgment is 1). Since in the \textit{Social Media Analysis} application GET requests dominates, the application performs better in N3R1W3 than in N3R2W2.

% --------------  Figure environment -------------------%
% -------------- End of figure environment ----------------------%

\textbf{Overhead of monitoring.} A weaker consistency model allows the application to increase the performance on a key-value store as illustrated above. To ensure correctness, a weaker consistency model needs monitors to detect violations and trigger rollback recovery when such violations happen. As a separate tool, the monitors are useful in debugging to ensure that the program satisfies the desired property throughout the execution. In all cases, it is desirable that the overhead of the monitors is small so that they would not curtail the benefit of weaker consistency or make the debugging cost expensive.

Figures \ref{fig:coloring-overhead-1} \ref{fig:coloring-overhead-2}, and \ref{fig:coloring-overhead-3} show the overhead of the monitors on different consistency settings in the \textit{Social Media Analysis} application. The overhead was between \SIrange{1}{2}{\percent}. At its peak, the number of active predicates being monitored reached \num{20 000} predicates. Thus, the overhead remains reasonable even with monitoring many predicates simultaneously.

\subsection{Analysis of System and Application Factors}

\textbf{Impact of workload characteristics.} In order to evaluate the impact of workload on our algorithms we ran the \textit{Weather Monitoring} application where the proportional of PUT and GET was configurable. 
The number of servers was 5 and the number of clients was 10. The machines hosting the servers and clients were in the same AWS region (North Virginia, U.S.) but in 5 different availability zones. We choose machines in the same region to reduce the latency (to less than \SI{2}{\ms}), thus increasing the throughput measure and stressing the servers and the monitors. If we put the clients and servers in different regions (e.g., Frankfurt Germany, Oregon USA, Ohio USA) then the throughput for 15 clients is low. To stress it further, we would have to add hundreds of clients which is very expensive. Hence, for the stress test, we put the servers and clients in the same region.

%choose machines in regions similar to the experiments with \textit{Social Media Analysis} above, in order to achieve the same level of aggregated throughput of this section (cf. Figure \ref{fig:graph-aws}), we would have to increase the number of clients 100 times. 

\begin{figure*}[th]
    \begin{center}
        \subfigure[PUT=25\%. Benefit=18\%]{%
            \label{fig:graph-aws-1-benefit}
            \includegraphics[width=0.3\textwidth]{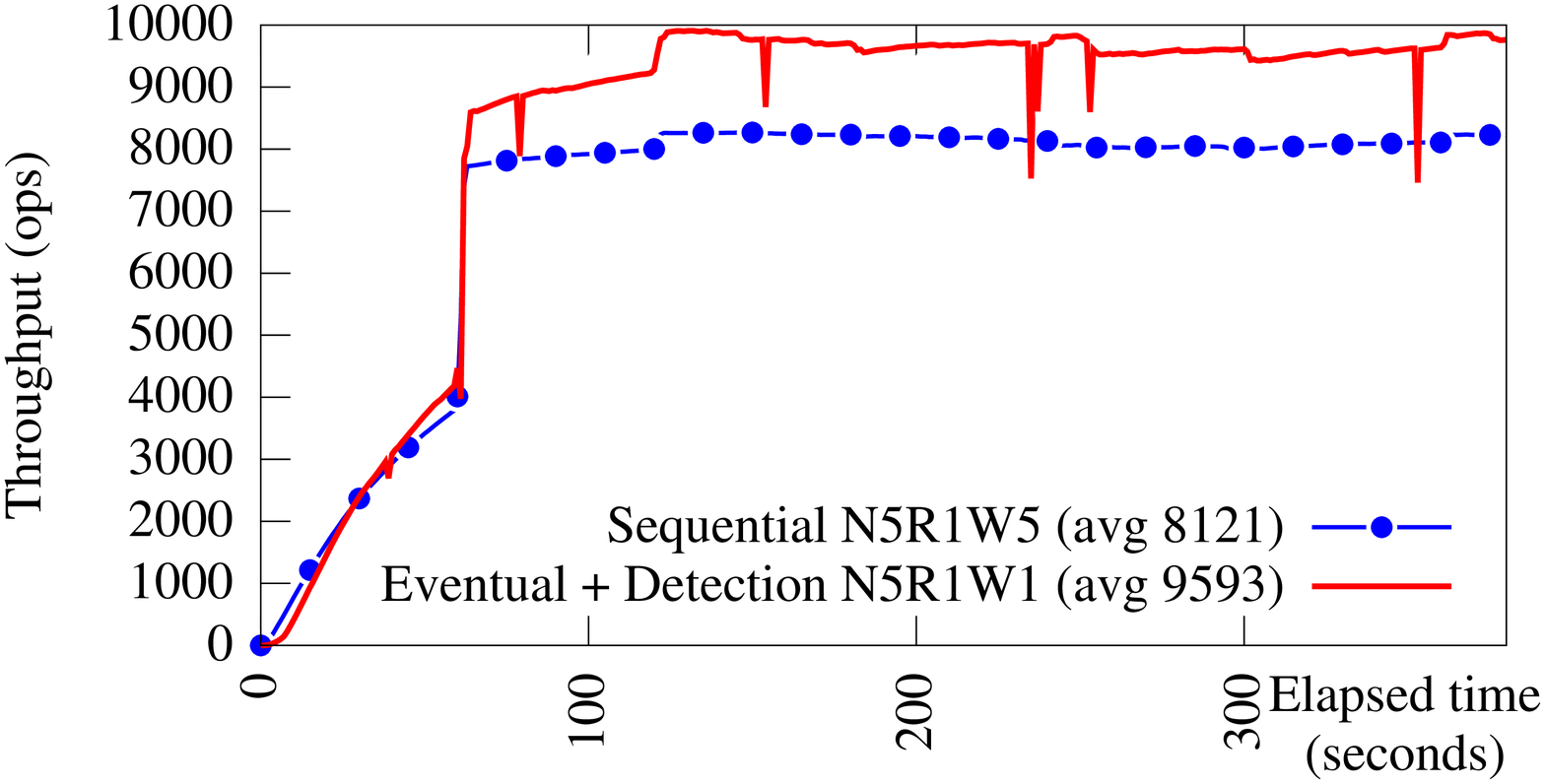}
        }%
        \subfigure[PUT=50\%. Benefit=37\%]{%
           \label{fig:graph-aws-2-benefit}
           \includegraphics[width=0.3\textwidth]{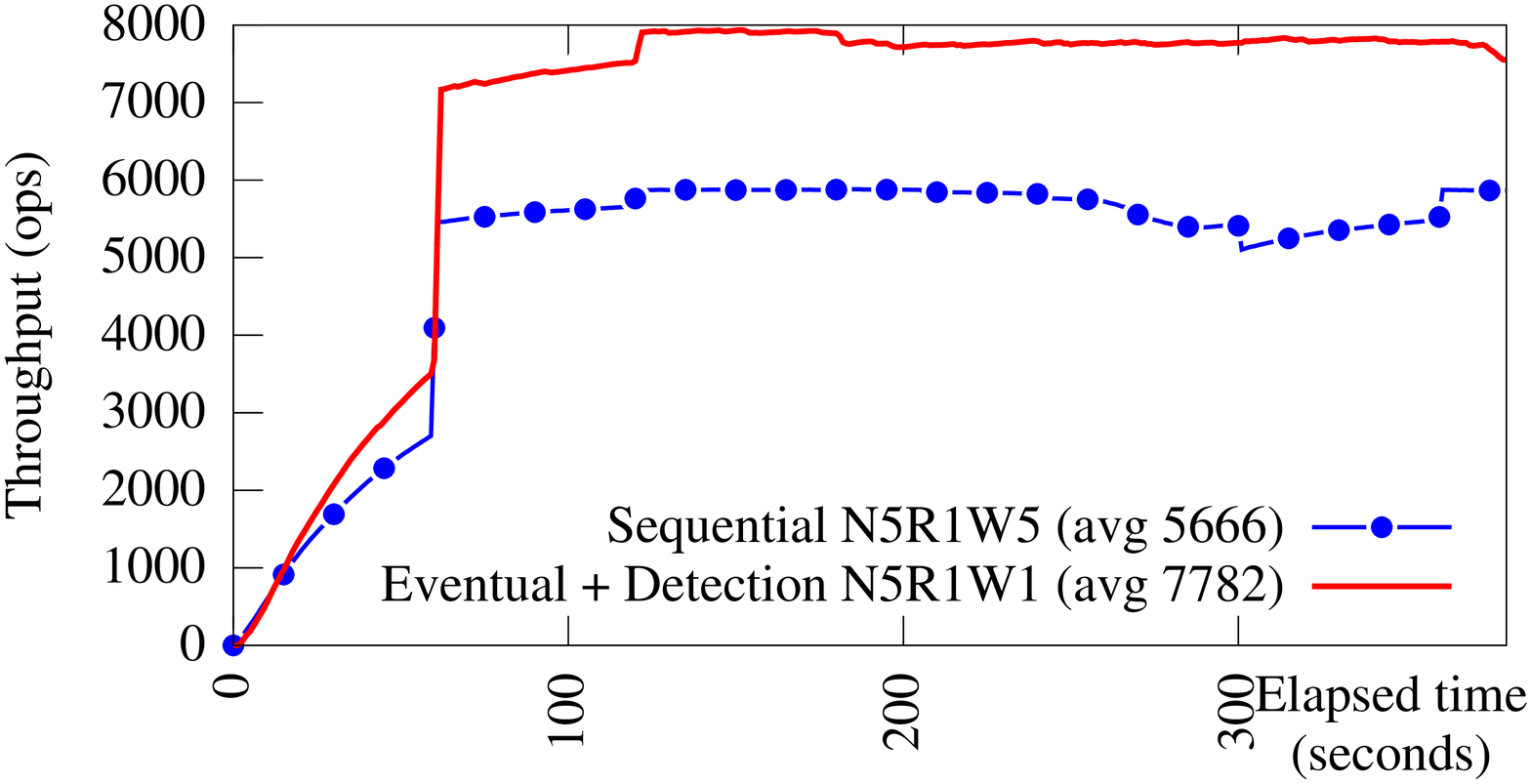}
        }
        \subfigure[PUT=50\%. Overhead=4\%]{%
            \label{fig:graph-aws-3-overhead}
            \includegraphics[width=0.3\textwidth]{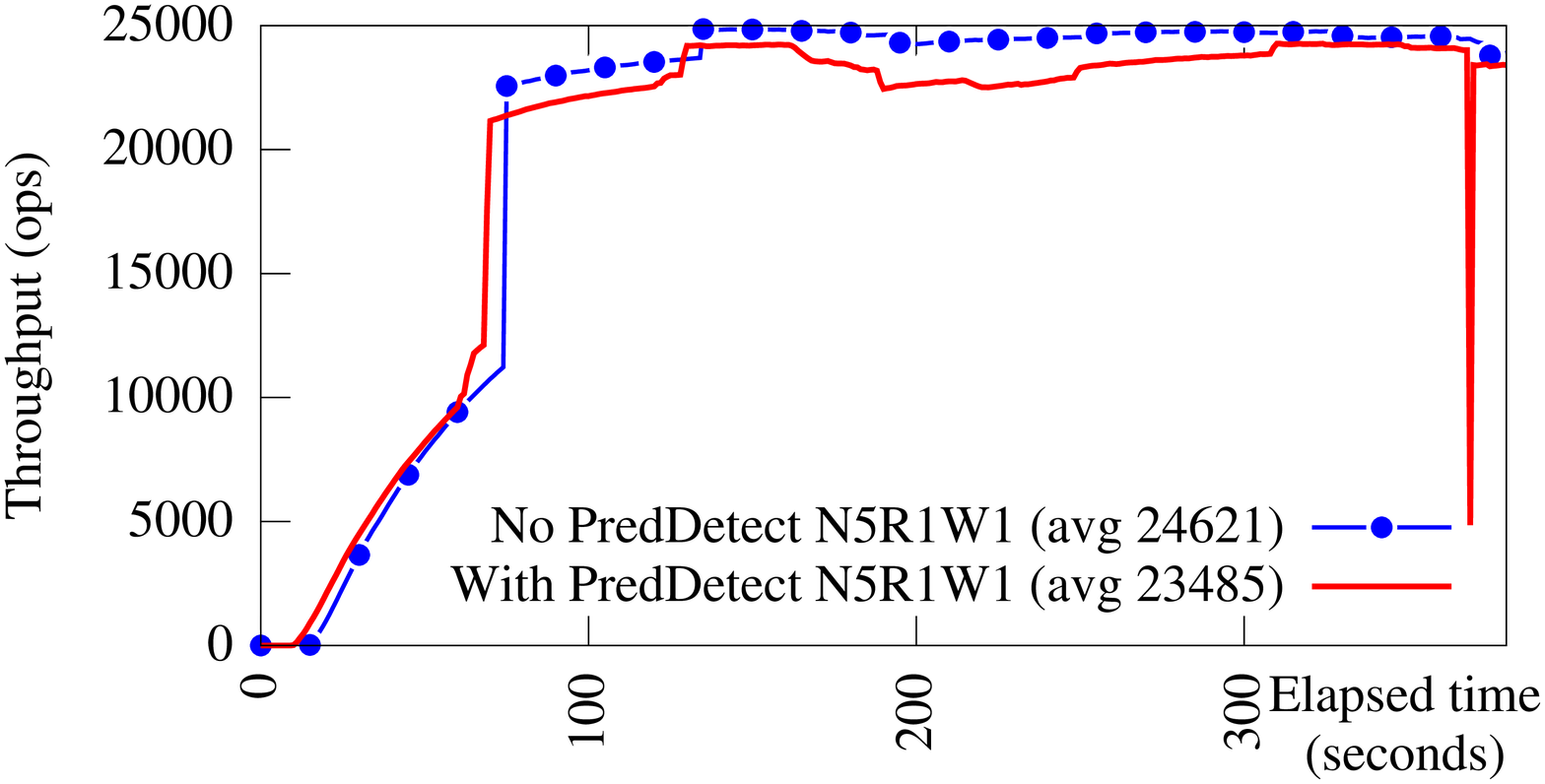}
        } %\\ %  ------- End of the first row ----------------------%
    \end{center}
    \caption{Benefit and overhead of monitors in \textit{Weather Monitoring} application. Percentage of PUT requests is 25\% and 50\% Number of servers =$5$. Number of clients = $10$. Machines are on the AWS North Virginia region but in different availability zones.}
    \label{fig:graph-aws}
\end{figure*}

From Figures \ref{fig:graph-aws-1-benefit} and \ref{fig:graph-aws-2-benefit}, we find that when the percentage of PUT request increased from \SIrange{25}{50}{\percent}, the benefit over sequential consistency (N5R1W5 in this case) increased from \SIrange{18}{37}{\percent}. 
This is because the cost for a PUT request is expensive in N5R1W5 as a PUT request is successful only when it is confirmed by all 5 servers. Thus, when the proportion of PUT increases, the performance of N5R1W5 decreases. In such cases, sequential settings that balance R and W (e.g. N5R3W3) will perform better than sequential settings that emphasize W (e.g. N5R1W5). When GET requests dominate, it is vice versa (cf. Figure \ref{fig:coloring-benefit}). 
We also observe that, when PUT percentage increased and other parameters were unchanged, the aggregated throughput measured at clients decreased. That is because a PUT request consists of a GET\_VERSION request (which is as expensive as a GET request) and an actual PUT request, therefore a PUT request takes a longer time to complete than a GET request does.

Regarding overhead, Figure \ref{fig:graph-aws-3-overhead} shows that the overhead was \SI{4}{\percent} when PUT percentage was \SI{50}{\percent}. Note that in \textit{Weather Monitoring} application, the number of predicates being monitored is proportional to the number of clients. Thus, the overhead remains reasonable even when monitoring several predicates simultaneously and the servers are stressed.

The number of violations detected in this experiment was only one instance in executions with a total time of \SI{18 000}{\ms}. The violation was detected within \SI{20}{\ms}. 

%This observation again supports that although the violation of mutual exclusion in eventual consistency is a theoretical possibility, it is quite rare.

\textbf{Impact of network latency.}
We ran experiments on the local lab network (cf. Section \ref{subsec:experiment-setup}) where the one-way latency within a region (cf. Figure \ref{fig:local-lab-network}) was \SI{1}{\ms} and one-way latency between regions varied from \SIrange{50}{100}{\ms}. 
The number of clients per each server varied between $10$ and $20$. The values in sub-columns ``server'' and ``app'' are the aggregate throughput measured at the servers and at the applications (unit is \si{ops}).
In Table \ref{table:local-overhead-benefit}, the overhead is computed by comparing server measurements when the monitors are enabled and disabled. The benefit is computed by comparing application measurements on sequential consistency without monitoring to those on eventual consistency with monitoring. 
For example, when one-way latency is \SI{50}{\ms}, if we run the \textit{Weather Monitoring} application on N3R1W3 without monitoring, the aggregate server throughput is \SI{649}{ops} (Table \ref{table:local-overhead-benefit}, column 12 (N3R1W3 $\rightarrow$ server) and row 6 (\SI{50}{\ms} $\rightarrow$ \textit{Weather Monitoring} $\rightarrow$ Monitor = no)) and the aggregate client throughput is \SI{313}{ops}.
If we run the same application on N3R1W3 with monitoring, the server throughput is \SI{628}{ops} (Table \ref{table:local-overhead-benefit}, column 12 and row 5). The overhead of monitoring \textit{Weather Monitoring} application on N3R1W3 is $(649-628)/649 = 3.2\%$. The client throughput when run the same application on N3R1W1 with monitoring is \SI{454}{ops} (Table \ref{table:local-overhead-benefit}, column 7, row 5). Thus, the benefit of eventual consistency with monitoring vs. sequential consistency N3R1W3 is $(454-313)/313 = 45\%$.
%For example, when one-way latency is \SI{50}{\ms}, if we run the \textit{Weather Monitoring} application on N3R1W3, the overhead of monitoring is $(649-628)/649 = 3.2\%$. If we run the same application on eventual consistency N3R1W1 with monitoring, the benefit (compared to running on N3R1W3 without monitoring) is $(454-313)/313 = 45\%$.
%

From Table \ref{table:local-overhead-benefit}, as latency increases, the benefit of eventual consistency with monitoring vs. sequential consistency increases. For example when one-way latency increased from \SIrange{50}{100}{\ms}, in \textit{Social Media Analysis} application, the benefit of eventual consistency with monitoring vs. sequential consistency R1W3 increased from \SIrange{47}{60}{\percent}. In the case of R2W2, the increase was from \SIrange{65}{80}{\percent}.
This increase is expected because when latency increases, the chance for a request to be successful at a remote server decreases. Due to strict replication requirement of sequential consistency, the client will have to repeat the request again. On the other hand, on eventual consistency, requests are likely to be successfully served a local server and the client can continue regardless of results at remote servers. Hence, as servers are distributed in more geographically disperse locations, the benefit of eventual consistency is more noticeable. Regarding overhead, it was generally less than \SI{4}{\percent}. In all cases, the overhead was at most \SI{8}{\percent}.

\begin{table*}[htbp]
%\vspace{-10pt}
\caption{Overhead and benefit of monitors in local lab network. For \textit{Conjunctive} and \textit{Weather Monitoring}, PUT percentage is 50\%.}
\vspace{-10pt}
\begin{center}
\begin{tabular}{|c|l|c|c|r|l|r|r|l|r|l|r|l|r|l|}
\hline
\multirow{2}{*}{\parbox{0.8cm}{Latency (ms)}} & \multirow{2}{*}{Application}  & \multirow{2}{*}{\parbox{0.8cm}{Client/ Server}}  & \multirow{2}{*}{\parbox{0.8cm}{Monitor}} & \multicolumn{ 3}{c|}{N3R1W1  } & \multicolumn{ 4}{c|}{N3R2W2  } & \multicolumn{ 4}{c|}{N3R1W3  } \\ \cline{5-15}
&  &  &  & \multicolumn{1}{l|}{server} & {\parbox{0.9cm}{overhead}} & \multicolumn{1}{l|}{app} & \multicolumn{1}{l|}{server} & {\parbox{0.9cm}{overhead}} & \multicolumn{1}{l|}{app} & benefit & \multicolumn{1}{l|}{server} & {\parbox{0.9cm}{overhead}} & \multicolumn{1}{l|}{app} & benefit \\ \hline
\multirow{6}{*}{50} & \multirow{2}{*}{Conjunctive} & \multirow{2}{*}{20} & yes & 821 & \multicolumn{1}{r|}{-0.2\%} & 470 & 842 & \multicolumn{1}{r|}{0.6\%} & 375 & \multicolumn{1}{r|}{25.3\%} & 588 & \multicolumn{1}{r|}{3.3\%} & 337 & \multicolumn{1}{r|}{40.7\%} \\ \cline{4-15}
 & &  & no & 819 &  & 470 & 847 &  & 375 &  & 608 &  & 334 &  \\ \cline{2-15}
 & \multirow{2}{*}{\parbox{1.5cm}{Weather Monitoring}} & \multirow{2}{*}{20} & yes & 924 & \multicolumn{1}{r|}{0.2\%} & 454 & 795 & \multicolumn{1}{r|}{7.1\%} & 345 & \multicolumn{1}{r|}{27.2\%} & 628 & \multicolumn{1}{r|}{3.2\%} & 312 & \multicolumn{1}{r|}{45.0\%} \\ \cline{4-15}
 &  &  & no & 926 &  & 453 & 856 &  & 357 &  & 649 &  & 313 &  \\ \cline{2-15}
 & \multirow{2}{*}{\parbox{1.5cm}{Social Media Analysis}} & \multirow{2}{*}{10} & yes & 560 & \multicolumn{1}{r|}{0.2\%} & 258 & 367 & \multicolumn{1}{r|}{0.5\%} & 156 & \multicolumn{1}{r|}{65.4\%} & 344 & \multicolumn{1}{r|}{7.8\%} & 174 & \multicolumn{1}{r|}{47.4\%} \\ \cline{4-15}
 &  &  & no & 561 &  & 267 & 369 &  & 156 &  & 373 &  & 175 &  \\ \hline
\multirow{6}{*}{100} & \multirow{2}{*}{Conjunctive} & \multirow{2}{*}{20} & yes & 476 & \multicolumn{1}{r|}{0.4\%} & 270 & 491 & \multicolumn{1}{r|}{-0.2\%} & 218 & \multicolumn{1}{r|}{23.3\%} & 354 & \multicolumn{1}{r|}{0.0\%} & 191 & \multicolumn{1}{r|}{42.1\%} \\ \cline{4-15}
 &  &  & no & 478 &  & 271 & 490 &  & 219 &  & 354 &  & 190 &  \\ \cline{2-15}
 & \multirow{2}{*}{\parbox{1.5cm}{Weather Monitoring}} & \multirow{2}{*}{20} & yes & 544 & \multicolumn{1}{r|}{0.7\%} & 266 & 500 & \multicolumn{1}{r|}{1.0\%} & 209 & \multicolumn{1}{r|}{28.5\%} & 371 & \multicolumn{1}{r|}{0.8\%} & 176 & \multicolumn{1}{r|}{49.4\%} \\ \cline{4-15}
 &  &  & no & 548 &  & 273 & 505 &  & 207 &  & 374 &  & 178 &  \\ \cline{2-15}
 & \multirow{2}{*}{\parbox{1.5cm}{Social Media Analysis}} & \multirow{2}{*}{10} & yes & 287 & \multicolumn{1}{r|}{0.0\%} & 135 & 236 & \multicolumn{1}{r|}{0.0\%} & 74 & \multicolumn{1}{r|}{80\%} & 185 & \multicolumn{1}{r|}{-0.5\%} & 86 & \multicolumn{1}{r|}{60.7\%} \\ \cline{4-15}
 &  &  & no & 287 &  & 133 & 236 &  & 75 &  & 184 &  & 84 &  \\ \hline
\end{tabular}
\end{center}
%\vspace{-10pt}
\label{table:local-overhead-benefit}
\end{table*}

\subsection{Analysis of Violations and Detection Latency}

\textit{Detection latency} is the time elapsed between the violation of the predicate being monitored and the time when the monitors detect it. 
In our experiment with \textit{Social Media Analysis} applications on eventual consistency (N3R1W1), in several executions of total \SI{9 000}{\s}, we detected only 2 instances of mutual exclusion violations. Detection latency for those violations were \SIlist{2 238;2 213}{\ms}. So for \textit{Social Media Analysis} application, violations could happen on eventual consistency every \SI{4 500}{\s} on average. %This rate is small enough so that the cost of general rollback recovery is outweighed by the performance benefit of eventual consistency. 

In order to evaluate the detection latency of monitors with higher statistical reliability, we need experiments where violations are more frequent.
In these experiments, the clients ran \textit{Conjunctive} application in the same AWS configuration as \textit{Weather Monitoring} application above.
The monitors have to detect violations of conjunctive predicates of the form $P = P_1 \wedge P_2 \wedge \cdots P_{10}$. Furthermore, we can control how often these predicates become true by changing when local predicates are true. In these experiments, the rate of local predicate being true ($\beta$) was \SI{1}{\percent}, which was chosen based on the time breakdown of some MapReduce applications \cite{RRPBK07HPCA, BPERST10SIGMOD}. 
The PUT percentage was \SI{50}{\percent}. The \textit{Conjunctive} application is designed so that the number of predicate violations is large and to stress the monitors. We considered both eventual consistency and sequential consistency. 
Table \ref{table:response-time-freq} shows detection latency distribution of more than \num{20 000} violations recorded in the \textit{Conjunctive} experiments. Predicate violations are generally detected promptly. Specifically, \SI{99.93}{\percent} of violations were detected in \SI{50}{\ms}, \SI{99.97}{\percent} of violations were detected in \SI{1}{\s}.
There were rare cases where detection latency was greater than ten seconds. Among all the runs, the maximum detection latency recorded was \SI{17}{\s}, the average was \SI{8}{\ms}.

%The average detection latency is $8\ ms$, which is much smaller than the average detection latency of $2226\ ms$ in Section \ref{subsec:result-aws-main} experiment. Note that the average network latencies in two Sections are $2 \ ms$ and $114 \ ms$.

\begin{table}[htbp]
\caption{Response time in $20,647$ conjunctive predicate violations}
\begin{center}
\begin{tabular}{|l|l|l|}
    \hline
    Response time (milliseconds) & Count & Percentage \\
    \hline
    $< 50$ & 20,632 & $99.927\%$ \\
    \hline
    $50 - 1,000$ & 6 & $0.029\%$ \\
    \hline
    $1,000 - 10,000$ & 3 & $0.015\%$ \\
    \hline
    $10,000 - 17,000$ & 6 & $0.029\%$ \\
    \hline
\end{tabular}
\label{table:response-time-freq}
\end{center}
\end{table}

Regarding overhead and benefit, the overhead of monitors on N5R1W1, N5R1W5, and N5R3W3 was \SIlist{7.81; 6.50; 4.66}{\percent}, respectively. The benefit of N5R1W1 over N5R1W5 and N5R3W3 was \SIlist{27.90; 20.16}{\percent}, respectively.
%
%Observing Figures \ref{fig:graph-aws-3-overhead}, we find that there are a few moments where the aggregated throughput of all servers drops down. This is happening because some or all servers are spending a significant computation for the local predicate detection module. Such moments are infrequent. Furthermore, since each M5.large server used in our experiment has only two Voldemort server threads, when one of the thread is running the predicate detection module, the aggregated throughput would be clearly affected. In a typical setting, each server runs a large number of server threads. Thus, when a server process is running predicate detection module, the decrease in aggregated throughput would be less noticeable.

%In results presented so far, we have considered experiments under various factors but network latency.
%In Section \ref{subsec:result-aws-main}, AWS machines are located on different continents with long latency. In Section \ref{subsec:result-aws-additional}, AWS machines are located in the same region with small latency. However, the applications are different in those sections. 
%In the next section, we consider the impact of network latency on the monitors.

%\subsection{Experimental Results on Local Lab Network}\label{subsec:result-local}

%\subsection{Dealing with Livelocks}
\subsection{Evaluating Strategies for Handling Livelocks}
%\todo{How about this title?}

%\todo{S: This section only talks about social media. What about weather monitoring?>}

\begin{figure*}[t]
    \begin{center}
        \subfigure[]{%
           \label{fig:fig:livelock-social}
           \includegraphics[width=0.46\textwidth]{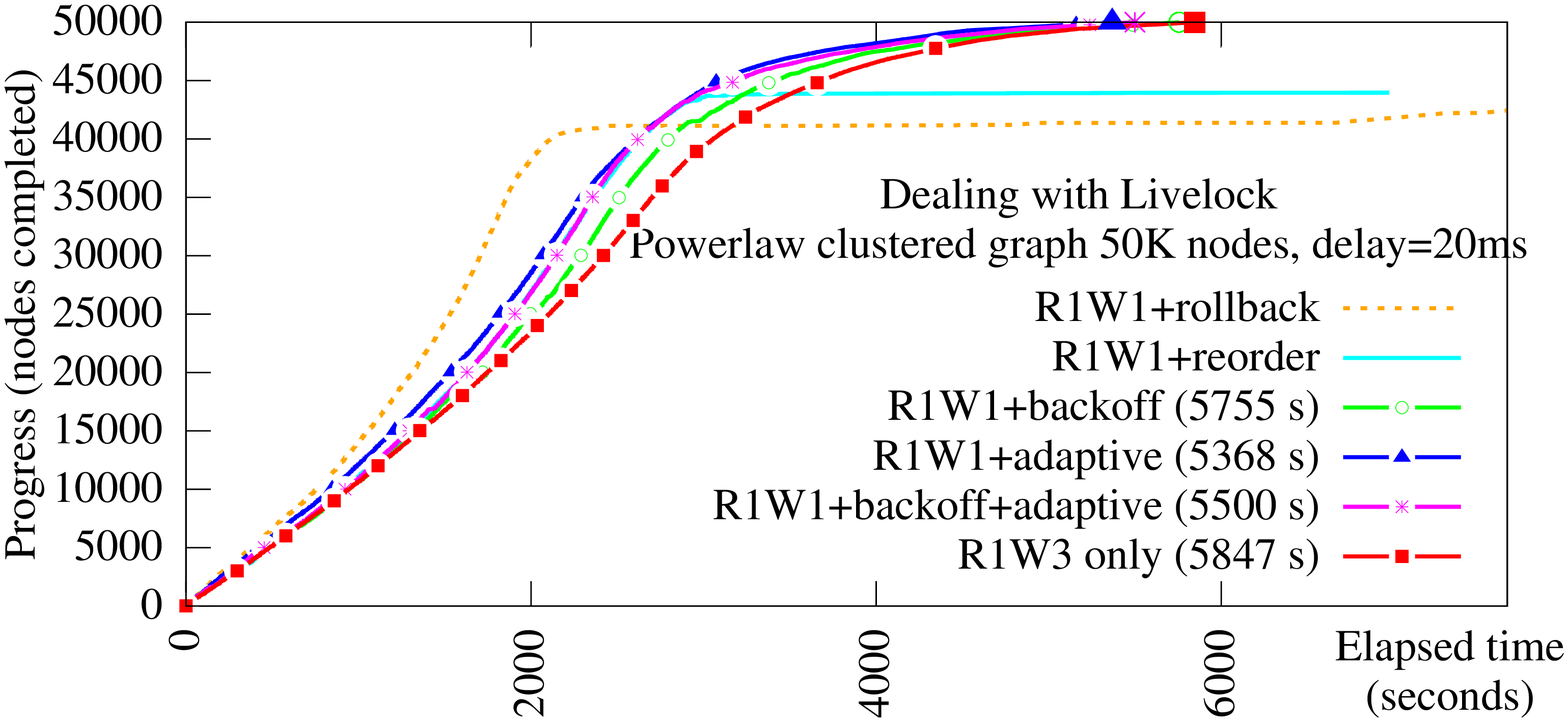}
        }
        \subfigure[]{%
           \label{fig:livelock-weather}
           \includegraphics[width=0.46\textwidth]{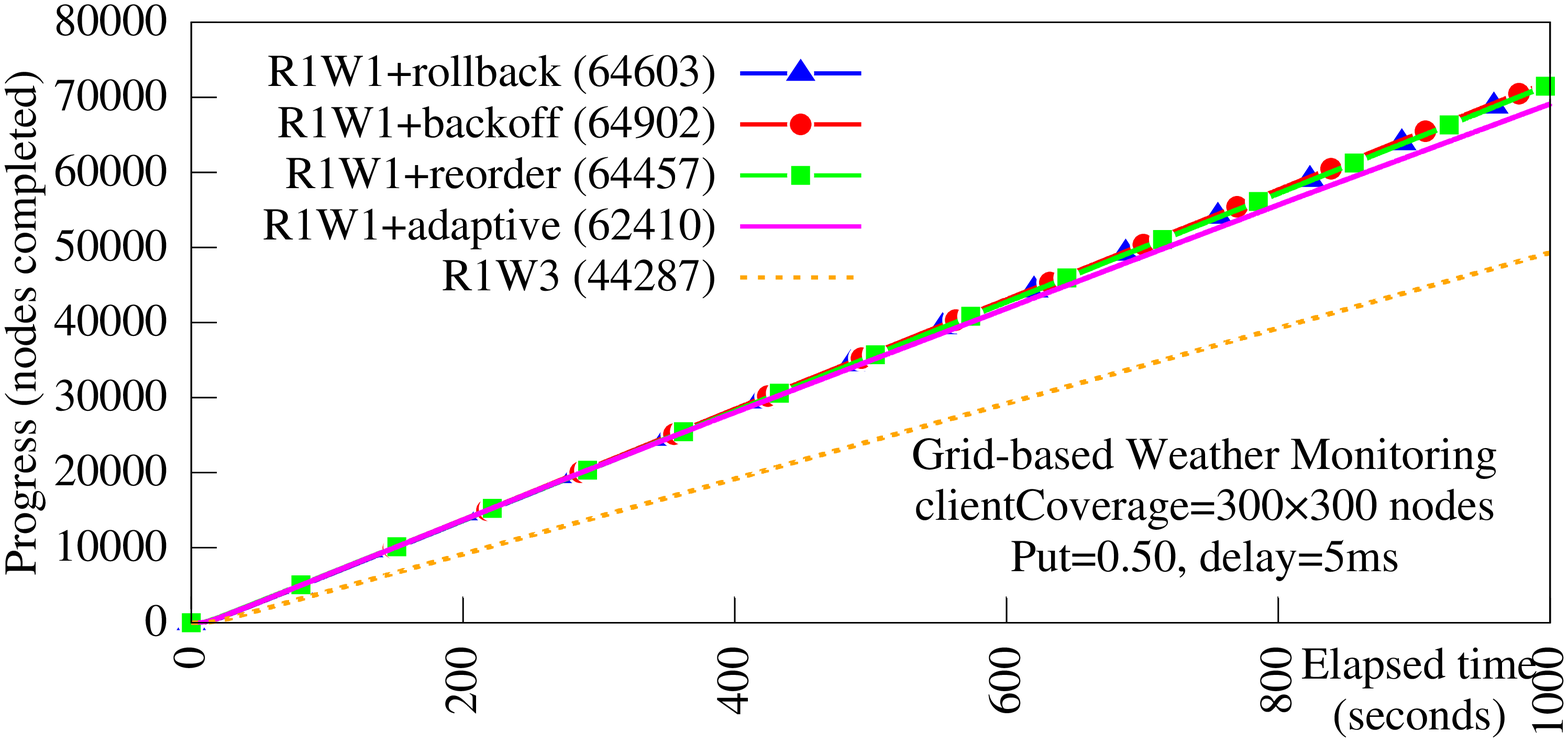}
        }
    \end{center}
    \caption{Effectiveness of livelock handling mechanisms. Number of servers=3, number of clients=30. We observed that adaptive mechanism worked best for \textit{Social Media Analysis} (Figure \ref{fig:fig:livelock-social}), and backoff mechanism worked best for \textit{Weather Monitoring} (Figure \ref{fig:livelock-weather}).}
    \label{fig:livelock}
\end{figure*}

In this section, we evaluate the effect of rollback mechanisms. We consider the evaluation of the \textit{Social Media Analysis} with a power-law graph and \textit{Weather Monitoring} with grid-based graph (cf. Section \ref{subsec:experiment-setup} for description of the graphs). We consider the execution with sequential consistency, eventual consistency with rollback but no mechanism for dealing with livelocks, and eventual consistency with one or more mechanism for dealing with livelocks. The results are shown in Figure \ref{fig:livelock}.

From this figure, we observe that the impact of livelocks is not the same in different applications. In particular, for terminating applications like \textit{Social Media Analysis}, if the livelock issue is ignored, the computation does not terminate. Likewise, computation does not terminate with the mechanism of reordering of remaining tasks upon rollback. This is anticipated, in part, because recurrence of rollback happens in end-stages where the number of remaining tasks is low. 
On the other hand, for non-terminating application like \textit{Weather Monitoring}, livelocks do not cause the computation to stall. Except for adaptive consistency, the effectiveness of different livelock handling strategies are almost similar. 
From Figure \ref{fig:livelock}, we observe that rollback with adaptive consistency works best for terminating applications, and rollback with backoff works best for non-terminating applications. 
Therefore, we choose these mechanisms to handle livelocks in the detailed analysis of applications in Section \ref{subsec:applications}. 

% \begin{figure}[tbhp]%[h]
% % \vspace{-10pt}
% \centering
% \includegraphics[width=0.46\textwidth]{figures-delay/{node-progress-graph-20ms-livelock}.eps}
%     \caption{Effectiveness of different solutions for livelocks in \textit{Social Media Analysis} application. Number of servers=3, number of clients=30}
%     \label{fig:livelock}
% \end{figure}

%shows that using backoff can help clients to overcome livelocks and finish the computation a little faster ($2\%$) than sequential consistency. On the other hand, reordering the node does not help with livelocks since with a few high degree nodes at the final phase, the chance of conflicts between any pair of tasks is high.

\subsection{Analysis of Applications}
\label{subsec:applications}

In this section, to illustrate the benefit of our approach, we run the recovery algorithm described in section \ref{subsec:rollback-mechanism} for two applications: \textit{Weather Monitoring} and \textit{Social Media Analysis}. We do not consider \textit{Conjunctive}, as it was designed explicitly to cause too many violations for the purpose of detecting latency of violations. The analysis was performed in our local lab network with the round-trip latency varying between \SIrange{5}{50}{\ms}. We use the approach in Section \ref{subsec:experiment-setup} to add additional delays to evaluate the behavior of the application in a realistic setting where replicas are not physically co-located. 
In order to deal with livelocks, we utilize the backoff mechanism for \textit{Weather Monitoring} application, and adaptive mechanism for \textit{Social Media Analysis} application.
The number of servers was 3 and the number of clients was 30.

\begin{figure*}[ht]
    \begin{center}
        \subfigure[]{%
           \label{fig:node-progress-weather-line-put0.50-delay10ms-lineSize1000}
           \includegraphics[width=0.45\textwidth]{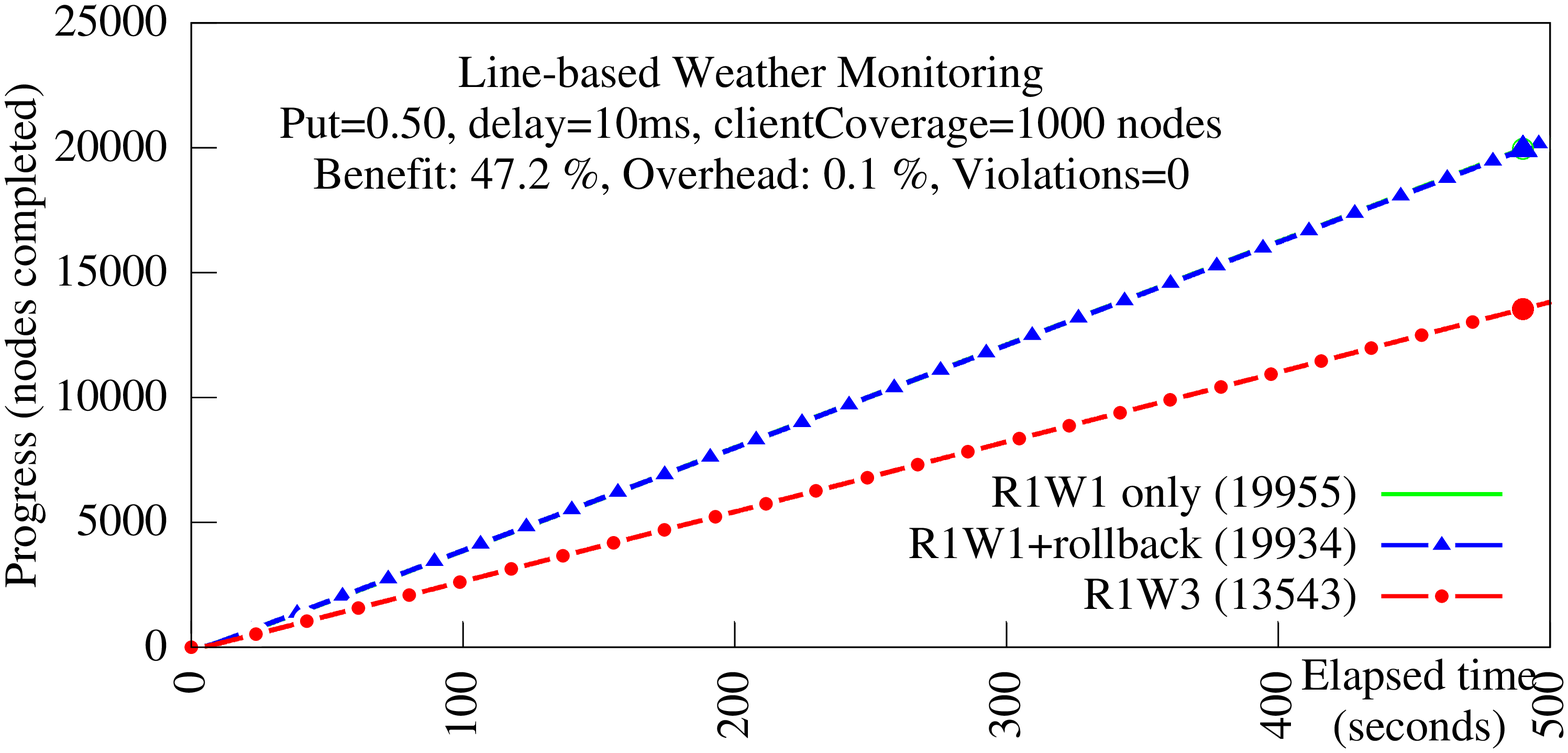}
        }
        \subfigure[]{%
           \label{fig:node-progress-weather-line-put0.50-delay10ms-lineSize500}
           \includegraphics[width=0.45\textwidth]{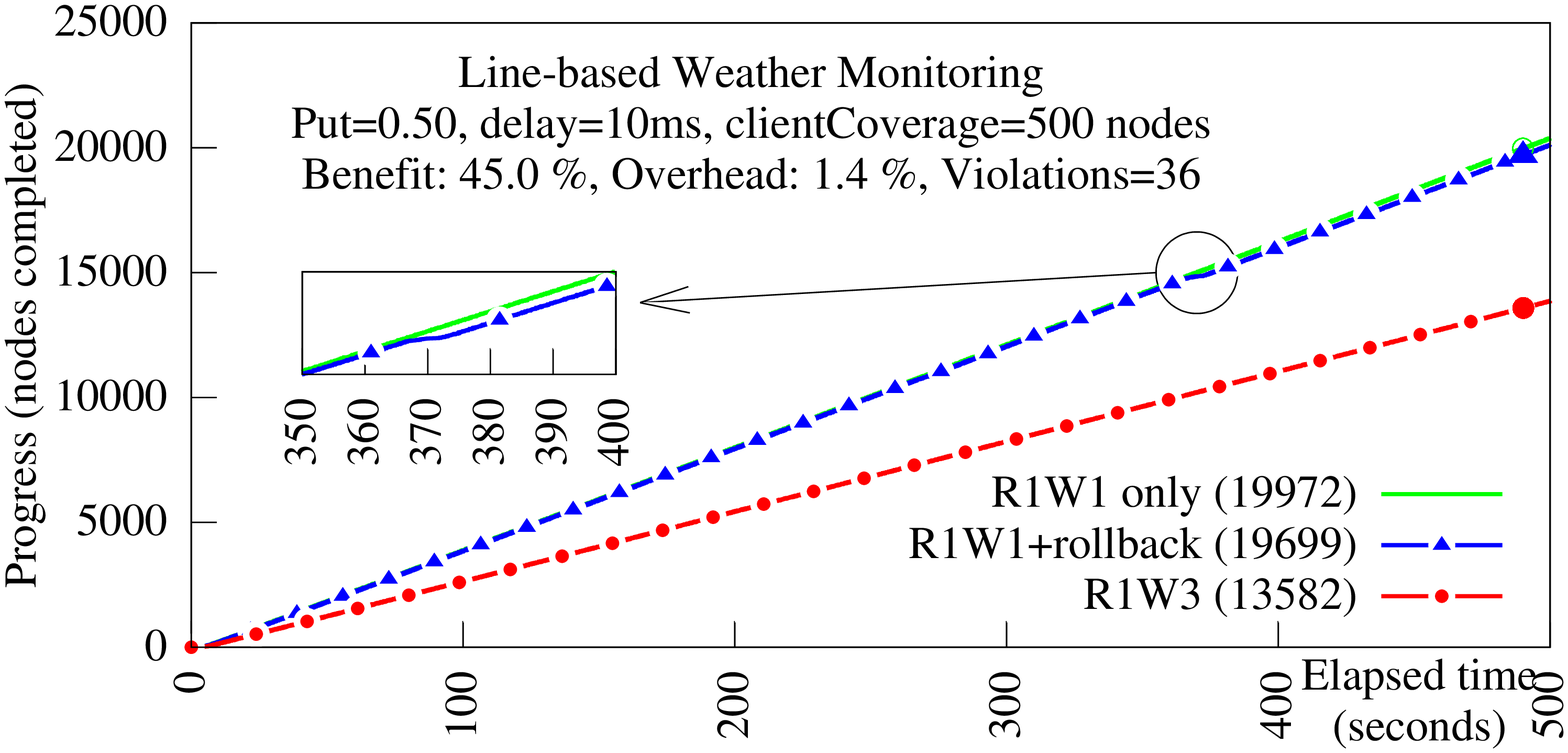}
        } \\
        \subfigure[]{%
            \label{fig:node-progress-weather-grid-gridSize500}
            \includegraphics[width=0.45\textwidth]{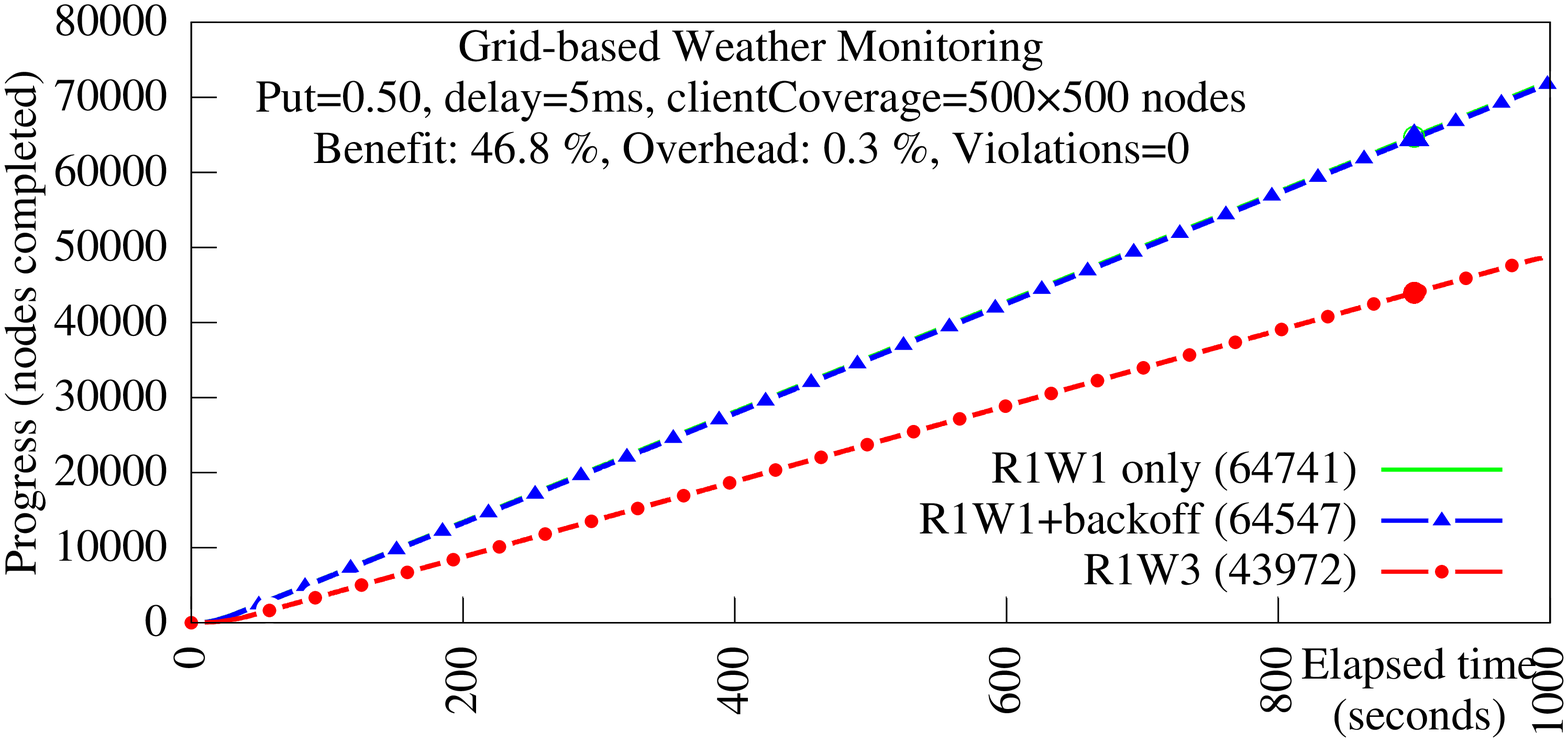}
        }
        \subfigure[]{%
            \label{fig:node-progress-weather-grid-gridSize300}
            \includegraphics[width=0.45\textwidth]{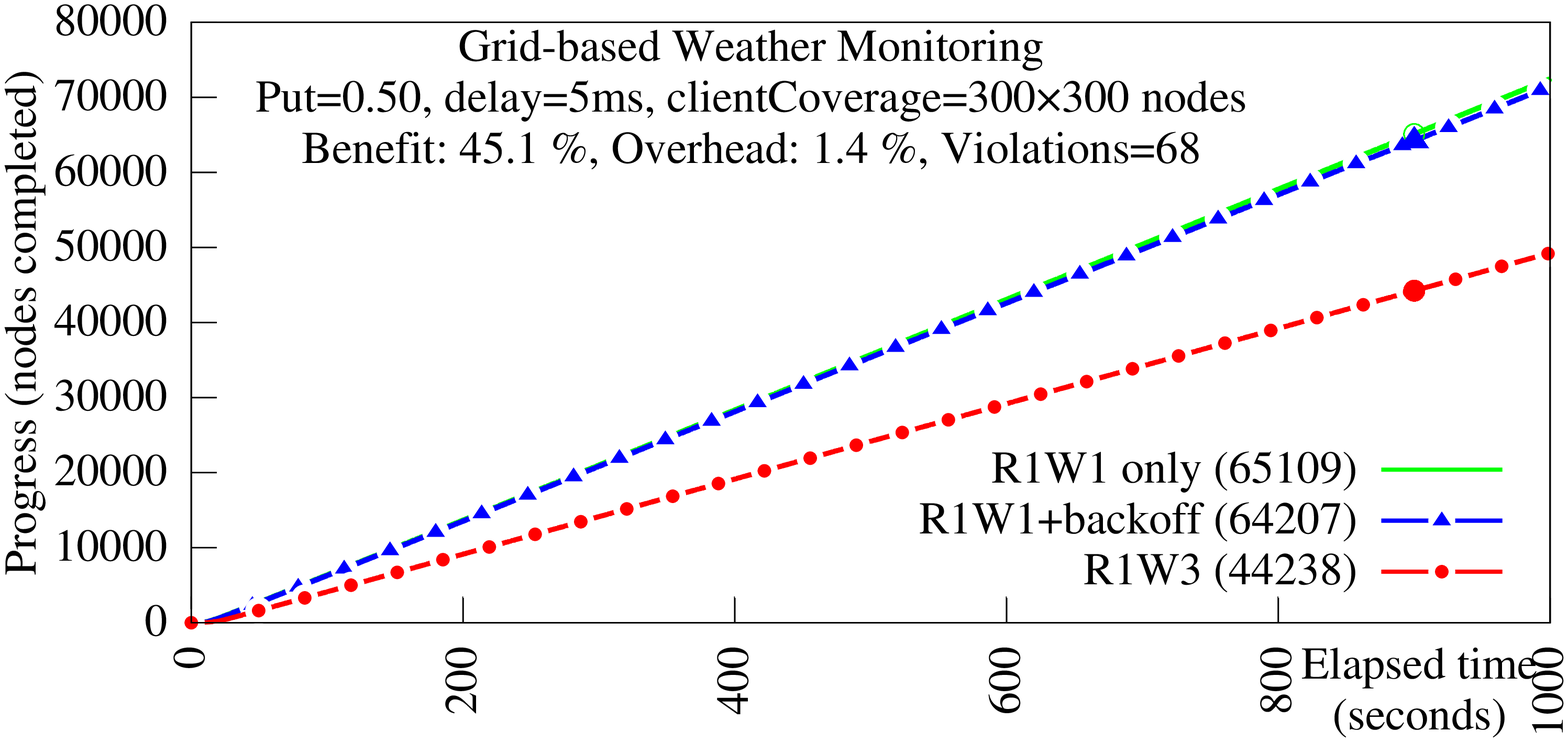}
        }
    \end{center}
    \caption{The benefit and overhead of Eventual consistency+Rollback vs. Sequential consistency in \textit{Weather Monitoring} application.  The inset figure within Figure \ref{fig:node-progress-weather-line-put0.50-delay10ms-lineSize500} is a close-up view showing the impact of rollback. The larger points near the end of each data sequence are where we choose the representative values for the data sequences.}
    \label{fig:node-progress-weather}
\end{figure*}

\textbf{Weather Monitoring.} 
When running the \textit{Weather Monitoring} application with eventual consistency, first, we consider the nodes organized in a line. In this case, the application progressed \SI{47.2}{\percent} faster than running on sequential consistency (cf. Figure \ref{fig:node-progress-weather-line-put0.50-delay10ms-lineSize1000}).
Even if we extend it to a grid graph, the results are similar. In Figure \ref{fig:node-progress-weather-grid-gridSize500}, we find that in the grid graph, the application progressed \SI{46.8}{\percent} faster under eventual consistency than in sequential consistency. In both of these executions, no violations were detected in the \SIlist{500; 1000}{\s} window, respectively.

To evaluate the effect of rollbacks, we increase the chance of conflicts by reducing the coverage of each client (i.e. the number of nodes in the graph assigned to each client) so that the clients work on bordering nodes more frequently. In that setting, on a line graph, eventual consistency still progressed about \SI{45}{\percent} faster than running on sequential consistency (cf. Figure \ref{fig:node-progress-weather-line-put0.50-delay10ms-lineSize500}), even though we had a substantial number of rollbacks (\num{36} in \SI{500}{\s}). The detection latency for violation was on average \SI{18}{\ms}. The worst case detection latency was \SI{55}{\ms}. We note that the application motivated by \textit{Weather Monitoring} is a non-terminating application which keeps running without termination. Hence, the number of nodes processed measured in stable phase reflects the overall progress of the application. For this reason, in order to compare the progress of different experiment configurations, we measure the progress made by the clients after the same execution duration. For example, in Figure \ref{fig:node-progress-weather-line-put0.50-delay10ms-lineSize1000}, the larger points on each line are where we measure the progress after the execution has run for \SI{490}{\s}.
Figure \ref{fig:node-progress-weather-line-put0.50-delay10ms-lineSize500} also considers the progress made by the application on eventual consistency without rollback or monitoring. 
%In this part, violations are detected but ignored {\color{red} (Indeed we do not run the monitors at all. So the difference is the cost of monitors and rollbacks)}. 
Thus, the resulting answer may be incorrect. The reason for this analysis is to evaluate the cost of monitoring and rollback. As shown in Figure \ref{fig:node-progress-weather-line-put0.50-delay10ms-lineSize500}, the cost of rollback is very small. Specifically, with rollback, the number of nodes processed decreased by about \SI{1.4}{\percent}.

In grid-based graphs, eventual consistency progressed \SI{45.1}{\percent} faster than sequential consistency did (cf. Figure \ref{fig:node-progress-weather-grid-gridSize300}) even though it had to rollback a number of times (68 times in \SI{1000}{\s}). The detection latency was \SI{10}{\ms} on average, and \SI{41}{\ms} in the worst case. The cost of rollback was \SI{1.4}{\percent}.
%Regarding the cost of rollback, the number of nodes processed under eventual consistency with rollback is 1.4\% slower than under eventual consistency without rollback.

%Conjunctive: perpetual application. The goal is to stress the monitors, and we do not perform rollback since the number of violations is too large

\textbf{Social Media Analysis. }
Since the \textit{Weather Monitoring} task is a non-terminating task, its behavior remains the same throughout the execution. Hence, to evaluate the effect of termination, we evaluate our approach in the \textit{Social Media Analysis} application. Terminating computation suffer from the following when compared with non-terminating computations: (1) At the end, some clients may have completed their task thereby reducing the level of concurrency, and (2) The chance of rollback resulting in the same conflict increases, as the tasks remaining are very small. Therefore, the computation after rollback is more likely to be similar to the one before the rollback. In other words, the conflict is likely to recur.

%We consider Social Media Analysis as a finite task with termination. Hence we use the completion time as the metric for comparison.

%% old text: only has the beginning of the computation, not the complete computation
%We run this recovery algorithm on eventual consistency for the \textit{Social Media Analysis} application in our local lab network with the round-trip latency of $100 \ ms$. There are nine recovery instances reported during the experiment. As shown in Figure \ref{fig:computation-progress}, even with the cost of monitoring and recovery from violations, the computation on eventual consistency progresses $60\%$ to $86\%$ faster than that on sequential consistency.

\begin{figure*}[ht]
    \begin{center}
        % \subfigure[]{%
        %     \label{fig:node-progress-graph-20ms-adaptive}
        %     \includegraphics[width=0.3\textwidth]{figures-delay/{node-progress-graph-20ms-adaptive}.eps}
        % }
        \subfigure[]{%
            \label{fig:node-progess-50ms}
            \includegraphics[width=0.45\textwidth]{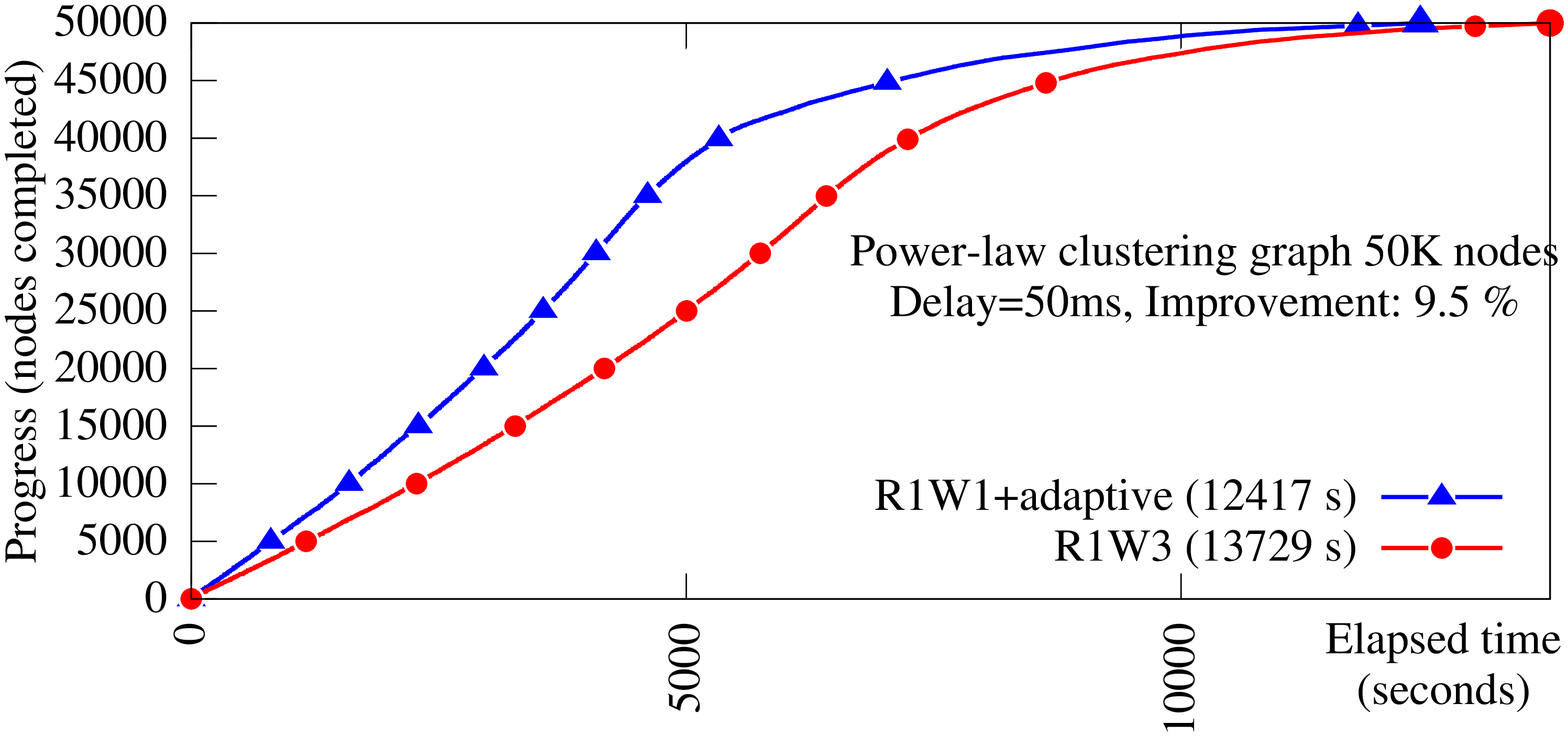}
        }
        \subfigure[]{%
           \label{fig:node-progress-regular-graph-50ms}
           \includegraphics[width=0.45\textwidth]{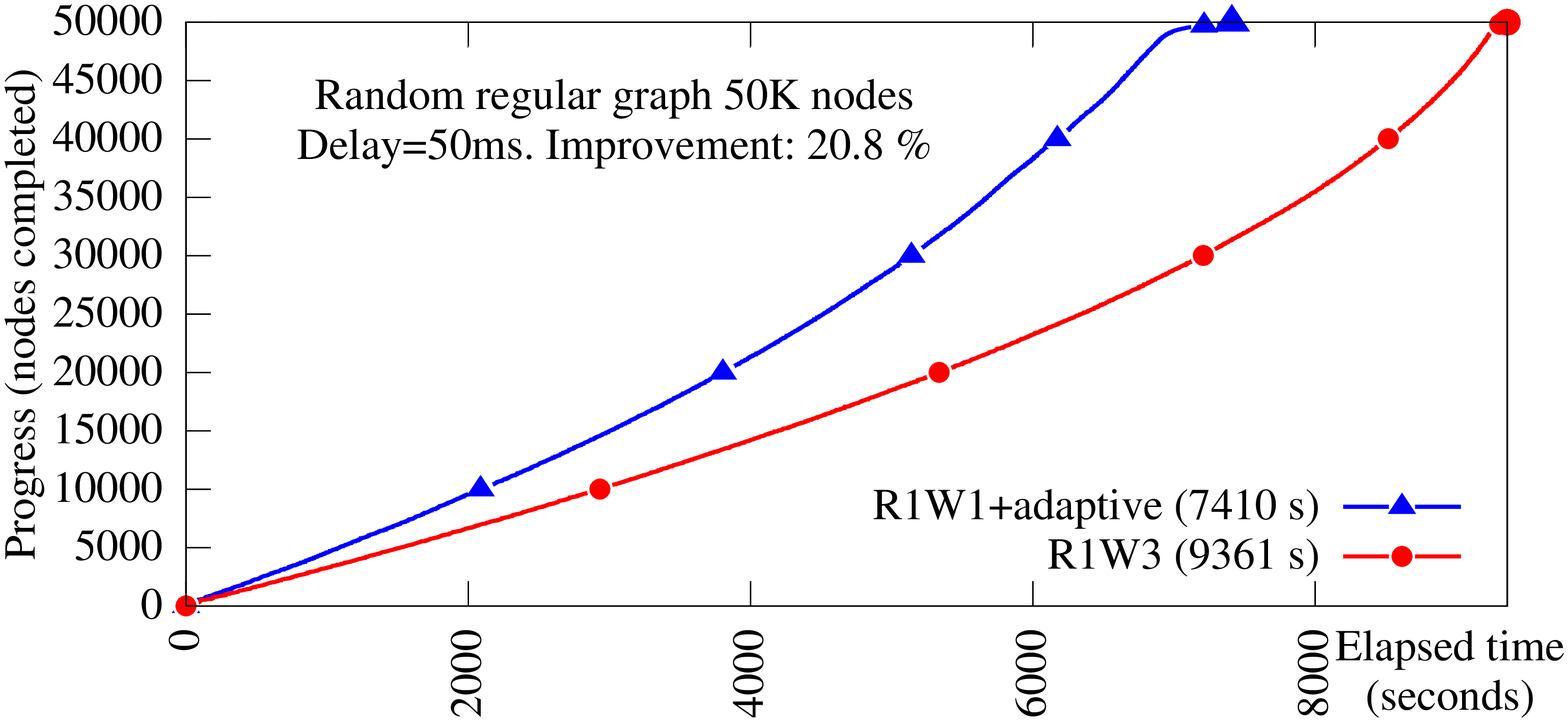}
        } 
    \end{center}
    \caption{Comparing the completion time of Sequential Consistency (R1W3) vs. Eventual Consistency with rollback and adaptive consistency (R1W1+adaptive) in \textit{Social Media Analysis} application. On a power-law clustering graph, before 90\% of the nodes are processed, R1W1+adaptive progresses about 18\% faster than R1W3. Overall, R1W1+adaptive is $9.5\%$ faster than R1W3. On a regular random graph, the benefit before $90\%$ of the nodes are processed is $26\%$ and the overall benefit is $20.8\%$.}
    \label{fig:computation-progress}
\end{figure*}

We evaluate the effect of termination in two types of graph: (1) Power-law clustering (cf. Figure \ref{fig:node-progess-50ms}), and (2) Regular graphs (cf. Figure \ref{fig:node-progress-regular-graph-50ms}) where degrees of all nodes are \textit{close}. (The details of these graphs is given in Section \ref{subsec:experiment-setup}.)  

On power-law clustering graphs, as shown in Figure \ref{fig:node-progess-50ms}, before the execution reached $90\%$ completion of the work, eventual consistency -- even with the cost of monitoring and rolling back -- progressed about $18.5\%$ faster than sequential consistency. However, in the remaining $10\%$ of the work, when there were a few nodes to be colored, the chance of conflict increased. Furthermore, the same conflict occurred after rollback as well. Hence, in the final phase, execution under eventual consistency almost stalled due to frequent rollbacks. When the clients utilized adaptive consistency then they could make progress through the final phase and finished about $9.5\%$ faster than sequential consistency. We note that the decline in computation rate in the final phase is also true for sequential consistency, and that is related to a property of power-law cluster graph that some nodes are high degree nodes. In regular random graph, we do not observe this decline as shown in Figure \ref{fig:node-progress-regular-graph-50ms}. The main reason for this is that the likelihood of conflict in the power-law graph is high since there are several nodes with a high degree. Furthermore, it is difficult to distribute the workload of power-law clustering graph to the clients evenly. Therefore, in the final phase, some clients have completed before the others, thus reducing the parallelism. By contrast, in the regular graph, the likelihood of conflict in end stages remains the same and the workload can be evenly distributed among the clients. On a regular graph, eventual consistency with monitoring and rollback was $26\%$ faster than sequential consistency before $90\%$ of the nodes were processed, and $20.8\%$ faster overall (cf. Figure \ref{fig:node-progress-regular-graph-50ms}).

% \begin{figure}[tbp]%[h]
%  \vspace{-10pt}
% \centering
% \includegraphics[width=0.45\textwidth]{figures-delay/{Benefit-task-progress-coloring-N3R1W3-N3R1W1-C30-put0.50-uniform-getAllfalse-prepHighDegreetrue}.eps}
% \caption{Comparing computation progress of sequential consistency vs. eventual consistency with monitoring and recovery}
% \vspace{-10pt}
% \label{fig:computation-progress}
% \end{figure}

%% bar graph
% \begin{figure}[tbhp]%[h]
%  \vspace{-10pt}
% \centering
% \includegraphics[width=0.45\textwidth]{figures-delay/{result_summary_20ms}.eps}
%     \caption{Comparison of Social Media Analysis application progress between Eventual consistency + adaptive vs. Sequential consistency.}
%     \label{fig:social-complete-percent-50ms}
% \end{figure}

% \subsection{NEW CONTENT TO BE ADDED. BLUE TEXT COULD BE IN A NEW SECTION}

% {\bf{Analytical}} If the error/violation rate is $r_{err}$, the average completion time without recovery in eventual consistency is $T_{free}^{eventual}$, then the expected completion time with recovery in eventual consistency with error rate $r_{err}$ is
% $$T_{err}^{eventual} = \dfrac{T_{free}^{eventual}}{1 + r_{err}}$$

% {\bf{Experiments} (3/4-1 page)}
% \begin{itemize}
% \item Verify the Analytical prediction about the expected completion time where violation rate is $r_{err}$
% \item Impact of latency in our approach:
% a graph where x-axis is latency (controlled by proxy), y-axis is the completion time.

% A graph where the x-axis is latency (controlled by proxy), the y-axis is the speedup (like the speedup measurement in Figure \ref{fig:computation-progress}.

% \item Extend the results in table \ref{table:local-overhead-benefit} into a graph: x-axis is network latency. the y-axis is the benefit. For each value of network latency (x-value), we have 3 columns corresponding to the benefits measured in three applications (Conjunctive, Weather monitoring, Social Media Analysis).

% \end{itemize}

\subsection{Discussion}
\label{subsec:discussion}

In this section, we consider some of the questions raised by this work including questions raised by the reviewers of LADC 2018 and JBCS.

\textbf{What is the likely effect of the number of clients on the probability of rollback?}
First, we note that for linear predicates (e.g. conjunctive predicates), when the number of clients increases, the number of violations decreases as it is less likely to find a consistent snapshot where the local predicate at every client is true \cite{YNVSD16RV}. As a result, if we increase the concurrency level, the probability of rollback decreases.
Hence in the following discussion, we limit the context to semi-linear predicates (e.g. mutual exclusion).

In general, the probability of rollback depends upon the probability that two clients are updating conflicting data. Thus, if the number of clients is too large when compared with the size of the graph (i.e. the coverage of a client is too small), the probability of conflict/rollback is high. We have validated this with experimental analysis of applications motivated by \textit{Weather Monitoring} in section \ref{subsec:applications}. However, in a typical deployment, the coverage of a client is usually large enough (e.g. thousands of nodes) that the chance of two clients concurrently working on neighboring nodes is small. Furthermore, when working on neighboring nodes, clients utilize mutual exclusion mechanism such as Peterson locks to prevent conflicts. Conflicts/rollbacks only happen when there is some data inconsistency related to the mutual exclusion mechanism that causes the clients concurrently updating neighboring nodes. In eventual consistency, data inconsistencies exist but are rare \cite{dynamo} and usually involve hardware and/or network failures. Hence, from our analysis, we anticipate that the probability of rollback is small given that each client is assigned a reasonable workload.

\textbf{How do the observations in this paper relate to the CAP theorem?}
%
%\emph{In the very beginning, the CAP theorem is mentioned, but not referred to anymore in the paper. A more detailed discussion about how it is tackled with should be included.}
%
%$\Rightarrow$ 
When latency increases, we are simulating pseudo network partition. In this case, which consistency level is better depends upon the configuration of the Voldemort servers. 

As an illustration, consider the example where we have 5 servers and we use R2W4. Furthermore, suppose that one of the servers is partitioned from the other 4 servers.

With sequential consistency, the 4 servers and their associated clients can still make progress correctly. The partitioned server and its clients would not make progress. Hence, they could be considered as being dead. By some detection mechanism, we can detect such partitioning and assign the tasks of dead clients to other clients. And the computation could progress to the end.

With eventual consistency, all 5 servers and their clients make progress but the clients will process based on stale data. When the network is recovered, the data inconsistencies will invalidate the computation results of both sub-networks.
%(i.e. if we abort all computation results of one partition, the result of the other partition is still correct). 
%In that case, the server in the smaller partition will rollback its execution and copy data from servers in the larger partition. 
%
%If the monitor can detect such partition early, the server in the smaller partition can decide to wait until the network is resumed to normal condition, thus reducing the cost of rollback.
%
When the monitor detects such partition and inconsistencies, we will have to rollback the whole systems, including the 4 servers and their clients (even though their results are correct, given that the partitioned server has been rolled back).

While the above discussion applies to R2W4, if the system used R1W5 then neither eventual consistency nor sequential consistency could make progress. This is because eventual consistency would result in inconsistently updated replicas. These inconsistencies would be resolved based on the implementation of Voldemort (e.g., latest write wins, minority replicas follow the majority replicas, etc). However, this conflict resolution may not be consistent with the needs of the application. And, in sequential consistency, no write operation would succeed. 
However, if the nodes are not partitioned but rather suffer from a high delay (but no partition), eventual consistency may be able to make progress. However, it would need to rollback frequently. By contrast, in sequential consistency, it is likely that most write operations fail (as they take too long to complete). Consequently, sequential consistency will not be able to make progress. 

What the above discussion suggests is that when the delays are very high, the above approach would work for some configurations (e.g., R2W4) but not for others (e.g. R1W5). 
Hence, one of the future work in this area is to allow only certain clients to rollback while allowing others to continue without rollback.

\textbf{Applications that cannot be rolled back.}
In this paper, we assume the application has exclusive access to its data. Specifically, before the application finishes, other applications will not read this application results. If the data is shared and used by multiple applications, then the rollback approach is not suitable since it is almost impossible to rollback other applications. For instance, the results of computing shortest paths, routing information can be produced by one application and used by other applications. In this case, other approaches such as self-stabilization can be useful.

%% file: related.tex
%Related Works

\subsection{Predicate Detection in Distributed Systems}

Predicate detection is an important task in distributed debugging. An algorithm for capturing consistent global snapshots and detecting stable predicates was proposed by Chandy and Lamport \cite{CL85TCS}.
A framework for general predicate detection is introduced by Marzullo and Neiger \cite{MN91WDAG} for asynchronous systems, and Stollers \cite{Stoller00DC} for partially synchronous systems.
These general frameworks face the challenge of state explosion as the predicate detection problem is NP-hard in general \cite{CG98DC}.
However, there exist efficient detection algorithms for several classes of practical predicates such as unstable predicates \cite{GW92ICFSTTCS, GW94TPDS, GW96TPDS}, conjunctive predicates \cite{GC95ICDCS, GCMK95HICSS}, linear predicates, semilinear predicates, bounded sum predicates \cite{CG98DC}.
Some techniques such as partial-order method \cite{SUL00CAV} and computation slicing \cite{MG05DC, CGNM13SRDS} are also approaches to address the NP-Completeness of predicate detection.
Those works use vector clocks to determine causality and the monitors receive states directly from the constituent processes. Furthermore, the processes are static. \cite{WMHG09ICPADS, WMH10ICPADS} address the predicate detection in dynamic distributed systems. However, the class of predicate is limited to the conjunctive predicate.
In this paper, our algorithms are adapted for detecting the predicate from only the states of the servers in the key-value store, not from the clients. The servers are static (except failure), but the clients can be dynamics. The predicates supported include linear (including conjunctive) predicates and semilinear predicates.

In \cite{VK2018SIROCCO, VYKTD2017RV}, the monitors use Hybrid Logical Clock (HLC) to determine causality between events in a distributed execution. HLC has the advantage of low overhead but suffers from false negatives (some valid violations are not detected). In contrast, we use hybrid vector clocks to determine causality in our algorithms. In \cite{YNVSD16RV}, the authors discussed the impact of various factors, among which is clock synchronization error, on the precision of the monitors. In this paper, we set epsilon at a safe upper bound for practical clock synchronization error to avoid missing potential violations. In other words, a hybrid vector clock is practically a vector clock. Furthermore, this paper focuses on the efficiency and effectiveness of the monitors. 
Bloom Clock \cite{Ramabaja2019CoRR} is another alternative to vector clock. Due to the overhead of the counting Bloom filter, the benefit of Bloom clock only payoffs on very large distributed systems.

\subsection{Distributed data-stores}
Many NoSQL data-stores exist on the market today, and a vast portion of these systems provide eventual consistency. The eventual consistency model is especially popular among key-value and column-family databases. The original Dynamo~\cite{DHJKLPSVV07SOSP} was one of the pioneers in the eventual consistency movement and served as the basis for Voldemort key-value store. Dynamo introduced the idea of hash-ring for data-sharding and distribution, but unlike Voldemort it relied on server-side replication instead of active client replication.  Certain modern databases, such as Cosmos DB and DynamoDB ~\cite{CosmosDB, DynamoDB} offer tunable consistency guarantees, allowing operators to balance consistency and performance. This flexibility would enable some applications to take advantage of optimistic execution while allowing other applications to operate under stronger guarantees if needed.  However, many data-stores ~\cite{CDEFFFGGHH13TOCS,CDGHWBCFG08TOCS} are designed to provide strong consistency and may not benefit from optimistic execution module. 

Aside from general purpose databases, a variety of specialized solutions exist. For instance, TAO ~\cite{BACCDDFGKL13ATC} handles social graph data at Facebook. TAO is not strongly consistent, as its main goal is performance and high scalability, even across datacenters and geographical regions. Gorilla ~\cite{PFTCHMV15VLDB} is another Facebook's specialized store. It operates on performance time-series data and highly tuned for Facebook's global architecture. Gorilla also favors availability over consistency in regards to the CAP theorem. 
Crail-KV \cite{BCCBK2018IPCC} is Samsung's extension for Apache Crail data storage system \cite{STPSMIK2017DEB} that leverages recent advances in hardware technology, especially key-value solid state drive, to provide higher I/O performance for distributed data store.

Various consistency models in distributed system are presented in the survey \cite{ADMG2019Corr}. In \cite{KKW2019Corr}, the authors introduce the notion of \textit{Fluctuating Eventual Consistency} which is the mix of eventual consistency and strong consistency in order to provide stronger guarantee for eventual consistency. However, this correctness property is not suitable for the adaptive behaviour of application since it is not sufficient to prevent violations as sequential consistency does, and it has more extra synchronization effort than eventual consistency. Consistify \cite{SMG2019ICDCN} is a framework that supports tuning the consistency level of a distributed data store. However, Consistify has to statically analyzes the semantics of the application.

\subsection{Snapshots and Reset}
The problem of acquiring past snapshots of a system state and rolling back to these snapshots has been studied extensively. Freeze-frame file system ~\cite{SGBCX16SoCC} uses Hybrid Logical Clock (HLC) to implement a multi-version Apache HDFS. Retroscope ~ \cite{CADK17ICDCS} takes advantage of HLC to find consistent cuts in the system\textquotesingle s global state by examining the state-history logs independently on each node of the system. The snapshots produced by Retroscope can later be used for node reset by simple swapping of data-files.
Eidetic systems ~\cite{DCDFC14OSDI} take a different approach and do not record all prior state changes. Instead, the eidetic system records any non-deterministic changes at the operating system level and constructing a model to navigate deterministic state mutations. This allows the system to revert the state of an entire machine, including the operating system, data and applications, to some prior point. 
Certain applications may not require past snapshots and instead need to quickly identify consistent snapshots in the presence of concurrent requests affecting the data. VLS ~\cite{CSHF16ICDE} is one such example designed to provide snapshots for data-analytics applications while supporting high throughput of requests executing against the system.

\subsection{Distributed Data Processing}
MapReduce \cite{MapReduce2008CACM}, DataFlow \cite{DataFlow2015VLDB} are general-purpose distributed data processing frameworks. In the realm of distributed graph processing, many frameworks are available such as Pregel \cite{Pregel2010SIGMOD}, GraphLab \cite{GraphLab2012VLDB}, GraphX \cite{GraphX2014OSDI}, and PowerGraph \cite{PowerGraph2012OSDI}. In those works, data is persisted in semi-structural storages such Google File System, Hadoop Distributed File Systems \cite{GFS2003ACM}, BigTable \cite{BigTable2008TOCS}, or in in-memory storage such as Spark \cite{Spark2012Usenix}. Our work focuses on the no-structure key-value stores and the impact of different consistency models on key-value store performance. Our approach's usefulness is also not limited to graph applications.

%% file: conclusion.tex
Due to limitations of the CAP theorem and the desire to provide availability/good performance during network partitions (or long network delays), many key-value stores choose to provide a weaker consistency such as eventual or causal consistency. This means that the designers need to develop new algorithms that work correctly under such weaker consistency models. An alternative approach is to run the algorithm by ignoring that the underlying system is not sequentially consistent but monitoring it for violations that may affect the application. For example, in the case of graph-based applications (such as those encountered in \textit{Weather Monitoring}, \textit{Social Media Analysis}, etc.), each client operates on a subset of nodes in the graph. It is required that two clients do not update two neighboring nodes simultaneously. In this case, the predicate of interest is that the local mutual exclusion is always satisfied.

We demonstrated the usage of this approach in the Voldemort key-value store. We considered two types of predicates: conjunctive predicates and semi-linear predicates (such as that required for local mutual exclusion). 
We evaluated our approach using Amazon AWS for graph applications motivated by \textit{Social Media Analysis} and \textit{Weather Monitoring}. 
Our approach improved the client throughput performance by $50\%$ -- $80\%$.
%our approach improves the throughput performance of the computation from 50\% to 80\%. 
Furthermore, we find that the number of violations of predicates of interest was infrequent. Violations were also detected promptly. When all clients and servers were in the same region, the violations were detected within \SI{50}{\ms} whereas if they were in different regions, time for detection was higher. For example, in a network where clients and servers were located in Frankfurt Germany, Ohio USA, and Oregon USA, violations were detected in less than \SI{3}{\s}.
In this context, the time required for a client to work on one task was at least \SI{22}{\s} and was on average \SI{45}{\s}. Thus, detection latency was significantly lower than the time for processing a task.

%In a regional network, violations are detected within $50 \ ms$ while in a global network, they can be detected within $5 \ s$.\todo{fix}

%Thus, the amount of work wasted due to rollback would be very small especially if one utilizes techniques such as Retroscope \cite{CADK17ICDCS} that allows one to roll back the system to an earlier state on-demand. For graph processing applications such as \textit{Social Media Analysis} or \textit{Weather Monitoring}, if we defer clients' updates until the end of a task, the recovery can be achieved by restarting the task without state rollback.

We developed an efficient rollback algorithm for graph-based applications with the assumption that all violations are detected quickly enough. Our rollback algorithm has mechanisms to handle livelocks (i.e. multiple rollbacks caused by a recurring violation) such as back-off and adaptive consistency where clients switch from eventual consistency to sequential consistency if violations are frequent. We observe that livelocks occur at the end of terminating computation. This is due to the fact that, in a graph processing application, when the computation is about to terminate, there are only a few nodes of the graph that need to be processed. Hence, if a conflict occurs between two clients $C1$ and $C2$, computation after their rollback is likely to have the same conflict again, as each client has only a very small set of nodes to be processed. In this case, without a livelock mechanism, eventual consistency will fail to process all the nodes. Adaptive consistency is also useful in scenarios where the network condition is unstable for an extended period of time. In this case, data inconsistencies are likely to happen and the clients process stale information and produce incorrect results. By switching to sequential consistency, some clients can make progress while some other clients those do not make progress also do not produce conflicting data.
Since Voldemort uses active replication (where clients are responsible for replication), such an adaptive approach can be implemented by clients alone without any changes to the underlying server architecture. If passive replication were used, implementation of an adaptive approach would require servers to perform such a change.

We demonstrated the benefit of using eventual consistency with monitoring and rollback. On non-terminating applications such as those motivated by \textit{Weather Monitoring}, our approach was $45\%$ -- $47\%$ faster than running the application on sequential consistency, even occasional rollbacks occurred during the execution. Furthermore, the cost of the monitors and rollback was as low as $1.4\%$.
On terminating applications such as those motivated by \textit{Social Media Analysis}, adaptive consistency is required as eventual consistency fail to process all nodes. For this reason, the overall benefit is reduced. Specifically, when 90\% of the nodes were processed, the benefit was $19\%$ -- $26\%$. However, since it needed to switch to sequential consistency at the end due to excessive recurring violations, the final benefit was reduced to $10\%$ -- $20\%$.

There are several possible future extensions of this work. Currently, the adaptive solution switches from eventual consistency to sequential consistency based on the feedback from monitors. It is possible that the increase in conflicts is temporary due to network issues. When the condition is resumed to normal, it would be beneficial to run in eventual consistency again. However, in sequential consistency, monitors are not required and, thus, there is no feedback mechanism to determine when using eventual consistency is reasonable. One needs to develop new techniques to permit this possibility. 

%It is possible that the increase in conflicts is temporary due to network issues. When the condition is resumed to normal, it would be beneficial to loosen the client mode back to eventual consistency. It is one of our future work to identify some effective mechanism for determining whether it is safe to have clients running on eventual consistency. We note that we may not be able to use the monitor feedback as input for the mechanism because in sequential consistency, no conflict/violation would happen, except false alarms. One possibility is probation in which after escalating to sequential consistency, the clients wait for some time $T_{wait}$ and try eventual consistency. If violations recur quickly, clients will double the wait time before the next try.

Another issue is that the monitors used in this work suffer from false positives, i.e., they initiate rollback when it was not absolutely necessary. One possible reason for false positives is that the clients, say $C1$ and $C2$, involved in rollback had only read from the key-value store. In this case, one of the clients can continue the execution without rollback. However, in our implementation, as each client rolls back independently, both of them rollback. If this is prevented, it can not only reduce the wasted work, it can also potentially avoid re-occurrence of conflict between $C1$ and $C2$ after rollback.
%
%Tradeoff between cost and false positives
Another reason for false positives is the impedance mismatch in the synchrony assumptions made by the monitors and the applications \cite{YNVSD16RV}. In order to reduce or eliminate the false positives, we would have to augment the clients and servers with more information and the monitors would have to examine the candidates more extensively. Consequently, that would increase the cost of monitoring but reduce the need for performing rollback.

%Furthermore, it is also feasible to utilize these monitors for the sake of debugging as well. In particular, the overhead of the monitors is very low. The overhead is typically less than $4\%$ and in stressed experiments less than $8\%$.

%There are several possible future works in this area. This paper considered linear and semilinear predicates. In general, the problem of predicate detection is NP-complete. 
%Hence, we intend to evaluate the practical cost of these algorithms. 
%Hence, we intend to evaluate the practical cost of general predicate detection algorithms.
%We are also working on making these algorithms more efficient by permitting them to occasionally detect phantom violations. We are evaluating whether this increased efficiency would be worthwhile even though some unnecessary rollbacks may occur. 

The rollback algorithm proposed in this paper is specific for graph-based application and has the assumption on small detection latency. For a general application, we are investigating the possibility of integrating the monitor with Retroscope \cite{CADK17ICDCS} to automate the rollback and recovery.

%The problem of general predicate detection is NP-hard but can be weakened to other classes of predicates.

%Open problem of how to do the recovery.